\documentclass[12pt,epsf,amssymb,qsymbols]{article}
\usepackage{tabularx}
\usepackage{array}
\usepackage{graphics}
\usepackage{graphicx}
\usepackage{psfrag}
\usepackage{epsfig}
\usepackage{amsmath}
\usepackage{amssymb}
\usepackage{ulem}
\usepackage{setspace}
\usepackage{rotating}
\usepackage{colortbl}
\usepackage{tabularx}
\usepackage{longtable}
\usepackage{multirow}
\makeatletter

\usepackage{verbatim}

\setlongtables

\newcommand{\msbar}{{\overline{\rm MS}}}

\newcommand{\bea}{\begin{eqnarray}}
\newcommand{\eea}{\end{eqnarray}}
\newcommand{\beq}{\begin{equation}}
\newcommand{\eeq}{\end{equation}}
\newcommand{\ec}{\end{center}}
\newcommand{\bc}{\begin{center}}
\newcommand{\gev}{{\rm GeV}}

\newcommand{\pdir}{p\kern -5.2pt\raise 0.2ex\hbox {/}}
\newcommand{\one}{1\hspace*{-1.05mm} \hbox {I}}
\newcommand{\vdir}{v\kern -5.75pt\raise 0.15ex\hbox {/}}
\newcommand{\kdir}{k\kern -5.75pt\raise 0.15ex\hbox {/}}
\newcommand{\epsdir}{\epsilon\kern -5.0pt\raise 0.15ex\hbox {/}}
\newcommand{\bvdir}{\bar{v}\kern -5.75pt\raise 0.15ex\hbox {/}}
\newcommand{\Ddir}{D\kern -7.75pt\raise 0.20ex\hbox {/}}
\newcommand{\Adir}{A\kern -7.75pt\raise 0.20ex\hbox {/}}
\newcommand{\ldir}{l\kern -5.0pt\raise 0.2ex\hbox{/}}
\newcommand{\varepsdir}{\varepsilon\kern -5.5pt\raise 0.15ex\hbox{/}}

\newcommand{\m}[0]{\phantom{$-$}}

\newcommand{\n}[0]{\cellcolor[gray]{0.85}}

\newcommand{\nn}{\nonumber}
\makeatother

\begin{document}
\thispagestyle{empty} 
\begin{flushright}
\begin{tabular}{l}
LPT 09-33
\end{tabular}
\end{flushright}
\begin{center}
\vskip 3.0cm\par
{\par\centering \textbf{\LARGE  
\Large \bf On internal structure of  the heavy-light mesons }}\\
\vskip 1.25cm\par
{\scalebox{.87}{\par\centering \large  
\sc Damir Be\'cirevi\'c, Emmanuel~Chang and Alain Le Yaouanc}}
{\par\centering \vskip 0.75 cm\par}
{\sl 
Laboratoire de Physique Th\'eorique (B\^at.~210)~\footnote{Laboratoire de Physique Th\'eorique est une unit\'e mixte de recherche du CNRS, UMR 8627.}\\
Universit\'e Paris Sud, Centre d'Orsay,\\ 
F-91405 Orsay-Cedex, France.}\\
\vskip1.cm
 
{\vskip 0.35cm\par}
\end{center}

\vskip 0.55cm
\begin{abstract}
We compute the radial distributions of the vector and axial charge density as well as of the matter density in the heavy-light mesons on the lattice.  
The results for the lowest lying static heavy-light mesons and their first excited counterparts are obtained with $N_f=2$ dynamical quarks of Wilson (clover) type  and 
with the various improved static heavy-quark actions. From these distributions we were able to compute the corresponding charges and slopes of the form factors 
(i.e. $\left< r^2\right>$). Those results are presented, and briefly discussed too.
\end{abstract}
\vskip 2.2 cm 
\setcounter{page}{1}
\setcounter{footnote}{0}
\setcounter{equation}{0}
\noindent

\renewcommand{\thefootnote}{\arabic{footnote}}

\newpage
\setcounter{footnote}{0}
\section{Introduction}
Numerical simulations of QCD on the lattice are bound to provide us with information on nonperturbative dynamics among quarks and gluons.  
Not only do they allow us to compute the hadronic quantities with controllable systematic errors, they also give us access to studying the physics of confinement in more detail. 
In that respect the static limit of QCD is particularly attractive. Many studies of interaction between the two static (infinitely heavy) quarks lead to the results that helped us 
shape our understanding concerning the interquark potential (for a review see ref.~\cite{greensite} and references therein). The internal dynamics of heavy-light mesons in the static limit 
is far more complicated than that of the heavy-heavy ones, but it is  even more interesting with regard  to numerous phenomenological applications.  For example, one can calculate 
spatial distributions of the electro-magnetic charge and  of the light-quark densities, which have a very direct connection with the wave functions of quark models. Now, let us recall
 that the quark models are still a necessary tool in understanding hadronic spectra and various other properties of hadrons, including hybrid states. In fact, the lattice data show 
 a striking agreement with the wave functions predicted by certain firmly established quark models, as we shall discuss in our next paper~\cite{workinprogress} (namely, certain models 
 with Dirac equation and others formulated in the so-called Bakamjian-Thomas framework). On the other hand, this spatial distributions constitute a detailed testing ground 
 for the quark models, for their choice of parameters, as well as for the basic assumptions, and therefore may be a precious means to select among them and/or to improve them.

Recent progress in lattice QCD makes it possible to tackle this physics in more detail. In this paper we will compute the radial distribution of the matter, of the electric charge and of the axial charge 
by calculating the suitable hadronic matrix elements, which allow us to study these distributions as the light-quark is pulled away from the static source of color.
Of essential importance is to have a good behavior of the static heavy-quark propagator, i.e. that the statistical noise does not become overwhelming as the separation between the source operators is 
increased to  moderately large times. In recent years, substantial progress in that respect has been made. We will discuss the implementation of various improved static heavy-quark actions on 
the lattice, and comment on the benefit of each one of them.  In producing the heavy-light mesons we use the light-quarks with mass that is varied between $0.7\times m_s^{\rm phys}$ and $2\times m_s^{\rm phys}$, 
with $ m_s^{\rm phys}$ being the physical strange quark mass. The gauge field configurations produced by the CP-PACS Collaboration~\cite{CP-PACS} include the effect of $N_f=2$ dynamical quarks,  
the masses of which are equal to that of the valence quark whose dynamics we will be studying in this paper. Besides the usual ground states, i.e. heavy meson doublet with the quantum numbers $j^P=(1/2)^-$, 
we will also study the similar properties of their orbitally excited counterparts,  $j^P=(1/2)^+$. 

In the course of the following presentation we will occasionally make reference to some phenomenologically interesting information that comes out from our computation. 
However, a more detailed discussion concerning the comparison of our results with various quark models will be addressed in a separate paper. 

In Sec.~\ref{SecX}, we give a brief description of the lattices that we use in this work. We then discuss  various heavy (static) quark actions and discuss the results from the two-point heavy-light 
correlation functions. Description of the calculation of the  radial distribution, and the presentation of our results is made in Sec.~\ref{SecY}. There, we also present our results for the integrated 
distributions, and we finally conclude in Sec.~\ref{SecZ}. 

\section{Heavy-light mesons: two-point functions \label{SecX}}
In this section we will discuss various improvement schemes of the static heavy-quark action. We will then test the gain of such an improvement on the  two-point correlation functions involving 
the heavy-light meson source operators.  Before we embark onto that discussion we will first give a short description of the lattices that are used in this work. 
\subsection{About the lattices used in this work}
\setcounter{equation}{0}
We use the gauge field  configurations produced by the CP-PACS Collaboration, with $N_f=2$ mass-degenerate dynamical quarks:  more specifically, they are obtained by using the Iwasaki gauge 
field action and the ${\cal O}(a)$ improved Wilson quark action~\cite{CP-PACS}.  From the available data-samples we use the ones that are obtained on the $24^3\times 48$ lattices at $\beta=2.1$ 
for which the lattice spacing is $a\simeq 0.1$~fm (or $a^{-1}\simeq 2$~GeV).  From the large sample of configurations we choose the ones that are fully decorrelated. We actually take the configurations that are 
separated by  $80$ HMC trajectories. As a first step,  we reproduced the results obtained in ref.~\cite{CP-PACS}, for the fully unquenched combinations 
(i.e. those for which the mass of the sea and valence light-quarks are equal).  
The   pseudoscalar  masses and the corresponding decay constants are obtained from the fit to 
\bea \label{eeq}
\sum_{\vec x} \left< (\bar q \gamma_0 \gamma_5 q)_{\vec 0,0}  (\bar q  \gamma_0 \gamma_5 q)_{\vec x,t} \right> \longrightarrow {1\over 2} f_P^2 m_P \  \exp(-m_P t) \,,
\eea
at large time separations which are found for $t\in[10,23]$, like in ref.~\cite{CP-PACS}.  
\begin{table}[h!]\bc
\hspace*{-8mm}\begin{tabular}{|c|c|c|c|}
\hline
{\phantom{\Huge{l}}}\raisebox{-.1cm}{\phantom{\Huge{j}}}
{$\kappa_{\rm sea}= \kappa_{\rm val}$} &  {$r_0 m_P$ [ $m_P$ (GeV)]} &   {$m_P/m_V$ } &   {$r_0 f_P/Z_A$ [ $f_P/Z_A$ (GeV)]}\\
 \hline
{\phantom{\Huge{l}}}\raisebox{-.1cm}{\phantom{\Huge{j}}}
 $0.1357$ & $2.425(11)$ [$1.025(5)$]  & $0.802(4)$ &$0.649(8)$ [$0.274(3)$] \\
{\phantom{\Huge{l}}}\raisebox{-.1cm}{\phantom{\Huge{j}}}
 $0.1367$ & $2.101(9)$ [$0.888(4)$] & $0.751(5)$&  $0.599(8)$ [$0.253(4)$] \\
{\phantom{\Huge{l}}}\raisebox{-.1cm}{\phantom{\Huge{j}}}
  $0.1374$ & $1.800(10)$ [$0.761(4)$] & $0.690(8)$ & $0.576(17)$ [$0.243(7)$] \\
{\phantom{\Huge{l}}}\raisebox{-.1cm}{\phantom{\Huge{j}}}
  $0.1382$ & $1.332(10)$ [$0.563(4)$] & $0.472(12)$ & $0.525(18)$ [$0.222(8)$] \\
\hline 
\end{tabular}
\caption{\label{tab:0}
\footnotesize  The values for the light pseudoscalar masses and decay constants obtained from the standard exponential fit to the two-point correlation functions, on the configurations 
produced by CP-PACS, at $\beta=2.1$ with Iwasaki gauge and the Wilson clover ($c_{SW}=1.47$) action. }
\vspace*{-.3cm}
\ec
\end{table}
In Table~\ref{tab:0} we list those results together with the corresponding vector meson masses, which  
 we computed from the correlation function similar to that in eq.~(\ref{eeq}), after replacing $\gamma_0\gamma_5 \to \gamma_i$. 
 We used the values of $r_0/a$ from ref.~\cite{CP-PACS} and $r_0=0.467$~fm, to convert to physical units.

\subsection{Static heavy-quark on the lattice}
The notorious problem that troubled lattice practitioners for many years was the fact that the signal of the heavy-light correlation functions with infinitely heavy-quark was prohibitively noisy even 
at  moderately large time separations. To reach the ground state for a static heavy-light meson it was necessary to implement some kind of smearing 
and hope that the lowest lying state was indeed the one observed from the effective mass plots at short time separations between the smeared sources in two-point correlation functions. 
Checking whether or not the smearing procedure is efficacious, by confronting to the correlation functions computed with the local sources, was practically impossible. 

Over the past several years that problem has been overcome by various recipes to improve the static quark propagator, i.e. the Wilson line.  The implementation 
of the so-called ``FAT-link" procedure~\cite{fat-6}, and even more optimized versions with hyper-cubic blocking (HYP)~\cite{hasenfratz} resulted in an exponential improvement of the signal to noise ratio  of the static-light 
correlation functions~\cite{exponential}, making the corresponding physics results far more reliable than they were in the past. 
In the following we briefly describe these procedures  and illustrate the benefit in using them.

\begin{figure}[t!]
\begin{center}
\resizebox{90mm}{!}{\includegraphics{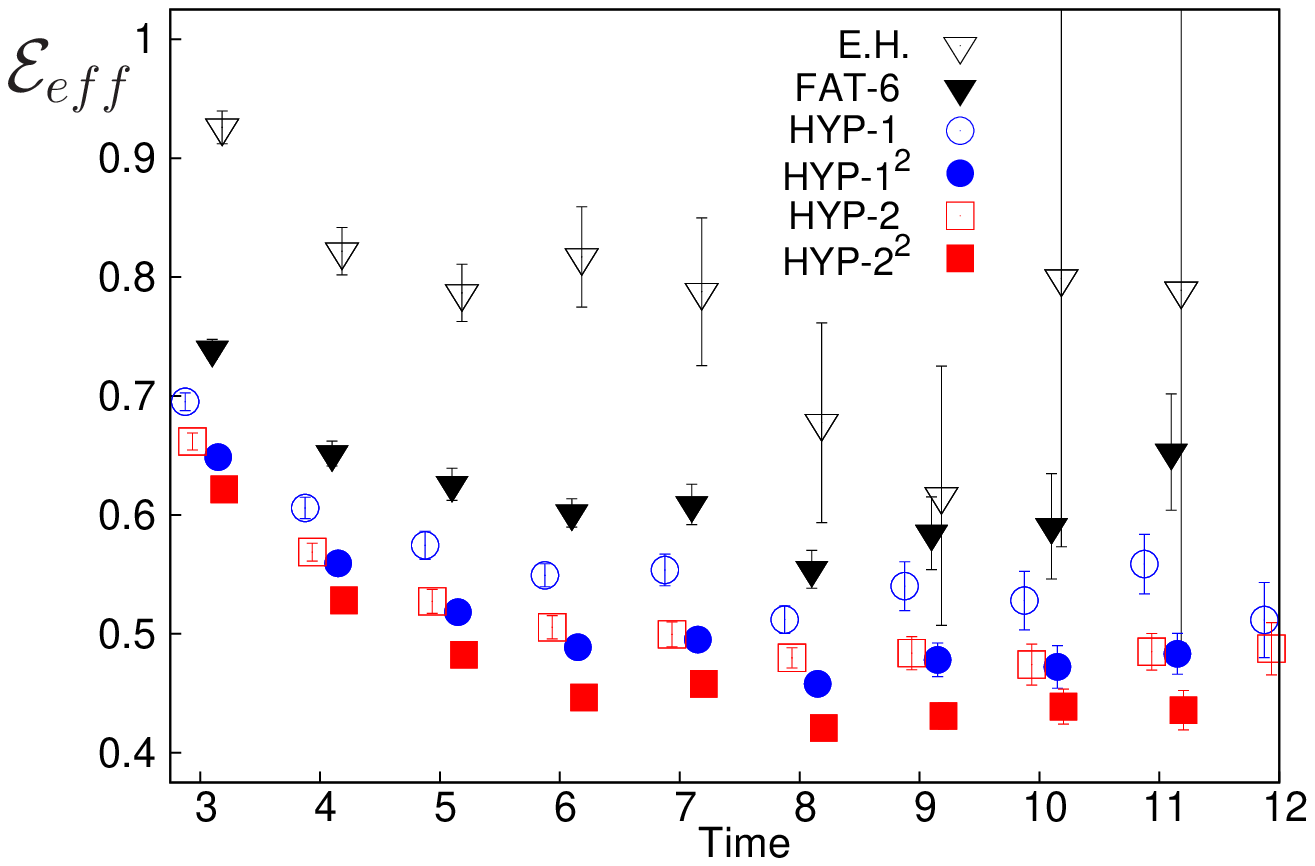}}\\
\vspace*{8mm}
\resizebox{90mm}{!}{\includegraphics{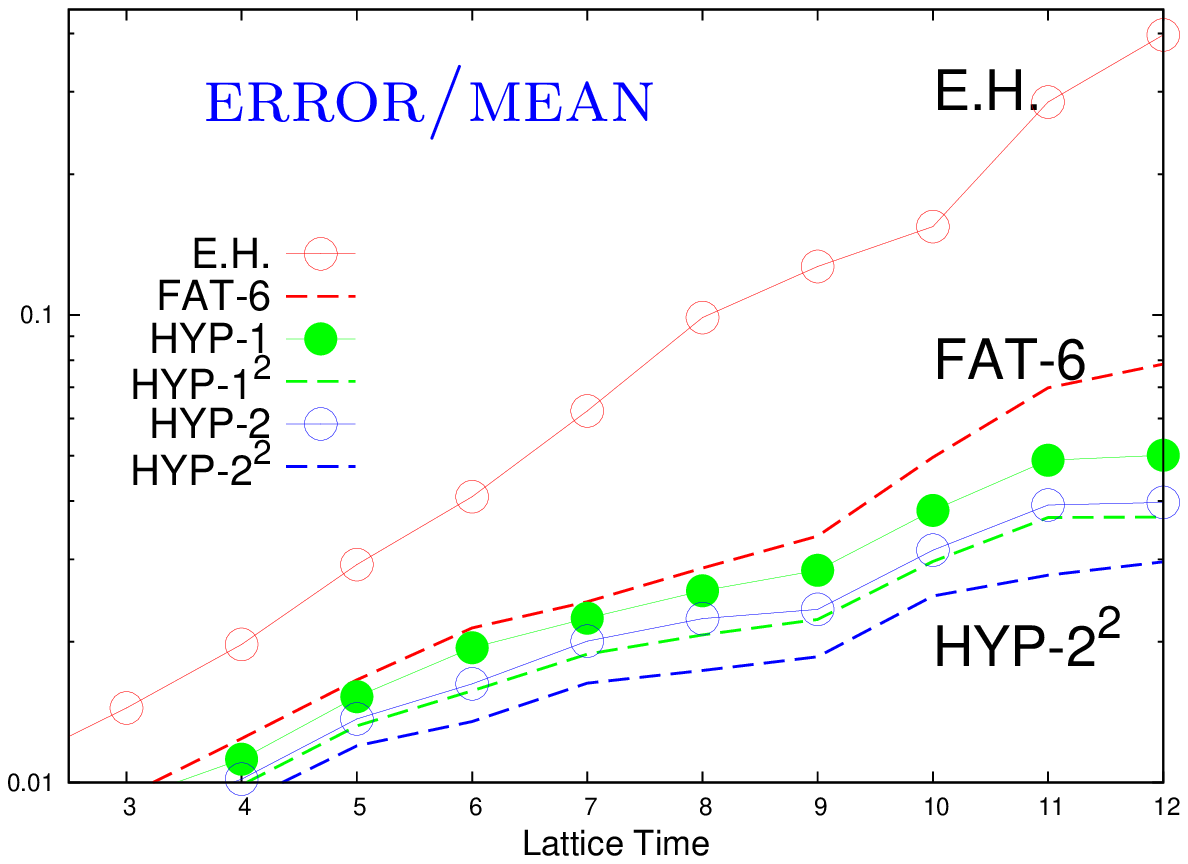}}
\caption{\label{fig:1}\footnotesize{\sl 
In the upper plot we show the effective mass from the static-light two-point correlation function as obtained by using 
6 different heavy-quark actions and the light-quark with $\kappa_q=0.1374$, which is somewhat lighter but close to the physical strange 
quark mass. The lower plot shows the (exponential) growth of the statistical error of the two-point function as the time separation between the sources is increased. Notice that the error on the correlation function with HYP-2 heavy-quarks grows in the same way as with the HYP-1$^2$ one. }} 
\end{center}
\end{figure}

The  latticized version of the static heavy-quark effective theory Lagrangian (HQET), in the frame in which the heavy-quark is at rest, reads~\cite{Eichten} 
\bea \label{hqet-lagr}
{\cal L}_{\rm HQET}&=&\sum_x \biggl\{  h^\dagger(x)\left[ h(x) - V_0(x-\hat 0)^\dagger h(x-\hat 0)\right] + \delta m(g^2) h^\dagger(x)h(x) \biggr\}\,,
\eea
where  $V_\mu(x)$ is the lattice link variable, and we include the power divergent term of the discretized HQET [$\delta m(g^2) h^\dagger(x)h(x)$]
which is a regularization artifact. For notational simplicity we omit it hereto,  but we will of course discuss its impact when appropriate. 
The static heavy-quark propagator is then
\bea
S_{ h} (\vec x, 0; \vec y, t)= {1+\gamma_0\over 2}\left( \prod_{n=0}^{t-1} V_0^\dagger (n)\right) \delta(\vec x- \vec y)\ \theta(t)\,,
\eea
with $V_0(n)\equiv V_0(\vec x, n )$. Concerning the Wilson line on the lattice we will distinguish the following cases:
\begin{itemize}
\item {\underline{The Eichten-Hill (EH) static quark propagator}} is the simplest form obtained by replacing~\cite{Eichten} 
 \bea  
V_0(x) \rightarrow  U_0^{\rm EH}(x)\equiv U_0(x)\,,
\eea
with $U_0(x)$ being the simplest (unimproved) temporal link on the lattice.
\item  {\underline{The FAT-link  static quark propagator}}  is the first improvement which involves neighboring staples and is obtained by averaging over the six staples 
that surround the link~\cite{fat-6}
\bea  
V_0(x) \rightarrow  U_0^{\rm FAT-6}(x) = 
\frac{1}{6} \sum_{\pm i} 
U^{\rm Staple}_i ( x,x+\hat{t})  \; .
\eea
By that modification the Wilson line becomes a thin flux tube. All the properties of the corresponding Wilson line remain the same so that the systematics of lattice calculations in principle is not changed.
\item {\underline{The HYP  static quark propagator}}  correspond to a more efficient choice of the lattice static quark action, in terms of improving the quality of signal for the heavy-quark propagator, 
and is  obtained by an optimized combination of links belonging to the hypercube surrounding a given link.  This so-called HYP procedure can be summarized by the following three steps:
\bea 
\label{hypstat} 
V_0 (x)\rightarrow U_0^{\rm HYP}(x)={\rm Proj_{SU(3)}} \left[ (1-\alpha_1) U_0 (x) + 
 \frac{\alpha_1}{6} \sum_{\pm i \neq 0} \widetilde V_{i;0}(x)  \widetilde V_{0;i}(x+i)  \widetilde V^\dagger_{i;0}(x+\hat 0 )  \right] \,,\nn
\eea
where the links $\widetilde V_{\mu;\nu}(x)$ are defined via
\bea \label{deco1}  
\widetilde V_{\mu;\nu}(x)={\rm Proj_{SU(3)}} \left[(1-\alpha_2) U_{\mu}(x)+ \frac{\alpha_2}{4}\sum 
\limits_{\pm \rho\neq \mu,\nu}\overline V_{\rho; \nu,\mu}(x) \overline V_{\mu; \rho\nu}(x+\rho) \overline V^\dagger_{\rho; \nu,\mu}(x+\mu)
\right] \;  ,\nn
\eea
and finally  $\overline V_{\mu; \nu,\rho}(x)$ is defined via
\bea \label{deco2}
\overline V_{\mu; \nu,\rho}(x)={\rm Proj_{SU(3)}} \left[(1-\alpha_3) U_{\mu}(x)+ \frac{\alpha_3}{2}
\sum\limits_{\pm \sigma\neq \mu,\nu\rho} U_\sigma (x) U_\mu(x+\sigma)U_\sigma^\dagger(x+\mu)\right] \; .
\eea
The optimal choice found in ref.~\cite{hasenfratz}, in which the HYP procedure was actually proposed, is $\vec \alpha \equiv (\alpha_1,\alpha_2,\alpha_3) =  (0.75, 0.6, 0.3)$, and we will refer to it as HYP-1. 
The other optimal choice, $\vec \alpha=  (1.0, 1.0, 0.5)$, was proposed in ref.~\cite{actions-dellamorte} after minimizing the noise to signal ratio. We will refer to that option as HYP-2. 
\end{itemize}
The choices for parameters $\vec \alpha$ made in refs.~\cite{hasenfratz,actions-dellamorte} were obtained from the numerical studies with the Wilson plaquette lattice-gauge action. In this paper 
we use the gauge field configurations generated with the Iwasaki gauge action. Instead of looking for yet another optimal choice for the coefficients $\vec \alpha$, we will explore the above four 
possibilities, study the impact of each of the improved action on the static-light quantities, and then choose the one that gives us the best results in terms of the quality of the signal and the associated 
statistical errors. Only then we will be ready to study the three-point functions.
\begin{table}[t!]
\begin{center} 
{\scalebox{.85}{
\begin{tabular}{|c|c|c|c|c|c|c|c|}
\hline
{\phantom{\Huge{l}}}\raisebox{-.1cm}{\phantom{\Huge{j}}}
{$\kappa_q$}& {} & \underline{EH}& \underline{FAT-6} & \underline{HYP-1}& \underline{HYP-${\rm 1}^2$}  & \underline{HYP-2}& \underline{HYP-${\rm 2}^2$} \\  \hline
$0.1357$  & ${\cal Z}_q^{S\ 2}$ & $104(27)$& $106(9)$   & $111(7)$& $111(5)$ & $111(5)$& $103(4)$ \\  
{\phantom{\Huge{l}}}\raisebox{-.1cm}{\phantom{\Huge{j}}}
                   & ${\cal E}_q$ & $0.770(36)$& $0.613(11)$   & $0.578(8)$& $0.534(6)$ & $0.548(6)$& $0.506(5)$ \\   
{\phantom{\Huge{l}}}\raisebox{-.1cm}{\phantom{\Huge{j}}}
                  & $\widetilde {\cal Z}_q^{S\ 2}$ & $49(29)$& $59(14)$   & $63(13)$& $62(9)$ & $65(10)$& $55(7)$ \\  
{\phantom{\Huge{l}}}\raisebox{-.1cm}{\phantom{\Huge{j}}}
                   & $\widetilde {\cal E}_q$ & $1.07(10)$& $0.911(40)$   & $0.877(34)$& $0.824(26)$ & $0.839(26)$& $0.786(22)$ \\   
 {\phantom{\Huge{l}}}\raisebox{-.1cm}{\phantom{\Huge{j}}}
                  & $\Delta_q$ & $0.285(90)$& $0.282(38)$   & $0.282(33)$& $0.280(25)$ & $0.280(25)$& $0.275(20)$ \\  
{\phantom{\Huge{l}}}\raisebox{-.1cm}{\phantom{\Huge{j}}}
                   & ${\cal Z}^L_q$ & $0.341(50)$& $0.256(12)$   & $0.257(9)$& $0.234(6)$ & $0.215(5)$& $0.213(4)$ \\                       \hline
{\phantom{\Huge{l}}}\raisebox{-.1cm}{\phantom{\Huge{j}}}
$0.1367$  & ${\cal Z}_q^{S\ 2}$ & $87(28)$& $110(7)$   & $108(6)$& $106(4)$ & $107(5)$& $94(3)$ \\  
{\phantom{\Huge{l}}}\raisebox{-.1cm}{\phantom{\Huge{j}}}
                   & ${\cal E}_q$ & $0.730(44)$& $0.586(10)$   & $0.543(8)$& $0.497(6)$ & $0.511(6)$& $0.465(5)$ \\   
{\phantom{\Huge{l}}}\raisebox{-.1cm}{\phantom{\Huge{j}}}
                  & $\widetilde {\cal Z}_q^{S\ 2}$ & $36(27)$& $57(11)$   & $62(10)$& $58(7)$ & $58(6)$& $52(6)$ \\  
{\phantom{\Huge{l}}}\raisebox{-.1cm}{\phantom{\Huge{j}}}
                   & $\widetilde {\cal E}_q$ & $0.96(12)$& $0.838(32)$   & $0.806(26)$& $0.745(20)$ & $0.751(19)$& $0.706(19)$ \\   
 {\phantom{\Huge{l}}}\raisebox{-.1cm}{\phantom{\Huge{j}}}
                  & $\Delta_q$ & $0.166(90)$& $0.241(26)$   & $0.249(22)$& $0.238(17)$ & $0.233(17)$& $0.230(16)$ \\  
{\phantom{\Huge{l}}}\raisebox{-.1cm}{\phantom{\Huge{j}}}
                   & ${\cal Z}^L_q$ & $0.303(54)$& $0.241(9)$   & $0.236(7)$& $0.214(5)$ & $0.199(5)$& $0.193(4)$ \\                       \hline
{\phantom{\Huge{l}}}\raisebox{-.1cm}{\phantom{\Huge{j}}}
$0.1374$  & ${\cal Z}_q^{S\ 2}$ & $85(28)$& $115(10)$   & $112(7)$& $101(5)$ & $100(5)$& $88(3)$ \\  
{\phantom{\Huge{l}}}\raisebox{-.1cm}{\phantom{\Huge{j}}}
                   & ${\cal E}_q$ & $0.714(47)$& $0.582(12)$   & $0.533(8)$& $0.475(6)$ & $0.486(7)$& $0.436(5)$ \\   
{\phantom{\Huge{l}}}\raisebox{-.1cm}{\phantom{\Huge{j}}}
                  & $\widetilde {\cal Z}_q^{S\ 2}$ & $36(44)$& $77(18)$   & $71(14)$& $65(8)$ & $63(8)$& $59(6)$ \\  
{\phantom{\Huge{l}}}\raisebox{-.1cm}{\phantom{\Huge{j}}}
                   & $\widetilde {\cal E}_q$ & $1.28(20)$& $0.833(38)$   & $0.772(30)$& $0.709(20)$ & $0.713(21)$& $0.676(16)$ \\   
 {\phantom{\Huge{l}}}\raisebox{-.1cm}{\phantom{\Huge{j}}}
                  & $\Delta_q$ & $0.477(181)$& $0.234(33)$   & $0.227(27)$& $0.223(18)$ & $0.214(18)$& $0.229(14)$ \\  
{\phantom{\Huge{l}}}\raisebox{-.1cm}{\phantom{\Huge{j}}}
                   & ${\cal Z}^L_q$ & $0.294(55)$& $0.246(12)$   & $0.236(8)$& $0.205(5)$ & $0.188(5)$& $0.181(4)$ \\                       \hline
{\phantom{\Huge{l}}}\raisebox{-.1cm}{\phantom{\Huge{j}}}
$0.1382$  & ${\cal Z}_q^{S\ 2}$ & $102(42)$& $100(9)$   & $94(6)$& $91(5)$ & $94(5)$& $82(5)$ \\  
{\phantom{\Huge{l}}}\raisebox{-.1cm}{\phantom{\Huge{j}}}
                   & ${\cal E}_q$ & $0.706(56)$& $0.535(12)$   & $0.485(9)$& $0.437(7)$ & $0.453(7)$& $0.403(7)$ \\   
{\phantom{\Huge{l}}}\raisebox{-.1cm}{\phantom{\Huge{j}}}
                  & $\widetilde {\cal Z}_q^{S\ 2}$ & $35(22)$& $77(21)$   & $72(14)$& $73(10)$ & $79(11)$& $66(8)$ \\  
{\phantom{\Huge{l}}}\raisebox{-.1cm}{\phantom{\Huge{j}}}
                   & $\widetilde {\cal E}_q$ & $0.80(10)$& $0.753(43)$   & $0.698(29)$& $0.655(22)$ & $0.673(22)$& $0.621(20)$ \\   
 {\phantom{\Huge{l}}}\raisebox{-.1cm}{\phantom{\Huge{j}}}
                  & $\Delta_q$ & $0.102(74)$& $0.197(37)$   & $0.188(27)$& $0.201(19)$ & $0.195(20)$& $0.209(17)$ \\  
{\phantom{\Huge{l}}}\raisebox{-.1cm}{\phantom{\Huge{j}}}
                   & ${\cal Z}^L_q$ & $0.277(62)$& $0.207(10)$   & $0.196(7)$& $0.179(5)$ & $0.168(5)$& $0.162(5)$ \\                       \hline
\end{tabular}  }}
\caption{\label{tab:1}
\footnotesize  List of our results obtained by using all six static heavy-quark actions :  binding energies of the static-light $j^p=(1/2)^-$ mesons (${\cal E}_q$), and of the $j^p=(1/2)^+$ ones 
($\widetilde {\cal E}_q$), the orbital mass splitting ($\Delta_q=\widetilde {\cal E}_q-{\cal E}_q$).  We also list the values for the couplings to the smeared sources, ${\cal Z}^S_q$, $\widetilde {\cal Z}^S_q$, as well as 
 the  bare matrix element of the local static-light axial current in heavy-quark effective theory,  ${\cal Z}^L_q =|\langle 0 |\bar b  \gamma_0\gamma_5 q|B_q\rangle|$. All results are given in lattice units.
}
\end{center}
\end{table}
To illustrate the benefit of the above-mentioned improvement we compute the propagator of the lowest lying static-light mesons, i.e. the correlation functions 
\bea\label{eq:2pts}
&&C_2(t) =\sum_{\vec y}  \left<  \bar h(x) \gamma_5 q (x)  \  \left(\bar h(y) \gamma_5 q (y)\right)^\dagger \right> =
{1\over 3} \sum_{i, \vec y}  \left<  \bar h(x) \gamma_i q_(x)  \  \left(\bar h(y) \gamma_i q_(y)\right)^\dagger \right>  \cr
&& =  {1\over 2} \biggl\{ 
\langle {\rm Tr} \left[ \gamma_5 S_{ h} (\vec x, 0; \vec y, t) \gamma_5 S_q(\vec y, t; \vec x, 0) \right] \rangle + 
\langle {\rm Tr} \left[ \gamma_5 S_{\bar h} (\vec x, 0; \vec y, - t) \gamma_5 S_q(\vec y, -t; \vec x, 0) \right] \biggr\} ,
\eea
where one source operator has been fixed at some $\vec x$ at $t=0$ and the other is free and runs over all the lattice indices. 
The first line in the above equation exhibits the heavy-quark spin symmetry while in the second we average over the charge conjugated correlators. 
To illustrate the benefit of improving the static quark action 
we consider the effective mass (binding energy) defined as
\bea
{\cal E}_{\rm eff} = \log\left[{C_2(t)\over C_2(t+1)}\right]\,,
\eea 
and plotted in Fig.~\ref{fig:1}. 
Even though the improved Wilson line allows the signal to exist at larger time separations, the statistical errors are still growing with time and 
a smearing procedure that would allow us to isolate the ground state at smaller time separations is desirable. It is particularly important 
for the studies of the three-point correlation  functions, which will be discussed in the next section.  
We use the same smearing procedure as the one used in ref.~\cite{PPP,boyle}, with the same set of parameters.  
In Table~\ref{tab:1} we present the results of the fit to 
\bea
C_2(t) ={\cal {Z}}_q^I {\cal {Z}}_q^{I^\prime} \times \exp\left( - {\cal E }_q t \right)\,, \qquad I,I^\prime  \in (L,S),
\eea
where $L$ and $S$ stand for the local or smeared source operators.  The fits are made in the interval $7 \leq t\leq 11$ for all the light-quarks. The reader should not be surprised that the binding energies 
differ for different static heavy-quarks. This is because  the residual mass term, $\delta m$, in eq.~(\ref{hqet-lagr}) is different for different  heavy-quark actions.  In Fig.~\ref{fig:1} we see that with the EH  action the (smeared) 
correlation functions are prohibitively noisy as soon as one reaches $t>6$. With improved actions the situation is much better and the signal is still present for $t>12$. Besides lowest heavy-light mesons 
we also studied  the correlation functions  of operators coupling to the orbitally excited static-light mesons, i.e. those with  $j^P=(1/2)^+$. They are obtained in a completely analogous way to what we just described,  
after replacing $\gamma_5 \to \one$. Those correlators and the results obtained from them will be denoted by an extra tilde symbol. 
The fits to  $ \widetilde C_2(t)$  are made in the interval $6 \leq t\leq 9$.  In addition, in that same interval we fit the ratio 
\bea
{C_2(t)\over  \widetilde C_2(t) }\sim \exp(\Delta_q t )\,,
\eea
where $\Delta_q = \widetilde {\cal E}_q - {\cal E}_q \equiv \Lambda_{1/2^+} -  \Lambda_{1/2^-}$, which is free from the power divergent term coming from  the HQET 
Lagrangian as written in eq.~(\ref{hqet-lagr}).   
In ref.~\cite{exponential} it was shown that the signal-noise ratio is exponentially improving with the improved Wilson lines. We found it interesting to check on 
the relative error of the signal as the time separation between the sources is increased. We therefore simply compute $[\delta C_2(t)]/C_2(t)$, with  $\delta C_2(t)$ being the  statistical error, and plot that quantity for various 
improved heavy-quark actions. This is shown in Fig.~\ref{fig:1}. We  indeed see the same phenomenon as in ref.~\cite{exponential}, namely, that the signal with improved Wilson lines remains stable longer 
than in the EH case, and its relative statistical error remains under  $10\%$ for quite large time separations. 

By ``HYP-1$^2$" and  ``HYP-2$^2$" we denote the static quark obtained by reiterating the hyper-cubic blocking procedure (applying  it twice). That procedure does not bring any gain if we consider the two-point 
static-light correlation functions alone. This can be viewed as an indication that the HYP procedure is well optimized.  From the same figure and from our experience with three-point correlation functions, 
however,  it seems that the HYP-2 procedure is somewhat better optimized and that for a better extraction of the signal with three-point correlation functions, the HYP-2$^2$ static quark action will be the most interesting 
for our purpose. As a side remark, we should mention that we tried to apply the HYP procedure thrice but we did not see any extra benefit in doing it.

Before closing this section we should also comment on our procedure to compute the light-quark propagator. We do not use the {\sl ``all-to-all"} procedure~\cite{all2all}, which would certainly be very convenient, 
but instead  we use the fact that we compute with the sources smeared  along each direction to at most $R_{\rm max}=5$ spatial points, so that in  eq.~(\ref{eq:2pts}) 
\bea
C_2(t)=\langle {\rm Tr} \left[{1+\gamma_0\over 2}  \prod_{n=0}^{t-1} V_0^\dagger ( n)  \gamma_5 S_q(\vec x, t; \vec x, 0)\gamma_5    \right] \rangle 
\eea
we choose  ``$\vec x$" to be anywhere within the smearing radius, i.e.  we invert each of our light propagators $31$ times [$2 \times 5$ along each spatial axis plus with $\vec x$ at the lattice origin $(0,0,0)$]. 

\section{Heavy-light mesons: three-point functions \label{SecY}}

We now discuss the three-point functions. Since the heavy-quark is infinitely heavy, i.e. static, we can study the radial distribution of the light degrees of freedom within the bound state, which we choose to be either 
the ground state doublet $J^P=[0^-,1^-]$, or its first orbital excitation $J^P=[0^+,1^+]$. 
In other words, we will compute the correlation functions by pulling the light-quark away from the static source of color by means of one of the local operators such as 
the scalar density ($\bar q\one q$), vector current ($\bar q\gamma_0 q$) or the axial-vector current ($\bar q\gamma_i \gamma_5 q$). In this way we will compute the 
matter, electric charge and the axial charge distributions. Such a study was first proposed in ref.~\cite{duncan}, and in recent years in ref.~\cite{green}.~\footnote{
In fact the earlier study in ref.~\cite{duncan} was rather focused in extracting the gauge dependent Bethe-Salpeter amplitude, which --broadly speaking-- also 
gives indication about the spatial distributions.} Preliminary 
results with an improved heavy-quark action (improved static Wilson line) have been reported in ref.~\cite{koponen}.  
In order to study the above-mentioned radial distribution we will study the following correlation functions:
\bea\label{3ptsSV}
&&C_\Gamma (-t_{s1}, t_{s2}; \vec r ) =
\sum_{\vec x,\vec y}  \left<  \bar h(x) \gamma_5 q (x)  \ \bar q(z+\vec r ) \Gamma q(z+\vec r) \  \left(\bar h(y) \gamma_5 q (y)\right)^\dagger \right>_{z - {\rm fixed}} \nn \\
&&= \sum_{\vec x,\vec y }  \left<  {\rm Tr}\left[ S_h(y,x) \gamma_5 S_q (x,z+\vec r) \  \Gamma \  S_q (z+\vec r, y) \gamma_5\right] \right>_{z - {\rm fixed}}\\
&& = \sum_{\vec x }  \left< {\rm Tr}\left[ {1+\gamma_0\over 2} \prod_{n=-t_{s1}}^{t_{s2}-1} V_0^\dagger (n) \ \gamma_5  S_q (\vec x, t_{s2}; \vec z + \vec r,0)\  \Gamma \  S_q (\vec z+\vec r, 0; \vec x,- t_{s1}) 
\gamma_5 \right]  \right> , \nn
\eea
where one source of the heavy-light mesons is fixed to $t_y=- t_{s1}$, the operator whose distribution we want to  compute is fixed at $t_z=0$, and $t_x=t_{s2}$ is free. In the practical calculations we also 
exchange the roles of $t_x$ and $t_y$, and fix $t_x$ while letting $t_y$ free. For $\vec z=\vec 0$ this would be the usual way we compute the correlation function. For $\vec z\neq 0$ and anywhere 
on the lattice the stochastic method of all-to-all propagators is needed. As we already mentioned we restrain our $\vec z$ to $31$ points around the lattice origin. In other words,  
we make $31$ inversions per light-quark $q$, per gauge field configuration.  Finally, the components of the vector $\vec r$ can assume any of the values $-L/2 \leq r_{x,y,z} < L/2$. We should 
also add that in the above expression $\Gamma = \gamma_0, \one$. For  large time separations in  eq.~(\ref{3ptsSV}) we then have
\bea\label{charge-density}
R_{\Gamma}(  \vec r ) = { C_\Gamma (-t_{s1}, t_{s2}; \vec r ) \over\left( {\cal Z}_q^{S}\right)^2 \times \exp[- {\cal E}_q \left( t_{s2}-  t_{s1}\right) ] }\; \stackrel{  |t_{s1}|,t_{s2}\gg 0 }{\xrightarrow{\hspace*{18mm}} } 
\; \langle B_q \vert \bar q \Gamma q (\vec r) \vert B_q\rangle \,,
\eea
where we divided the couplings of the lowest states to the (smeared) source operators (${\cal Z}_q^{S}$) evaluated from the study of the two-point functions. 
One of the criteria to fix the correct fitting interval to extract the matrix element was to make sure that the result coincides with what is obtained if the three-point 
function is directly divided by the two-point functions. 
From that analysis and for $t_{s2}-  t_{s1} \in \{ 10,11,12,13\}$ we observe good plateaus of the ratio $R_{\Gamma}(  \vec r )$, where one of the sources is fixed at $\vert t_{s2}\vert =5$ or at $\vert t_{s2}\vert =6$ away 
from the origin $t=0$,  at which is placed our light-light-quark operator.  
For the sake of parity invariance, the above expression is exactly equal to zero if $\Gamma = \gamma_i\gamma_5$ ($i=1,2,3$). Therefore, for the axial charge one of the sources should be the vector current, and we consider  
\bea\label{axial-density}
&& C_{\gamma_i \gamma_5} (-t_{s1}, t_{s2}; \vec r ) =
\sum_{\vec x,\vec y }  \left<  \bar h(x) \gamma_5 q (x)  \ \bar q(z+\vec r ) \gamma_i \gamma_5  q(z+\vec r) \  \left(\bar h(y) \gamma_i q (y)\right)^\dagger \right>_{z - {\rm fixed}}\,, \nn \\
&& R_{\gamma_i \gamma_5}(  \vec r ) = { C_{\gamma_i \gamma_5} (-t_{s1}, t_{s2}; \vec r ) \over\left( {\cal Z}_q^{S}\right)^2 \times \exp[- {\cal E}_q \left( t_{s2}-  t_{s1}\right) ] }\; \stackrel{  |t_{s1}|,t_{s2}\gg 0 }{\xrightarrow{\hspace*{18mm}} } 
\; \langle B_q \vert \bar q \gamma_i \gamma_5 q (\vec r) \vert B_q^\ast \rangle\ ,
\eea
where we obviously used the fact that the heavy-light vector and pseudoscalar mesons are degenerate. 
We considered all possible values of $r^2$. We verified that the signal for various combinations of the spatial indices corresponding to the same $r^2$ are compatible among themselves, so that we could 
average over all of them. 
For larger values of $r^2$ the signals deteriorate and indeed after $r \equiv \vert \vec r\ \! \vert \simeq 12$ the signal is poor and in some situations completely lost after $r \equiv \vert \vec r\ \! \vert \simeq 15$, i.e. 
fully dominated by statistical errors.~\footnote{There are $7150$ combinations of $(r_x,r_y,r_z)$ for $\vert \vec r\ \! \vert  \leq 12$. After averaging over the equivalent combinations we end up with $122$ points.}  
This is not a worry because each of the distributions that we measure here are the rapidly decreasing functions, which eventually die out at larger values of $r$ . Special attention should naturally be paid to  
lighter quarks for which the distributions are more spread out over the lattice. It appears that the quark masses that we deal with in this paper are not small enough to be sensitive to this kind of 
 finite volume effects.

The matrix elements defined in eqs.~(\ref{charge-density}) and~(\ref{axial-density}) are the scalar, vector, and axial densities, which we will denote 
by $f_\Gamma(r)$. We will follow the terminology of ref.~\cite{green} and call $f_{\gamma_0}(r)$  the charge density, 
$f_{\mathbb{I} }(r)$
the matter density, and finally the one with $f_{\gamma_i \gamma_5}(r)$  the axial charge density.

As we already mentioned, here we also study the properties of the orbitally excited mesons, namely, the  $(1/2)^+$-doublet. 
To do so we repeat the same procedure as the one sketched above, by replacing in eq.~(\ref{3ptsSV}) $\gamma_5 \to \one$,  thus obtaining $\widetilde C_\Gamma (-t_{s1}, t_{s2};  r )$ ($\Gamma = \gamma_0, \one$), in addition to 
\bea
&&\hspace*{-7mm} \widetilde C_{\gamma_i \gamma_5} (-t_{s1}, t_{s2};  r ) = \sum_{\vec x,\vec y}  \left<  \bar h(x) \one  q (x)  \ \bar q(z+\vec r ) \gamma_i \gamma_5  q(z+\vec r) \  \left(\bar h(y) \gamma_i \gamma_5 q (y)\right)^\dagger \right>_{z - {\rm fixed}}\,,
\eea
where $t_z=0$ and $\vec z = \vec 0$ or any other of $30$ points surrounding the lattice origin, as mentioned after eq.~(\ref{3ptsSV}).
Desired distributions are then extracted from the ratios
\bea
&&\hspace*{-7mm} \widetilde R_{\Gamma}( r ) = {  \widetilde C_{\Gamma} (-t_{s1}, t_{s2}; \vec r ) \over (  \widetilde {\cal Z}_q^{S} )^2 \times \exp[-  \widetilde {\cal E}_q \left( t_{s2}-  t_{s1}\right) ] }\; \stackrel{  |t_{s1}|,t_{s2}\gg 0 }{\xrightarrow{\hspace*{15mm}} } 
\; \langle B_{0 q}^\ast \vert \bar q \Gamma q (  r) \vert B_{0 q}^\ast  \rangle \equiv  \widetilde f_{\Gamma}(r)\ ,\cr
&&\hspace*{-7mm} \widetilde R_{\gamma_i \gamma_5}(  r ) = {  \widetilde C_{\gamma_i \gamma_5} (-t_{s1}, t_{s2}; \vec r ) \over ( \widetilde  {\cal Z}_q^{S} )^2 \times \exp[- \widetilde  {\cal E}_q \left( t_{s2}-  t_{s1}\right) ] }\; \stackrel{  |t_{s1}|,t_{s2}\gg 0 }{\xrightarrow{\hspace*{15mm}} } 
\; \langle B_{0 q}^\ast \vert \bar q \gamma_i \gamma_5 q ( r) \vert B_{1 q} \rangle\equiv  \widetilde f_{\gamma_i\gamma_5}(r)\ ,
\eea
where we used the standard spectroscopic notation according to which the states belonging to the $(1/2)^+$-doublet, are labeled as $B_0^\ast$ (scalar) and $B_1$ (axial) states. 
We also keep track of the light-quark flavor when labeling the states. We should stress that the $B$-states in this work refer to the infinite $b$-quark mass limit. 
As expected --from the discussion of the two-point correlation functions-- the signal with the external excited states is less good than with the lowest states [$(1/2)^-$]. 

\begin{figure}
\vspace*{-14mm}
\begin{center}
\hspace*{-11mm}
\begin{tabular}{ll}
{\resizebox{88mm}{!}{\includegraphics{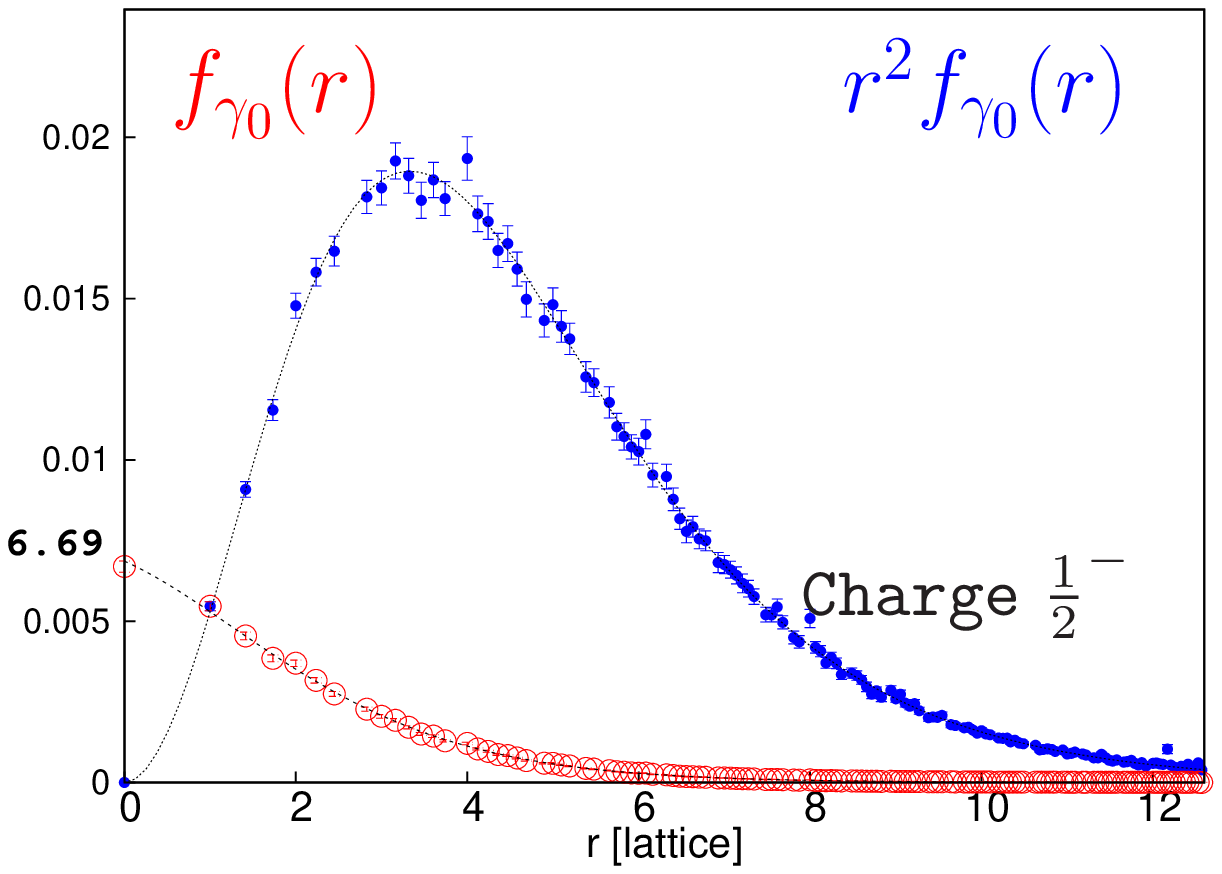}}} & {\resizebox{88mm}{!}{\includegraphics{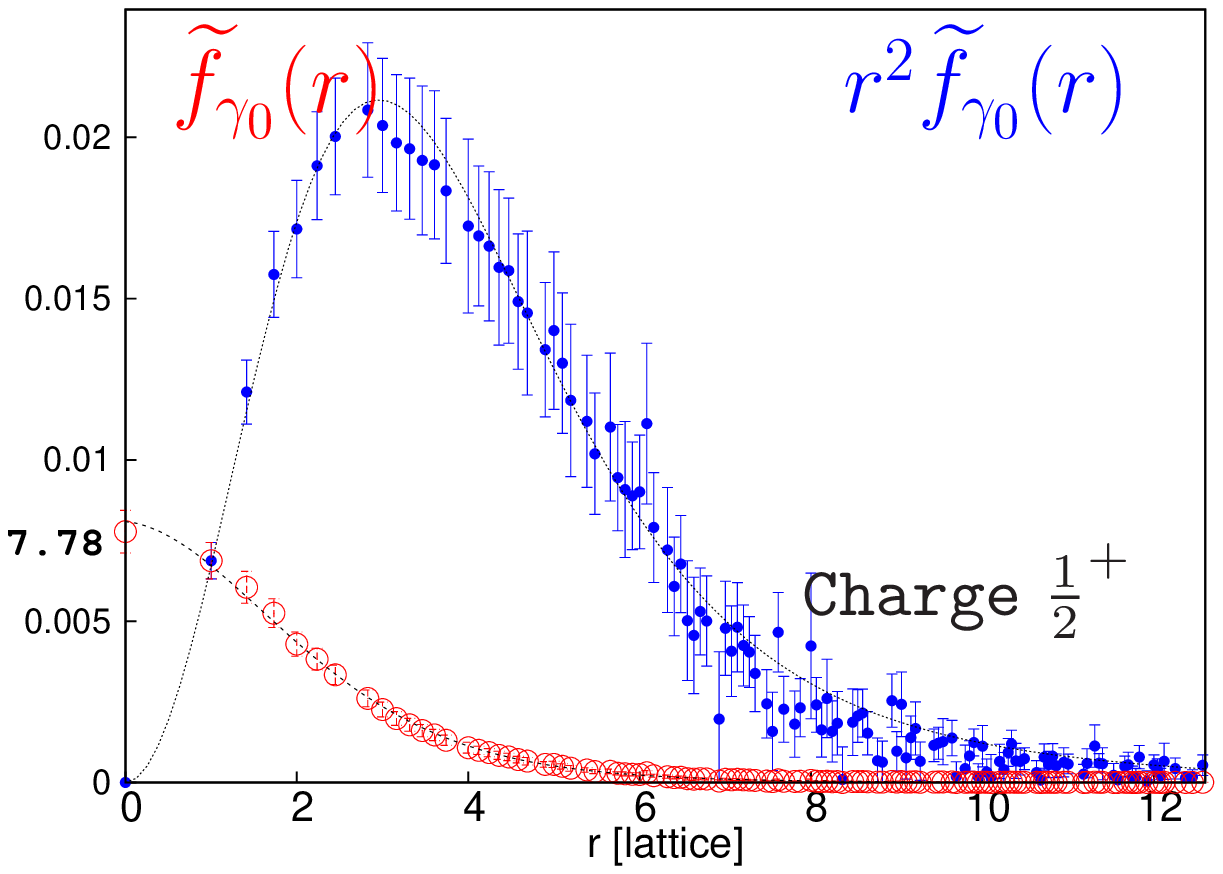}}}\\
{\resizebox{88mm}{!}{\includegraphics{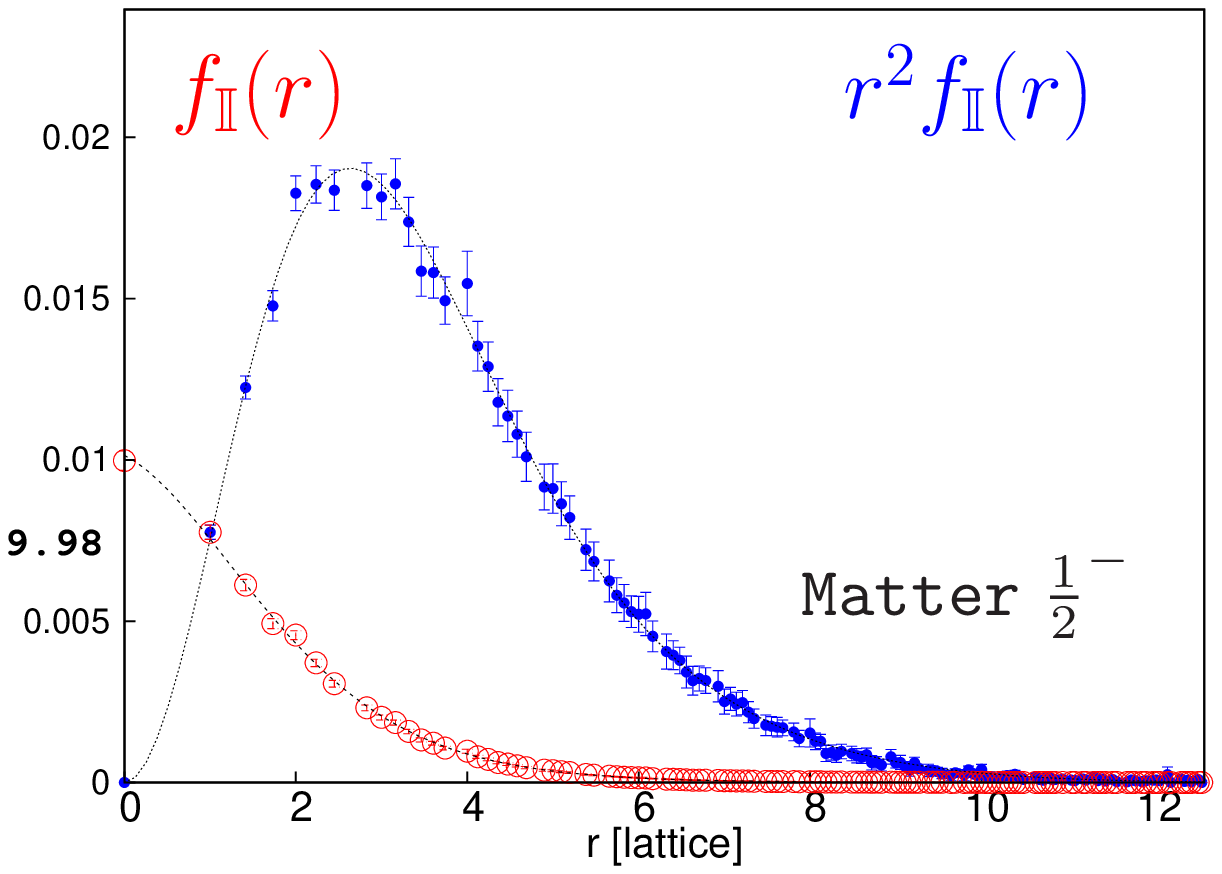}}} & {\resizebox{88mm}{!}{\includegraphics{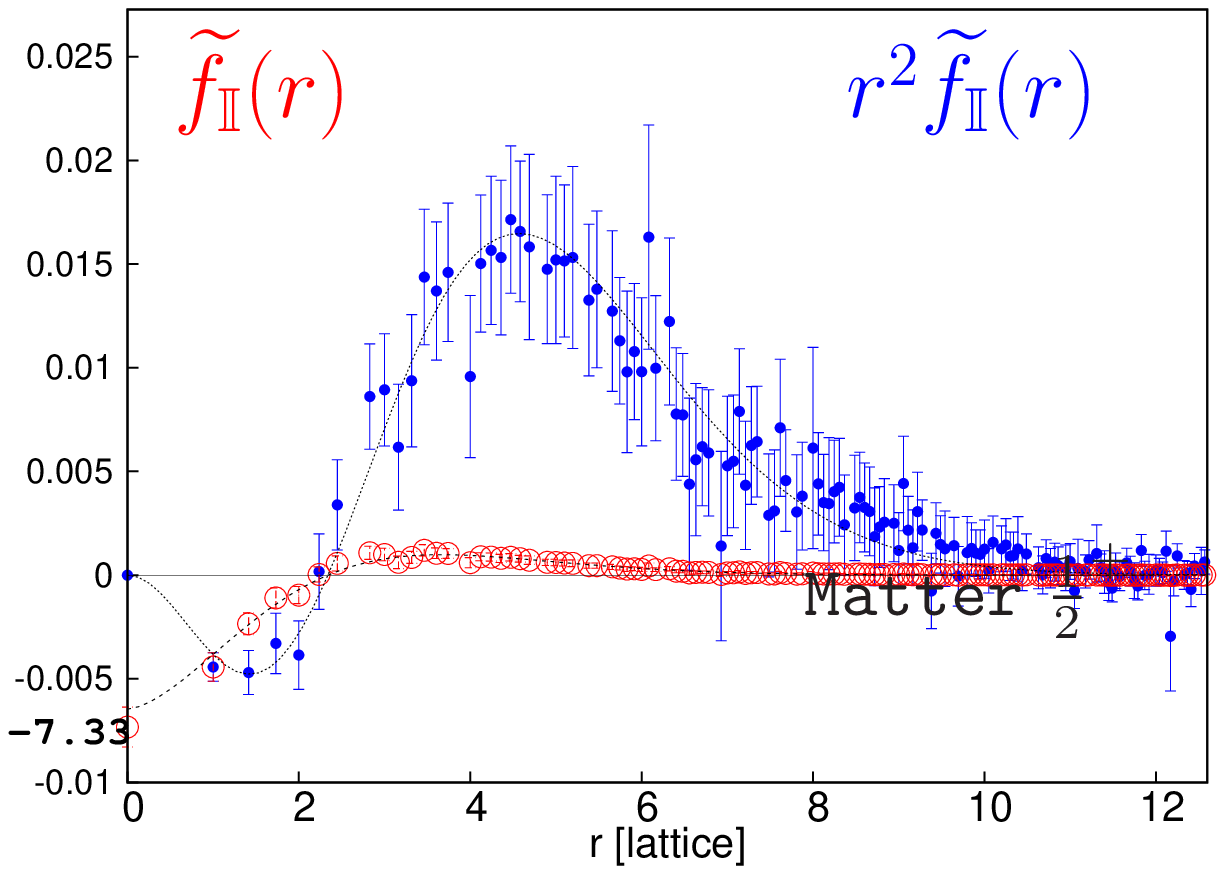}}}\\
{\resizebox{88mm}{!}{\includegraphics{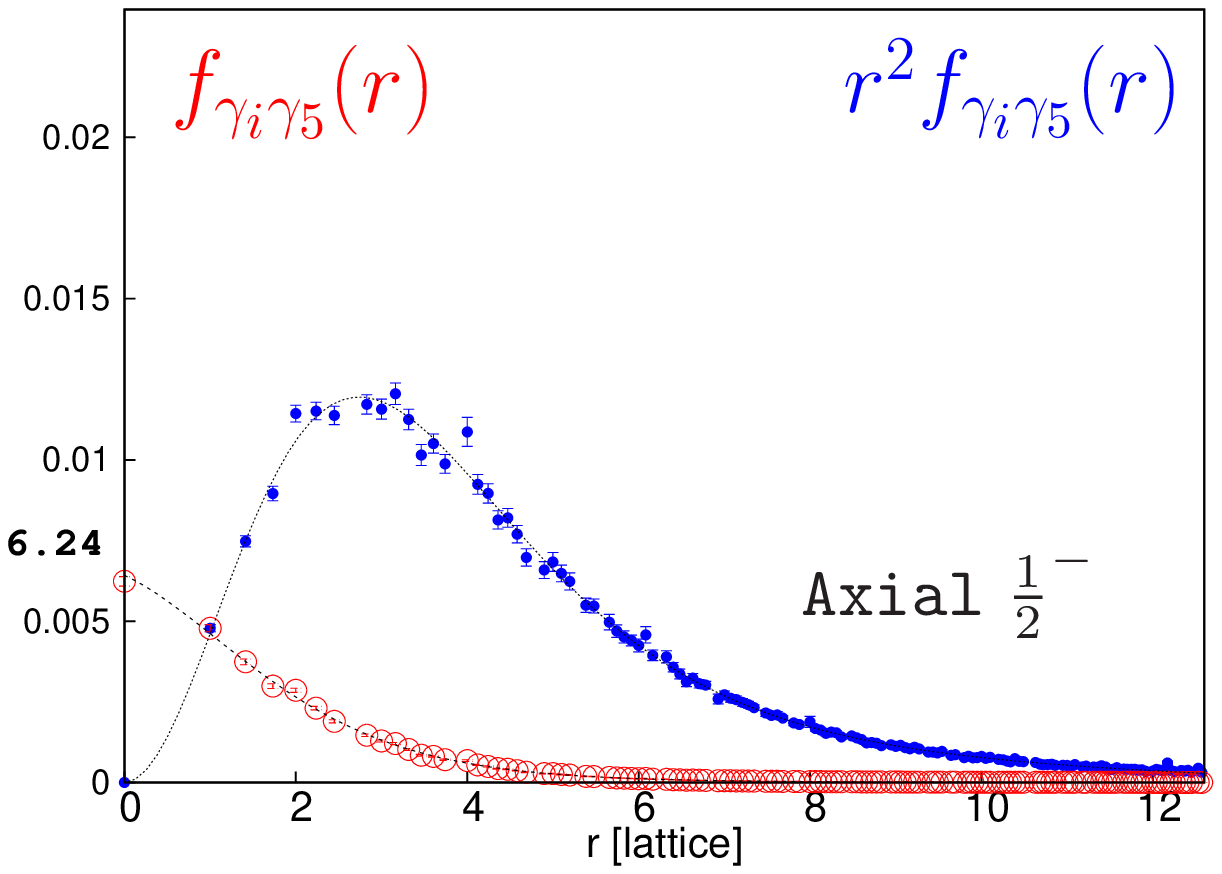}}} & {\resizebox{88mm}{!}{\includegraphics{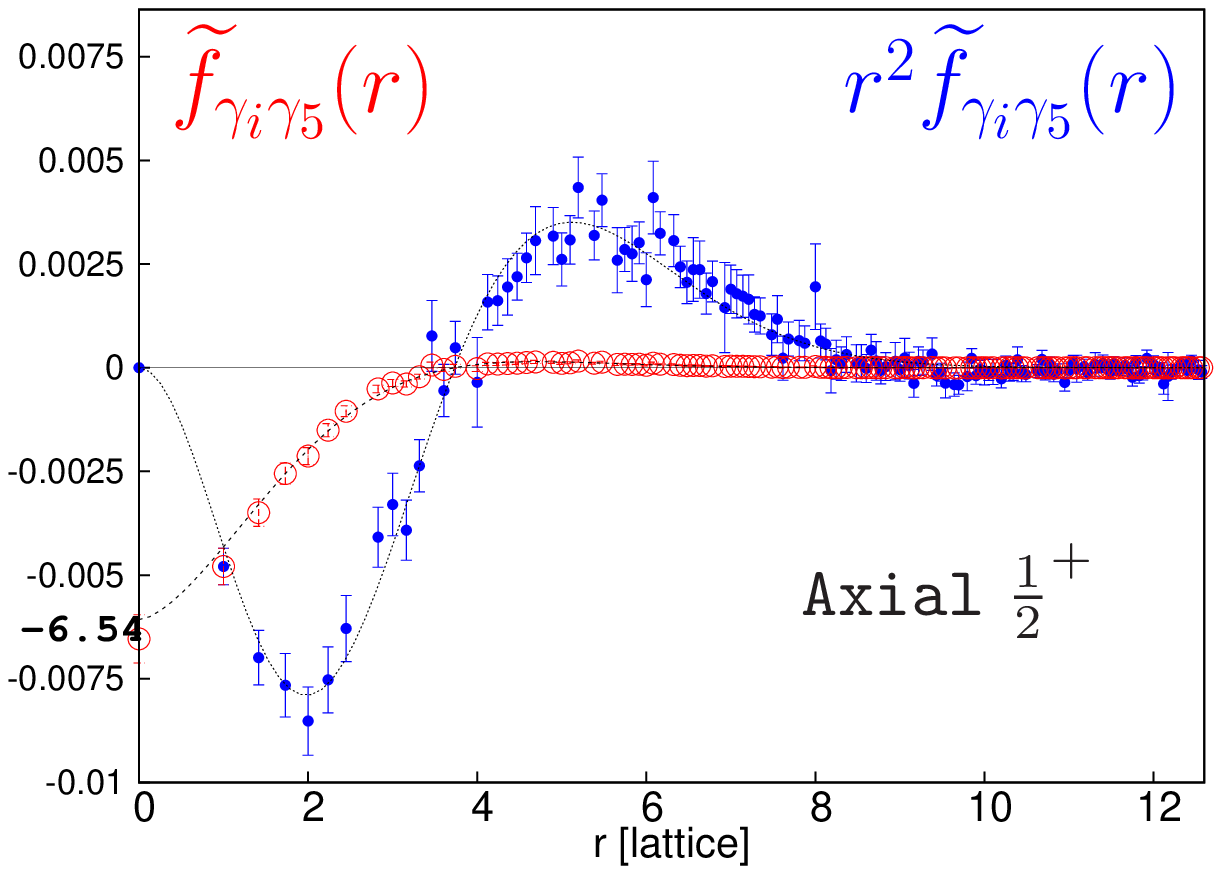}}}\\
 \end{tabular}
\caption{\label{fig:2}\footnotesize{\sl 
Radial distribution of the vector, axial and the scalar density within the static-light meson. On the left side the distributions are obtained with the external $(1/2)^-$ mesons, and those on the right with the $(1/2)^+$ 
ones. light-quark is $\kappa_q=0.1374$, which is around the physical strange quark mass. All results are displayed in lattice units. }} 
\end{center}
\end{figure}

At this point we dispose of quite a number of distributions in $r$: we have two kinds of external sources [$(1/2)^-$ and $(1/2)^+$] per light-quark mass, 
$4$ light-quark masses, and for each of these we have $3$ various densities 
that we are studying in this work. On the top of it we are computing everything with $5$ different heavy-quark actions 
(FAT-6, HYP-1, HYP-1$^2$, HYP-2 and HYP-2$^2$). That makes altogether $120$ distributions with $122$ points each, 
which would result in very cumbersome tables.~\footnote{All those  numbers can be obtained upon request from the authors.} 
Before we opt for one of the heavy-quark actions that we consider in this work, we need to make several important observations:
\begin{itemize}
\item In Fig.~\ref{fig:2} we show the typical example of distributions $f_{\gamma_0,\gamma_i\gamma_5, \mathbb{I}}(r)$. Since they are rapidly varying functions it is more informative to  multiply 
them by $r^2$. Also, those distributions  $r^2 f_{\gamma_0,\gamma_i\gamma_5, \mathbb{I}}(r)$  are shown in Fig.~\ref{fig:2}. Notice that the results are given in lattice units. 
We also explicitly show the value of each distribution at $r=0$. The heavy-quark action used to produce the results shown in  Fig.~\ref{fig:2} is HYP-2$^2$. 
Our observation is that by using various improved heavy-quark actions, the results are broadly compatible, but the quality of the signal with the HYP actions is much better than 
with the simple FAT-6 one.  The resulting distributions obtained with the action HYP-1$^2$ are better than those obtained with HYP-1 (the shapes of distributions are smoother, 
and the statistical errors are somewhat smaller).  The results obtained with HYP-2 action are as good as those obtained with HYP-1$^2$ (they are even numerically close). 
Furthermore, the results with HYP-2$^2$ are indistinguishable from the simple HYP-2 action. 
Therefore, it seems that with the Iwasaki gauge action the choice of parameters $\vec \alpha_{\mbox{\scriptsize HYP-2}} =(1,1,0.5)$ is more optimized than the HYP-1 action.  
\item The same observations are valid for the excited external states  $(1/2)^+$. Moreover, in this case the benefit of using the HYP-2$^2$ heavy-quark action is more obvious 
when applied to the data obtained with our two lightest quarks. For that reason in what follows we will restrain our attention to the results obtained by using the heavy HYP-2$^2$ action.
\item Before accounting for any renormalization factors we generically observe that for $r=0$, $f_{\mathbb{I}}(0) > f_{\gamma_0,\gamma_i\gamma_5}(0)$. In addition, still for unrenormalized distributions,  
we  confirm the observation made in ref.~\cite{green}, namely that  $f_{\mathbb{I}}(r)$ is falling faster with $r$ than  $f_{\gamma_0}(r)$. They cross at about  $r_{\rm cross}= 3$ for our least light-quark, the point that 
then slowly shifts to $r_{\rm cross} \simeq 3.3$, for our lightest quark. 
\item As we see from Fig.~\ref{fig:2}, all distributions $r^2 f_\Gamma (r)$ reach their maximum at some $r_m$ before asymptotically falling to zero at larger $r$'s. From our data we see that, as the light-quark mass is lowered, the maximum, 
\bea
{ \partial\over \partial r}\left[ r^2 f_\Gamma (r) \right]_{r=r_m}\!\!\!\!\! = 0\;,  
\eea
is reached between $2.4$ and $2.9$  for $\Gamma=\one$, and    $2.9\leq r_m\leq 3.7$  for $\Gamma=\gamma_0$. Finally for 
$\Gamma=\gamma_i\gamma_5$, we see that $r_m\approx 3$ for all our light-quark masses.
\end{itemize} 
For an easier use of our results, especially when comparing with various quark models, we fit the logarithm of our distributions to the polynomial of degree four, and the corresponding coefficients are 
given in Appendix~A.~\footnote{The only exceptions are the scalar and axial distributions with the excited external states, which change the sign and therefore taking their logarithm would not make sense.}
The explicit values for  all  distributions considered in this work, as  obtained with the action HYP-2$^2$, are given in  Appendix~B.

\subsection{Vector Charge}
In the following we explain how  we  compute the charge corresponding to each of the distributions discussed in this section so far. 
We checked that to a very good accuracy (indistinguishable by our error bars)
\bea
\sum_{r_x,r_y,r_z} f_\Gamma (r_x,r_y,r_z) = 4 \pi \int_0^{\infty} r^2 f_\Gamma (r) dr\,,
\eea
where on the left-hand side we have the sum over all of the lattice points. On the right-hand side, instead, to get the radial distributions we averaged over various combinations of $(r_x,r_y,r_z)$ which correspond 
to a same value of $r^2$.  Above equality is not in principle obvious since the right-hand side makes use of the spherical symmetry of the problem, and we then integrate over the angular dependence, 
while our calculations are made on the cubic lattice. The reason for the validity of the above equality lies in the fact that our distributions are negligibly small even before they reach the edges of the lattice. 
Notice also that the upper bound of the above integral  ``${\infty}$"  can be identified with $r=12$ (in lattice units), and further we go beyond that point we are picking the contributions from the lattice points that 
do not respect the spherical symmetry. We tested that the integration by including the points that go beyond $r=12$ never modify our integral by more than $1\%$.  
Finally,  concerning the error estimate when doing the integral we estimate them  by using the jackknife procedure. 

\begin{figure}
 \begin{tabular}{cc}
  \resizebox{80mm}{!}{\includegraphics{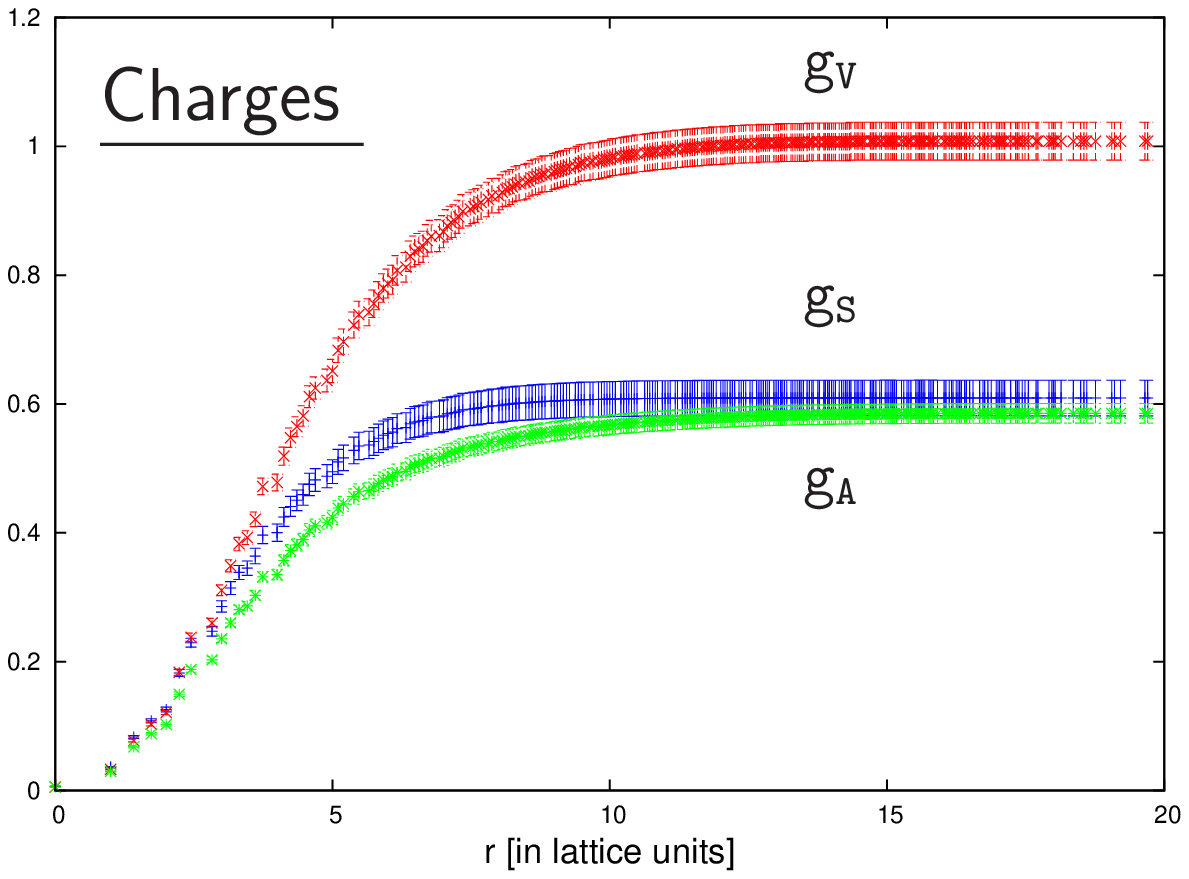}} & \resizebox{80mm}{!}{\includegraphics{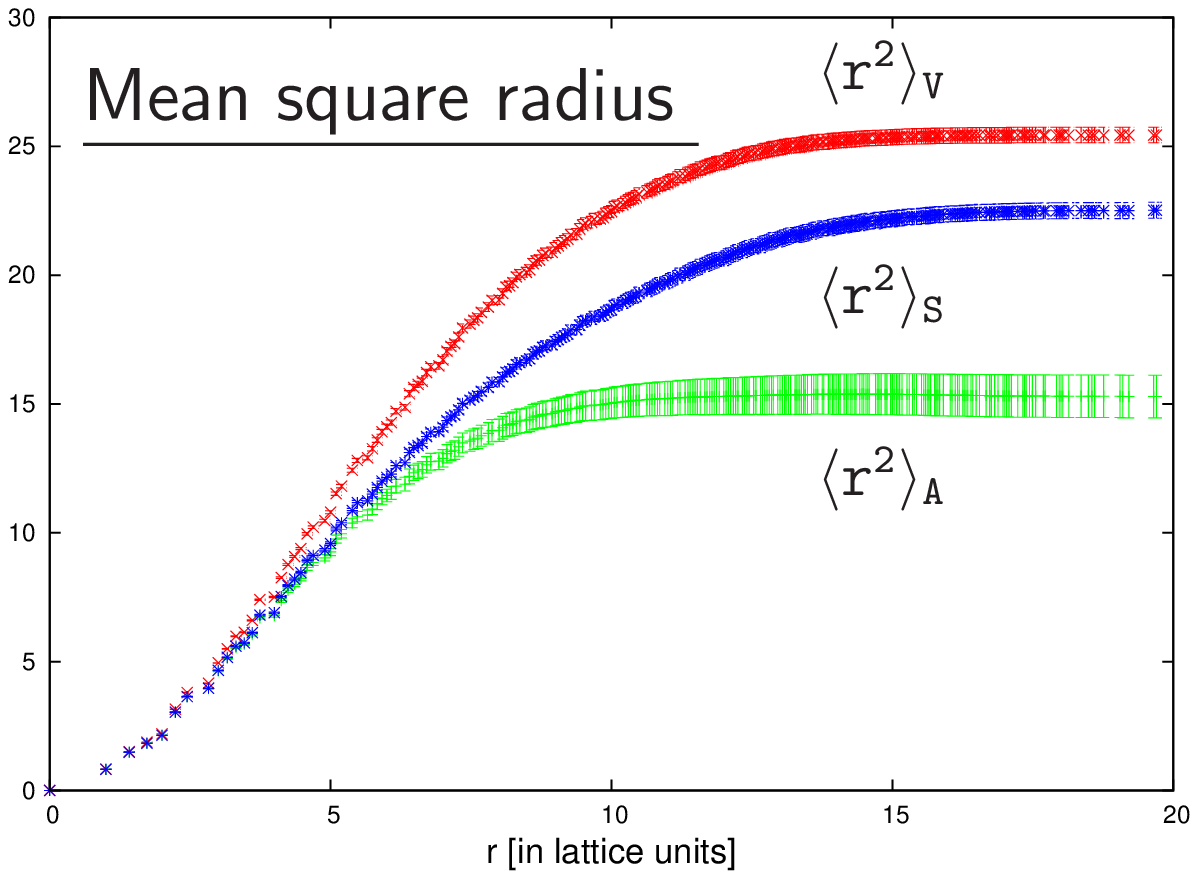}}\\
 \end{tabular}
\caption{\label{fig:3}\footnotesize{\sl 
Lattice sums with $\kappa = 0.1374$ static quark action $\mathtt{HYP2-}^2$.}}
\end{figure}

We now focus on the vector current and its charge. Since our heavy-quark is a spectator, the conservation of the light-quark vector current fixes the vector charges
\bea\label{int-V}
g_V^{(q)} =4 \pi \int_0^{\infty}  r^2 f^q_{\gamma_0} (r) dr \,, \quad {\rm and}\quad \widetilde g_V^{(q)} =4 \pi \int_0^{\infty}  r^2 \widetilde  f^q_{\gamma_0} (r) dr \,,
\eea
to unity, once we multiply them by the appropriate renormalization constant, i.e. $Z_V(g_0^2) g_V = Z_V(g_0^2) \widetilde g_V = 1$. This is therefore simply a method  to nonperturbatively 
evaluate the vector current renormalization constant $Z_V(g_0^2)$,  which we then confront to the perturbatively estimated values  to chose the suitable {\sl ``boosted"} coupling. 
Only  then we can use more reliably the perturbative formulas to renormalize other charges that will be discussed in what follows. In Table~\ref{tab:3} we list our results for both 
$g_V^{(q)}$ and $\widetilde g_V^{(q)}$ for all of our quark masses (i.e. $\kappa_q$). 
Moreover, we present those values for all kinds of improvement of  the heavy-quark propagator discussed in the previous section. We again see that the results with the action HYP-2$^2$  are 
indeed the best, and are compatible with those obtained by using other heavy-quark actions, as they should. It is also clear that the results with the excited heavy-light mesons $(1/2)^+$ are 
noisier, however compatible with what we obtain with the ground external states, $(1/2)^-$.  
We also checked the saturation of the sum/integral in eq.~(\ref{int-V}), and we found that the integral is already fully saturated for $r \leq 11$, for all of the quark masses discussed in this paper. 
We illustrate this saturation for one of our light-quark masses in Fig.~\ref{fig:3}.
To evaluate the renormalization constant we then linearly extrapolate $1/g_V^{(q)}$ in $am_q$ to the chiral limit.~\footnote{Here $am_q=(1/2)[1/\kappa_q-1/\kappa_{\rm crit}]$, where the critical 
hopping parameter has been evaluated in ref.~\cite{CP-PACS} and reads $\kappa_{\rm crit}=0.138984(13)$.} From our data with the HYP-2$^2$ action, 
the intercept is then  $Z_V^{(0)}(g^2_0)=0.739(40)$. We checked, however, that this result remains stable regardless of the heavy-quark action in the spectator line. 

This last result in fact helps us choosing a convenient boosted coupling. It turns out that with the simple replacement $g_0^2\to g_P^2= g_0^2/\langle P\rangle$, with the average plaquette being,  
$\langle P\rangle = 0.60220(6)$~\cite{CP-PACS},  we reproduce the nonperturbative result. Using the perturbative expressions reported in ref.~\cite{taniguchi}, the coupling $\alpha_s^P(a)=0.3775$ and   $c_{\rm SW}=1.47$, 
we obtain  $Z_V^{\rm bpt}(g_0^2)=0.780$. From now on to renormalize each of our currents we will use the boosted perturbative values for the renormalization and improvement constants, 
the values of which will be reported when necessary.
\begin{table}[t!]
\begin{center} 
{\scalebox{.8}{
\begin{tabular}{|c|c|c c|c c|}
\hline
{\phantom{\Huge{l}}}\raisebox{-.1cm}{\phantom{\Huge{j}}}
Heavy-quark & {$\kappa_q$}& $g_V^{(q)}$ & $\left< r^2 \right>_{{\frac{1}{2}}^-}$ & $\widetilde g_V^{(q)}$ & $\left< r^2 \right>_{{\frac{1}{2}}^+}$ \\  \hline
{\phantom{\Huge{l}}}\raisebox{-.1cm}{\phantom{\Huge{j}}} FAT-6       & 0.1357 &     
    1.01(5) &          17.1(2) &         1.22(31) &         17.3(18)\\
{\phantom{\Huge{l}}}\raisebox{-.1cm}{\phantom{\Huge{j}}}             & 0.1367 &     
    1.08(5) &          20.3(3) &         1.12(22) &         20.9(21)\\
{\phantom{\Huge{l}}}\raisebox{-.1cm}{\phantom{\Huge{j}}}             & 0.1374 &     
    1.30(8) &          23.2(3) &         1.31(29) &         17.2(25)\\
{\phantom{\Huge{l}}}\raisebox{-.1cm}{\phantom{\Huge{j}}}             & 0.1382 &     
    1.24(7) &          27.7(5) &         1.43(36) &         29.0(18)\\
\hline
{\phantom{\Huge{l}}}\raisebox{-.1cm}{\phantom{\Huge{j}}} HYP-1       & 0.1357 &     
    1.05(4) &          17.2(2) &         1.23(26) &         16.6(16)\\
{\phantom{\Huge{l}}}\raisebox{-.1cm}{\phantom{\Huge{j}}}             & 0.1367 &     
    1.09(4) &          20.5(3) &         1.24(20) &         21.0(17)\\
{\phantom{\Huge{l}}}\raisebox{-.1cm}{\phantom{\Huge{j}}}             & 0.1374 &     
    1.26(5) &          23.3(3) &         1.24(23) &         18.0(23)\\
{\phantom{\Huge{l}}}\raisebox{-.1cm}{\phantom{\Huge{j}}}             & 0.1382 &     
    1.22(6) &          28.2(4) &         1.36(27) &         29.4(17)\\
\hline
{\phantom{\Huge{l}}}\raisebox{-.1cm}{\phantom{\Huge{j}}} HYP-1{}$^2$ & 0.1357 &     
    1.08(4) &          17.7(2) &         1.26(21) &         17.7(13)\\
{\phantom{\Huge{l}}}\raisebox{-.1cm}{\phantom{\Huge{j}}}             & 0.1367 &     
    1.13(3) &          21.1(2) &         1.20(16) &         22.1(13)\\
{\phantom{\Huge{l}}}\raisebox{-.1cm}{\phantom{\Huge{j}}}             & 0.1374 &     
    1.23(4) &          23.8(3) &         1.15(16) &         18.0(20)\\
{\phantom{\Huge{l}}}\raisebox{-.1cm}{\phantom{\Huge{j}}}             & 0.1382 &     
    1.26(5) &          28.8(4) &         1.43(23) &         28.7(15)\\
\hline
{\phantom{\Huge{l}}}\raisebox{-.1cm}{\phantom{\Huge{j}}} HYP-2       & 0.1357 &     
    1.08(4) &          17.8(2) &         1.24(19) &         17.8(13)\\
{\phantom{\Huge{l}}}\raisebox{-.1cm}{\phantom{\Huge{j}}}             & 0.1367 &     
    1.12(4) &          21.2(2) &         1.15(16) &         22.7(13)\\
{\phantom{\Huge{l}}}\raisebox{-.1cm}{\phantom{\Huge{j}}}             & 0.1374 &     
    1.21(4) &          23.8(3) &         1.11(14) &         17.8(20)\\
{\phantom{\Huge{l}}}\raisebox{-.1cm}{\phantom{\Huge{j}}}             & 0.1382 &     
    1.26(6) &          28.8(4) &         1.35(22) &         28.7(16)\\
\hline
{\phantom{\Huge{l}}}\raisebox{-.1cm}{\phantom{\Huge{j}}} HYP-2{}$^2$ & 0.1357 &     
    1.10(3) &          18.4(2) &         1.25(18) &         18.4(12)\\
{\phantom{\Huge{l}}}\raisebox{-.1cm}{\phantom{\Huge{j}}}             & 0.1367 &     
    1.15(3) &          21.8(2) &         1.19(14) &         23.1(12)\\
{\phantom{\Huge{l}}}\raisebox{-.1cm}{\phantom{\Huge{j}}}             & 0.1374 &     
    1.22(4) &          24.4(3) &         1.17(14) &         18.9(17)\\
{\phantom{\Huge{l}}}\raisebox{-.1cm}{\phantom{\Huge{j}}}             & 0.1382 &     
    1.29(6) &          29.3(3) &         1.44(22) &         27.4(15)\\
\hline
\end{tabular}  }}
\caption{\label{tab:3}
\footnotesize  Vector charges obtained by integrating the corresponding vector current densities and for both the $(1/2)^-$ and $(1/2)^+$ external states. 
}
\end{center}
\vspace*{-.3cm}
\end{table}
One interesting by-product of our analysis is the charge radius of the heavy-light meson. We remind the reader that the charge radius provides us with information concerning the slope of 
the electromagnetic form factor [$F(q^2)$] of the pseudoscalar meson $P$ at $q^2=0$, i.e. 
\bea
\langle P(p^\prime )\vert J_\mu^{\rm em} \vert P(p)\rangle = F(q^2) \left( p + p^\prime \right)\,,\quad \left< r^2 \right> = 6 \left. {d F(q^2)\over dq^2}\right|_{q^2=0}\,,
\eea
where $q=p-p^\prime$, and $J_\mu^{\rm em}=e_Q \bar Q\gamma_\mu Q + e_q \bar q\gamma_\mu q$, with $ e_{Q(q)}$ being the electric charge of the heavy(light) quark. 
Similarly one can define the charge radius of the scalar meson.   
Since we already have radial distribution of the light-quark component of the electromagnetic form factor we can simply compute its moment, namely, 
\bea\label{r22}
 \left< r^2 \right>_{{\frac{1}{2}}^-} ={  \displaystyle{ \int_0^{\infty}  r^4 f_{\gamma_0} (r)\  dr}  \over \displaystyle{  \int_0^{\infty}  r^2 f_{\gamma_0} (r)\  dr } }\,, 
\eea
and the corresponding results obtained from our data are also listed in Table~\ref{tab:3}. Similarly for the $(1/2)^+$-states. As it could have been anticipated,  the saturation of the integral (lattice sum) 
in the numerator of eq.~(\ref{r22}) occurs at higher values of $r$,  since the extra $r^2$ in the sum enhances the importance of the values  of the rapidly falling distribution $f_{\gamma_0} (r)$ at larger radial distances. 
This is also illustrated in Fig.~\ref{fig:3}.
We do not attempt an elaborate chiral extrapolation here. We relegate that discussion to our future publication. We note, however, that  by simply linearly 
extrapolating to the chiral limit we obtain 
\bea\label{rrr}
&& \left< r^2 \right>_{{\frac{1}{2}}^-} = (31.6\pm 0.3)\times a^2\; \Longrightarrow  \left< r^2 \right>^{\rm lin.}_{{\frac{1}{2}}^-} =0.334(3)~{\rm fm}^2\,,\cr
&&  \left< r^2 \right>_{{\frac{1}{2}}^+} = (28.3\pm 1.6)\times a^2\; \Longrightarrow  \left< r^2 \right>^{\rm lin.}_{{\frac{1}{2}}^+} =0.280(16)~{\rm fm}^2\,,
\eea
where we fixed the lattice spacing to  $a=0.0995(4)$~fm, from  $r_0/a=4.695(18)$, as obtained on this same lattice~\cite{CP-PACS}, and by using $r_0^{\rm phys}=0.467$~fm.
It is interesting to note that these values are slightly lower but comparable to what one would obtain by applying the simple vector meson dominance, i.e. $ \left< r^2 \right>_{\rm VMD}=6/m_\rho^2 = 0.393\  {\rm fm}^2$, and it is considerably 
smaller than the results obtained from the quark model of ref.~\cite{charge-radius}.~\footnote{We would like to stress that this is a feature of that specific model. In our next paper we will show that other models are quite consistent with  $\left< r^2 \right>_{{\frac{1}{2}}^\pm}$  obtained on the lattice~\cite{workinprogress}. }
Another phenomenologically  interesting result can be obtained  by using our results~(\ref{rrr}) to fix the left-hand side of the Cabibbo-Radicati sum rule for heavy hadrons discussed in ref.~\cite{CRSR}
\bea\label{crsr}
\left. {d F(q^2)\over dq^2}\right|_{q^2=0} =  {1\over 8\alpha_{\rm em}  e_q^2 } \sum_{\rm exc} (2 J + 1) {\Gamma(B_{\rm exc}\to B\gamma )\over \vert \vec k_\gamma\vert^3 } \,,
\eea
where the $B^\ast$ contribution has been transferred from the left-hand side to the sum on the right-hand side.  By only keeping that very first term in the sum, we obtain a bound
\bea
\Gamma( B^{\ast \pm} \to B^\pm \gamma ) \leq {16\over 81}  \alpha_{\rm em}  \left< r^2 \right> \vert \vec k_\gamma\vert^3 = 1.16(1)~{\rm keV}\,.
\eea
Similarly, $\Gamma( B^{\ast 0} \to B^0 \gamma ) \leq 0.29$~keV, where we used $\vert \vec k_\gamma\vert = (m_{B^\ast}^2-m_B^2)/(2 m_{B^\ast}) =45.4$~MeV.  
It should be emphasized that this upper bound is probably very weak as we will show in our forthcoming paper where we verify that, in the quark models, the magnetic moment is expected to be much smaller than what is suggested by this bound~\cite{workinprogress}.

\subsection{Scalar Charge}
Concerning the scalar charge and the mean square radius of the corresponding scalar form factor, we integrate the matter density 
\bea\label{int-SS}
g_S^{(q)} =4 \pi \int_0^{\infty}  r^2 f^q_{\mathbb{I} } (r) dr \,, \quad {\rm and}\quad    \left< r^2 \right>^S_{{\frac{1}{2}}^-} ={  \displaystyle{ \int_0^{\infty}  r^4 f_{\mathbb{I}} (r)\  dr}  \over \displaystyle{  \int_0^{\infty}  r^2 f_{\mathbb{I}} (r)\  dr } } \,,
\eea
and similarly for the $(1/2)^+$ states which we --as before-- distinguish by an extra tilde. 
In this case we only present our results obtained with the improved heavy-quark action HYP-2$^2$, and they are listed in Table~\ref{tab:4}.
\begin{table}[t!]
\begin{center} 
{\scalebox{.9}{
\begin{tabular}{|c||cc|cc||cc|cc|}
\hline
{\phantom{\Huge{l}}}\raisebox{-.1cm}{\phantom{\Huge{j}}}
 {$\kappa_q$}& $g_S^{(q)}$ & $\left< r^2 \right>^S_{{\frac{1}{2}}^-}$ & $\widetilde g_S^{(q)}$ & $\left< r^2 \right>^S_{{\frac{1}{2}}^+}$ & $g_A^{(q)}$ & $\left< r^2 \right>^A_{{\frac{1}{2}}^-}$ & $\widetilde g_A^{(q)}$ & $\left< r^2 \right>^A_{{\frac{1}{2}}^+}$\\  \hline
{\phantom{\Huge{l}}}\raisebox{-.1cm}{\phantom{\Huge{j}}} 0.1357 &    0.938(33) &    
 12.6(3) &   0.91(17) &     26(3) &    0.695(18) &      15.1(2) & -0.077(42)
&   --\\
{\phantom{\Huge{l}}}\raisebox{-.1cm}{\phantom{\Huge{j}}} 0.1367 &    0.963(36) &    
 15.1(5) &   0.90(18) &     36(3) &    0.705(18) &      18.1(2) & -0.035(40)
&   -- \\
{\phantom{\Huge{l}}}\raisebox{-.1cm}{\phantom{\Huge{j}}} 0.1374 &    0.954(43) &    
 15.3(7) &   0.70(22) &     34(6) &    0.683(18) &      20.7(3) & -0.088(36)
&   -- \\
{\phantom{\Huge{l}}}\raisebox{-.1cm}{\phantom{\Huge{j}}} 0.1382 &      1.02(8) &    
18.9(2.2) &     1.01(40) &     50(8) &    0.657(25) &      24.6(6) &   -0.220(55)
&  -- \\ \hline
\end{tabular}  }}
\caption{\label{tab:4}
\footnotesize The scalar and the axial charges obtained by integrating the axial and matter densities as indicated in eqs.~(\ref{int-SS},\ref{int-A}). We also present the results for the corresponding mean square radius. The results for the 
external ground and the excited heavy-light states are displayed. The values presented in this Table refer to the static heavy spectator quark obtained by using the 
HYP-2$^2$ action.}
\end{center}
\vspace*{-.3cm}
\end{table}
The perturbative matching to the continuum includes the following constants: $Z_S^{(0) \msbar}(1/a)^{\rm bpt}=0.602$ and $b_S^{\rm bpt}=1.472$,~\footnote{As a side note we should state  that for the situations considered in this paper, 
no ${\cal O}(a)$ improvement of the bare current gives a non-zero contribution. The improvement of the renormalization constants includes $Z_{S,V,A}(g_0^2) = Z_{S,V,A}^{(0)}(g_0^2)[1 +   b_{S,V,A}^{(0)}(g_0^2) am_q ]$, with  $Z_{S,V,A}^{(0)}(g_0^2)$ being the renormalization constants in the chiral limit, while $b_{S,V,A}^{(0)}(g_0^2) $ are the linear mass counter-term coefficients. As we discuss in the text, these quantities are taken from boosted $1$-loop 
perturbative formulas~\cite{taniguchi}.} so that after linearly extrapolating to the chiral limit we obtain 
\bea\label{S-CHARGE}
&&g_S^{\msbar}(1/a) = 0.606(37),\;{\rm and}\quad   \left< r^2 \right>^S_{{\frac{1}{2}}^-} = 0.193(9)~{\rm fm}^2\,,\nn \\
&&\widetilde g_S^{\msbar}(1/a) = 0.45(18),\;{\rm and}\quad   \left< r^2 \right>^S_{{\frac{1}{2}}^+} = 0.54(6)~{\rm fm}^2\,,
\eea
where we notice a strong deterioration of the signal for the  $(1/2)^+$  external states. 

So far by ``$q$" we denoted a light-quark without specifying its flavor. It stands for  $u$ or $d$, the two degenerate quarks in our  configurations with $N_{\rm f}=2$  dynamical light-quarks. 
With the scalar density, however, one should be particularly careful. We combine our quarks in a flavor-doublet  $\psi = [u\,\ d]^T$, so that the neutral current can be either flavor singlet or flavor triplet, i.e.
\bea
J_S^{\rm sing.}=\bar \psi {\tau_0\over 2}\one \psi\,,\quad \, J_S^{\rm trip.}=\bar \psi {\tau_3\over 2}\one \psi\, ,
\eea
with $\tau_0={\rm diag}(1,1)$, and $\tau_3={\rm diag}(1,-1)$. In other words, 
\bea
\bar u\one u = J_S^{\rm sing.} +  J_S^{\rm trip.}\,,\qquad \bar d\one d = J_S^{\rm sing.} -  J_S^{\rm trip.}\,.
\eea
The problem is that the computation with the singlet current contains disconnected diagrams, which are particularly important in this case.
Their computation on the lattice requires the implementation of the {\sl ``all-to-all propagator''} technique~\cite{all2all}.  In this work we neglect the disconnected diagrams (in other words, we compute only the flavor 
nonsinglet part of scalar density). Beside the computational complications in evaluating the disconnected diagrams, an extra complication arise from the fact that, on the lattice, the $\bar q q$ operator  
mix with the cubically divergent  $(d_1/a^3)\one $ [and, away from the chiral limit, also with $(d_2 m_q/a^2)\one $, and $(d_3 m_q^2/a)\one $], where $d_{1,2,3}$ are the constants whose values should be 
determined and mixing subtracted away before multiplicative renormalization. 
To check on the importance of disconnected diagrams we can compare our results with the  light-quark mass dependence of the binding energy ${\cal E}_q$. In other words, we would like to verify if the following equality is satisfied:~\footnote{
This is also known as Feynman-Hellmann theorem, discussed in ref.~\cite{quigg} and references therein.}
\bea\label{FHT}
{\partial {\cal E}_q\over \partial m_q} = \langle B_q\vert {\partial {\cal H}_q\over \partial m_q}\vert B_q\rangle  = \langle B_q\vert \bar q q\vert B_q\rangle \stackrel{?}{=} g_S^{\msbar}(1/a)\,,
\eea
where ${\cal H}_q= m_u \bar u u + m_d \bar d d$, and either we choose $q=u$ or $q=d$ the evaluation of the left-hand side of the above equation will contain the disconnected diagrams.  
The last $g_S$ on the right-hand side is our result obtained in eq.~(\ref{S-CHARGE}) in which the disconnected diagrams were omitted.  
If we fit our values for ${\cal E}_q$, obtained with the heavy-quark action HYP-2$^2$ and presented in Table~\ref{tab:1}, as ${\cal E}_q = {\cal E}_0 + \rho m_q$, 
we obtain 
\bea
&& {\cal E}_0 = 0.369(7)\,, \quad \rho^{\msbar}(1/a) =1.02(7)\,, 
\eea
where we used the quark mass defined via the vector Ward identity, $m_q= [Z_S^{(0) \msbar}(1/a)]^{-1} m_q^{(0)}$ $\times ( 1 - m_q^{(0)} b_S/2 )$, and $m_q=1/2 (1/\kappa_q-1/\kappa_{\rm crit})$. 
This result is more than $60\%$ larger than our value for $g_S^{\msbar}(1/a) = 0.606(37)$. We also attempted using the quark mass defined via the axial Ward identity (the bare values of which are 
given in ref.~\cite{CP-PACS}, which we recomputed and confirmed). The advantage of using the quark mass defined via the axial Ward identity lies in the fact that the correction due to $b_A-b_P =-0.009$, 
is negligible. From that fit we obtain that the slope is $1.28(9)$, and after accounting for the proper renormalization factor (i.e. $[Z_P(1/a)/Z_A]^{\rm bpt}=0.692$), we have $\rho^\prime=0.89(6)$, hence still 
about $50\%$ larger than our $g_S^{\msbar}(1/a)$ in eq.~(\ref{S-CHARGE}). Therefore it appears that a contribution due to disconnected diagrams is indeed important and large. We checked that this conclusion remains as such if we fit around any of our directly accessed light-quark masses [i.e.  ${\cal E}_q = {\cal E}_1 + \rho_1 (m_q - m_1)$], since eq.~(\ref{FHT}) is valid for any light-quark.

The calculation of  disconnected diagrams is not only demanding  computationally. It also requires to solve a problem of mixing with the power divergent operator even in the chiral limit, 
which is therefore yet another  source of systematic error when computing the light-quark mass dependence. 
As it is well known the binding energy computed on the lattice (${\cal E}_q$)  cannot be directly compared with the continuum definition  ($\bar \Lambda_q$)  unless the linearly divergent contribution 
due to the lattice regularization is properly subtracted, i.e. ${\cal E}_q =  \bar \Lambda_q + {c(g_0^2)/a}$. 
In addition, when away from the chiral limit  the subtraction constant becomes $c(g_0^2) = c_0(g_0^2) + am_q c_1(g_0^2)$. By taking  $\bar \Lambda_q 
=\bar \Lambda_0 + \rho_\Lambda \times  m_q$, with $\bar \Lambda_0$ being the binding energy of the static-light meson in the chiral limit,  then 
\bea\label{laast}
{\cal E}_q =  \bar \Lambda_0   +  {c(g_0^2)\over a} + \left[ \rho_\Lambda +  c_1(g_0^2) \right]  m_q \qquad \Rightarrow \qquad {\partial {\cal E}_q\over \partial m_q} =\underbrace{ \rho_\Lambda +  c_1(g_0^2)}_{\displaystyle{\rho}}\,.
\eea
To check whether or not this latter artifact is (partly) responsible for not verifying the equality in eq.~(\ref{FHT}) it is indispensable to compute the disconnected diagram and see to what extent 
its contribution would saturate the equality~(\ref{FHT}). 
We should also add that eq.~(\ref{laast})  is one reason more why --in HQET on the lattice-- it  is better to compute the mass differences, such as $\Delta_q=\widetilde {\cal E}_q -{\cal E}_q$ 
because in that case the linear divergence cancels altogether.    
Linear fit to our data gives
\bea
\Delta_q = 0.187(17)  + 0.58(19)\times  m_q^{\msbar}(2 \ \gev)\ .
\eea 
In physical units this leads to $\Delta_0= 372(37)$~MeV. Notice however that the slope is (clearly) positive.  
Therefore, {\underline{at the masses considered in this work}}, the lattice results exhibit a behavior opposite to what is established experimentally in the charm spectrum, namely 
that  $\Delta_q$ increases when one passes from the strange quark ($q=s$)  closer to the chiral 
limit ($q=u,d$) (see discussions in~\cite{puzzles} and references therein). 

\subsection{Axial Charge}

In Table~\ref{tab:4}, we also presented our results for the axial charge and its corresponding mean square radius, defined analogously to what we did with the vector and scalar charges
\bea\label{int-A}
&&g_A^{(q)} =4 \pi \int_0^{\infty}  r^2 f^q_{A } (r) dr \,, \quad\qquad \quad  \widetilde g_A^{(q)} =4 \pi \int_0^{\infty}  r^2\widetilde  f^q_{A } (r) dr \,, \cr
&&  \left< r^2 \right>^A_{{\frac{1}{2}}^-} ={  \displaystyle{ \int_0^{\infty}  r^4 f_{A}^q (r)\  dr}  \over \displaystyle{  \int_0^{\infty}  r^2 f_{A}^q (r)\  dr } }  \,,
\quad\qquad  \quad    \left< r^2 \right>^A_{{\frac{1}{2}}^+} =
{  \displaystyle{ \int_0^{\infty}  r^4 \widetilde f_A^q (r)\  dr}  \over \displaystyle{  \int_0^{\infty}  r^2 \widetilde f_A^q (r)\  dr } }  \,,
\eea
These quantities are particularly important because they are related to the $P$-wave soft pion emission/absorption in the transitions between the mesons belonging to  the same doublet, 
as discussed in ref.~\cite{g-lattice}. In particular, 
\bea
\langle B_q\vert \bar q \vec \gamma \gamma_5 q \vert B_q^\ast \rangle = g_A^{(q)} \vec \varepsilon\,,\qquad 
\langle B_{0 q}^\ast \vert \bar q \vec \gamma \gamma_5 q \vert B_{1q}^\prime \rangle = \widetilde g_A^{(q)} \vec \varepsilon^\prime\,, 
\eea
where $\varepsilon (\varepsilon^\prime)$ is the polarization vector of the vector $B_q^\ast $ (axial $ B_{1q}^\prime $) static-light meson, and the axial current is now renormalized.  After accounting for the axial renormalization $Z_A^{\rm bpt}(g_0^2)=0.803$, and $b_A^{\rm bpt}(g_0^2)=1.346$,  the simple linear chiral extrapolation gives
\bea
g_A \equiv \hat g = 0.526(22)\,, \qquad \widetilde g_A \equiv \widetilde g = -0.157(48)\,.
\eea
If this extrapolation is guided by expressions derived in chiral perturbation theory (HMChPT)~\cite{jernej}, we get
\bea
\hat g_{\rm HMChPT} = 0.444(18)\,, \qquad \widetilde g_{\rm HMChPT} = -0.14(6)\,.
\eea
Furthermore, the  simple linear extrapolation of the mean square radius to the chiral limit results 
in $\langle r^2\rangle^A_{{\frac{1}{2}}^-} = 0.246(3)~{\rm fm}^2$, and it measures the deviation of the axial coupling 
to the pion that is pulled off its mass shell. In other words, we obtain that  $ \hat g (q^2) = \hat g(0) [ 1 + \langle r^2\rangle^A_{{\frac{1}{2}}^-}/6 \times q^2 ]$, which is steeper  than
the shape predicted by the axial-vector meson dominance, namely  $\langle r^2\rangle^A_{{\frac{1}{2}}^-} = 6/m_{a_1}^2 =0.154(8)~{\rm fm}^2$. 
Since we are considering the coupling to a pion (flavor nonsinglet) and not to the singlet states ($\eta$, $\eta^\prime$), the problem of disconnected diagrams is not present in this situation.  

We should also comment on the values of the axial coupling $\widetilde g_A^{(q)}$, which straddle around zero and takes a more pronounced nonzero value as the light-quark mass becomes lighter. 
From the distributions that are presented in the Appendix~A, and also the one already shown in Fig.~\ref{fig:2}, we can see that they change the sign and have zero at about $r=0.4$~fm, so that 
in the sum the short distance (negative)  and long distance (positive) 
contributions to the coupling $\widetilde g_A^{(q)}$ cancel each other to a large extent, which is one of the reasons why the errors on $\widetilde g_A^{(q)}$ are much larger than in the $(1/2)^-$-doublet case.

\section{Summary and conclusions\label{SecZ}}
In this paper we discussed various spatial (radial) distributions of the light degrees of freedom with respect to the static heavy-quark. We considered both the lowest lying heavy-light mesons, as well as the nearest 
orbitally excited ones. 
We used various improvement of the static heavy-quark action on the lattice. We observe the best efficiency achieved  when the HYP-2$^2$ action is used.
Concerning the (electromagnetic) charge distributions, we find that they have almost the same shape in both cases, i.e. when the external states are either $j_P=(1/2)^-$ or  $(1/2)^+$. The matter and the axial distributions  have different shapes in two cases: while for the external $(1/2)^-$ states they are strictly positive, for the $(1/2)^+$ states they change the sign: they start by being negative and 
change the sign at $r\approx  0.25$~fm for the matter distribution, and at $r\approx  0.4$~fm for the axial one. That zero moves slowly towards larger $r$, as we lower the light-quark mass.  Such a qualitative difference of
the internal structure of $(1/2)^+$-mesons and  the $(1/2)^-$ states is particularly 
evident in the case of axial density. As a result, the integrated characteristics (axial charge) in   $(1/2)^-$  and in $(1/2)^+$-states are completely different. These axial charges are also physically interesting because 
they describe the coupling of a soft pion to the $(1/2)^-$ and to the  $(1/2)^+$ states, respectively. They are  essential parameters in the heavy meson chiral perturbation theory. From our analysis we found it to be $\hat g_{\rm lin.} 
= 0.52(2)$,  and $\hat g_{\rm HMChPT} =0.44(2)$, depending on the extrapolation formula used to go to the chiral limit. For the soft pion coupling to orbitally excited states we obtain, instead, $\widetilde g_{\rm lin.} 
= -0.18(5)$,  and $\widetilde g_{\rm HMChPT} =-0.14(6)$.  This further corroborates the fact that the model of chiral doublings of heavy-light hadrons~\cite{doubletrouble} is so badly broken that even its qualitative features 
should be taken very cautiously.

For each of the distributions we computed the mean square radius, which is a new result. From the linear chiral extrapolation we obtain the values that are 
fully consistent with what one would naively obtain by using the nearest resonance approximation in the crossed channel. In particular by using our result for  $\left< r^2 \right>^V_{ {\frac{1}{2}}^-}=0.33$~fm$^2$, 
and the Cabibbo-Radicati sum rule for the heavy-light mesons~\cite{CRSR}  we obtain  $\Gamma( B^{\ast \pm} \to B^\pm \gamma ) \leq  1.16(1)$~keV. Our results also indicate  that the 
scalar charge and the matter density receive important contribution from the flavour singlet piece, namely,  the one that contains disconnected diagrams, which have not been computed in this work.

As a logical continuation of this project we plan to compare the distributions  computed and discussed in this paper to those obtained by using two specific classes of quark models~\cite{workinprogress}.  
An attempt to study the issue of disconnected diagrams, to complete our study of the matter density  and make the comparison with $\partial {\cal E}_q/\partial m_q$ more conclusive, would be highly welcome. 
In this paper we did not extend our analysis to the radially excited states.  Such an analysis may soon be possible thanks to recent improvement regarding the extraction of the properties of such
 states from the lattice~\cite{radial-lattice}.  A more elaborate study of chiral extrapolation requires the appropriate  expressions in HMChPT, which we also plan to compute. Finally, even though our 
 computations are made on relatively fine lattice and with ${\cal O}(a)$ improved Wilson quarks, it is necessary to check how if our result remain stable if even finer lattices are used. Such a study is underway.
  
Phenomenologically more interesting is the comparison of results presented in this paper to various quark models. One such study is the topic of our forthcoming paper~\cite{workinprogress}.
\section*{Acknowledgements}
We thank  the CP-PACS Collaboration for making their gauge field configurations publicly available,   the {\it Centre de Calcul de l'IN2P3 \`a Lyon}, for giving us access to their computing facilities
and the partial support of ``Flavianet"  (EU Contract No.~MTRN-CT-2006-035482), and of  the ANR (ÒDIAMÓ Contract  No.~ANR-07-JCJC-0031). We also thank V.~Lubicz for discussions and comments.

\begin{figure}
\vspace*{-1.7cm}
\hspace*{-18mm}
\begin{tabular}{c c c}
\resizebox{66mm}{!}{\includegraphics{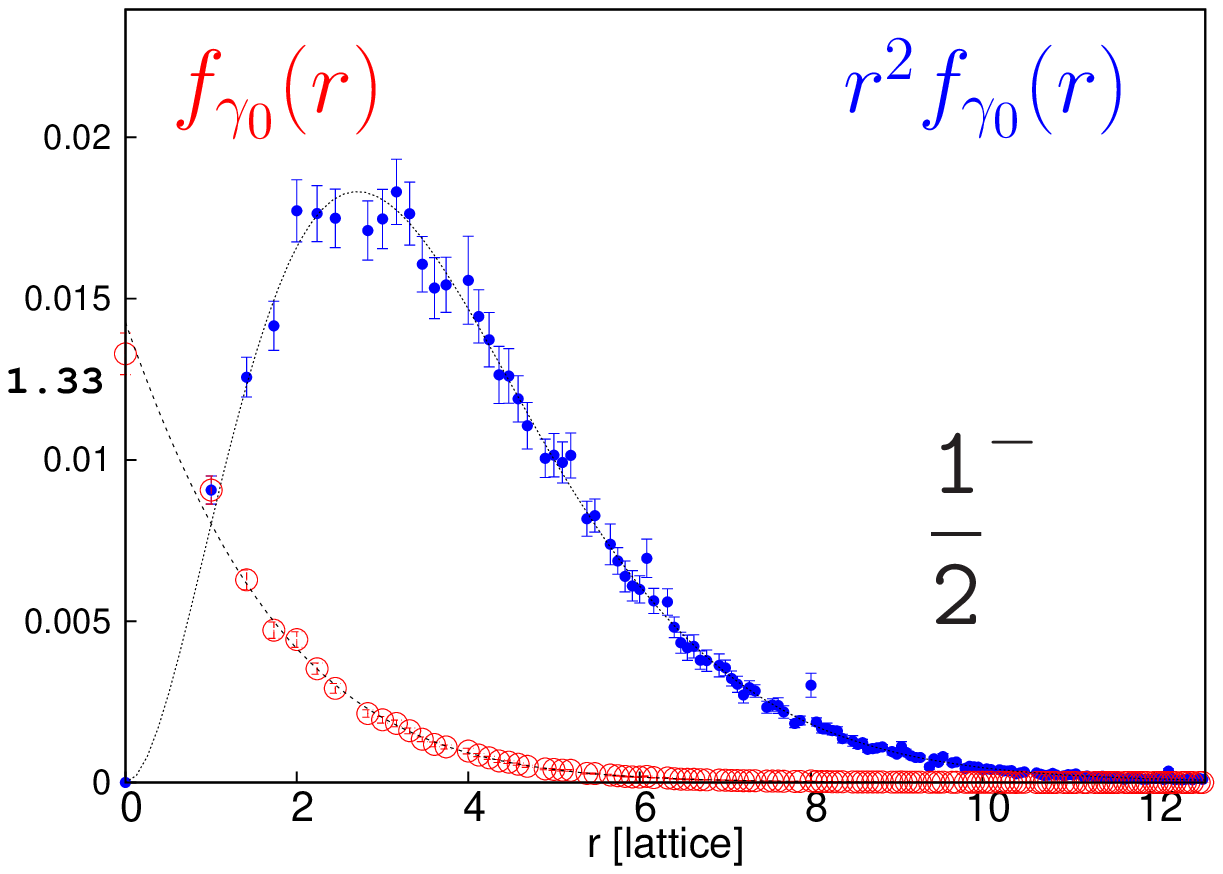}} & \resizebox{66mm}{!}{\includegraphics{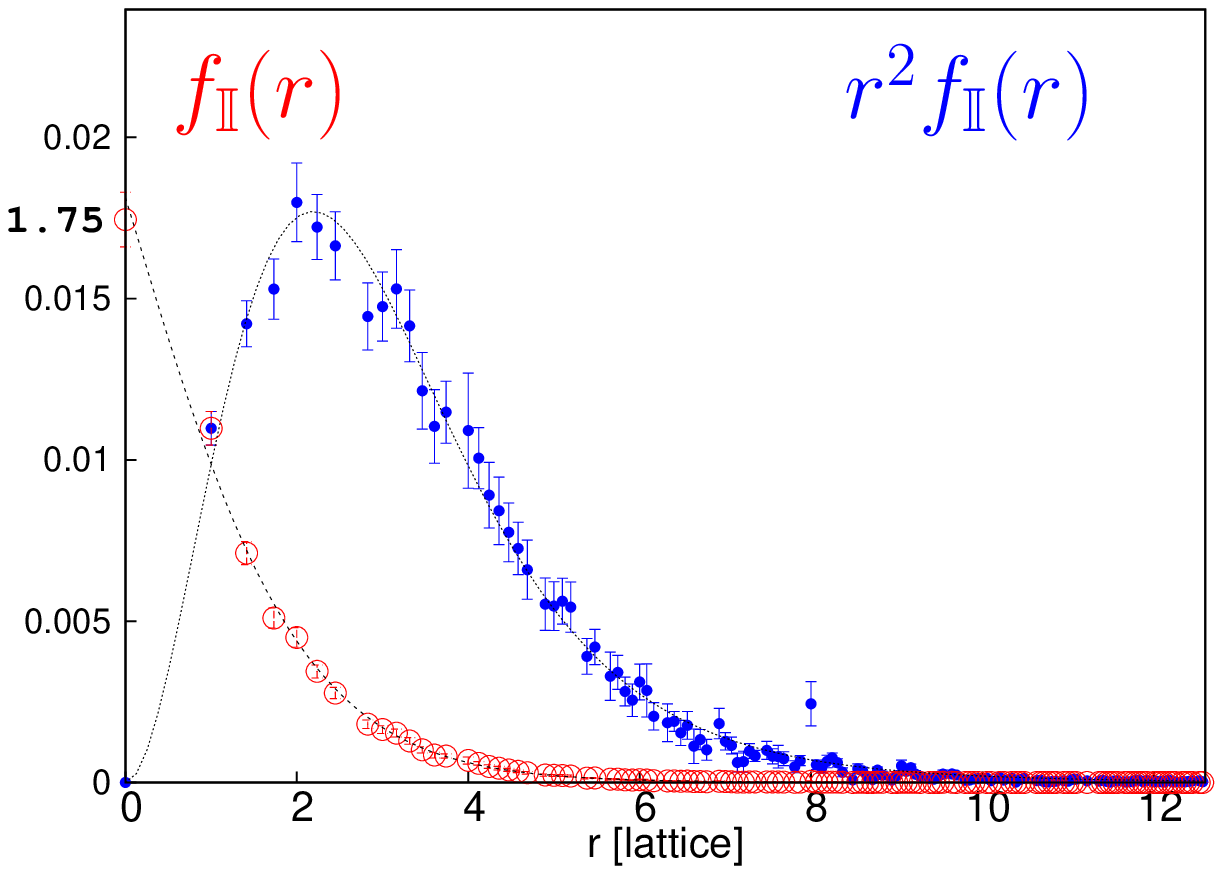}}& \resizebox{66mm}{!}{\includegraphics{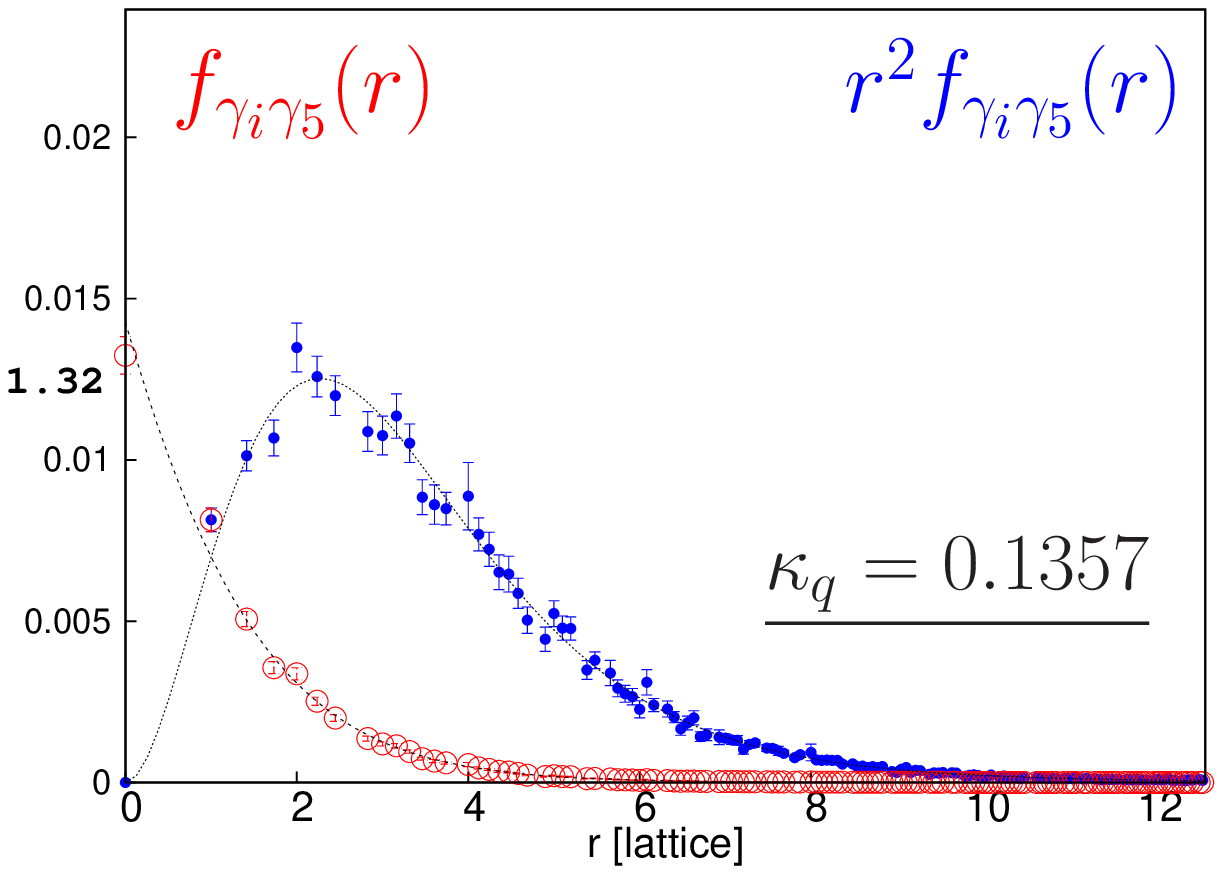}}\\
\resizebox{66mm}{!}{\includegraphics{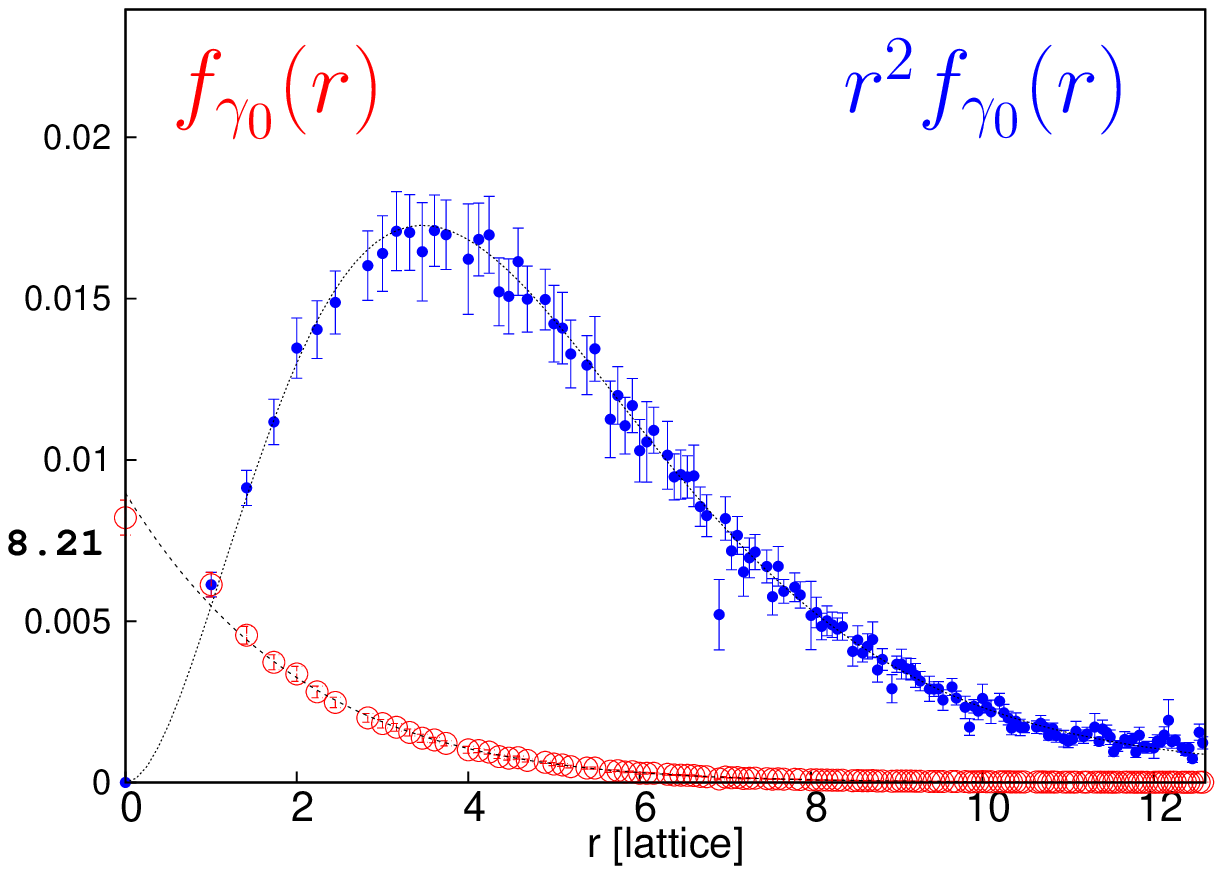}} & \resizebox{66mm}{!}{\includegraphics{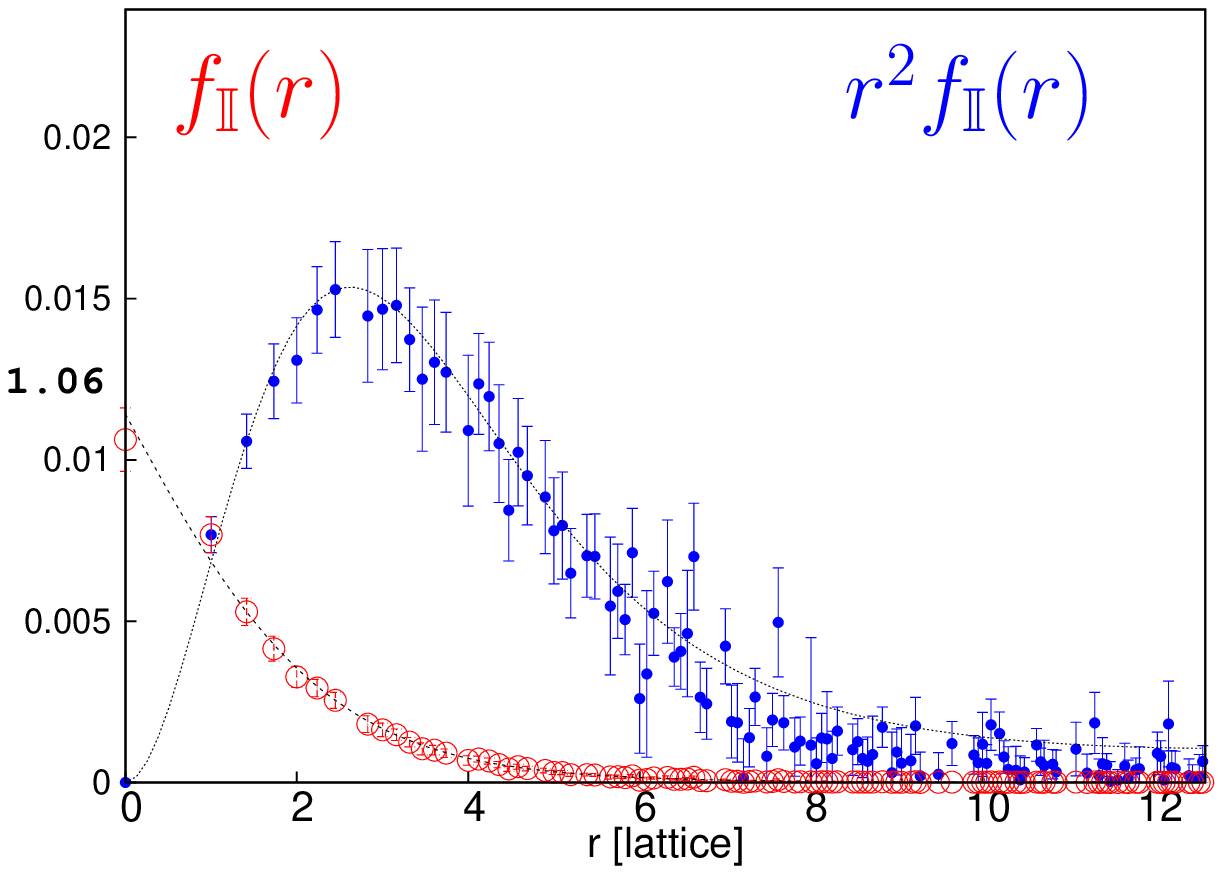}}& \resizebox{66mm}{!}{\includegraphics{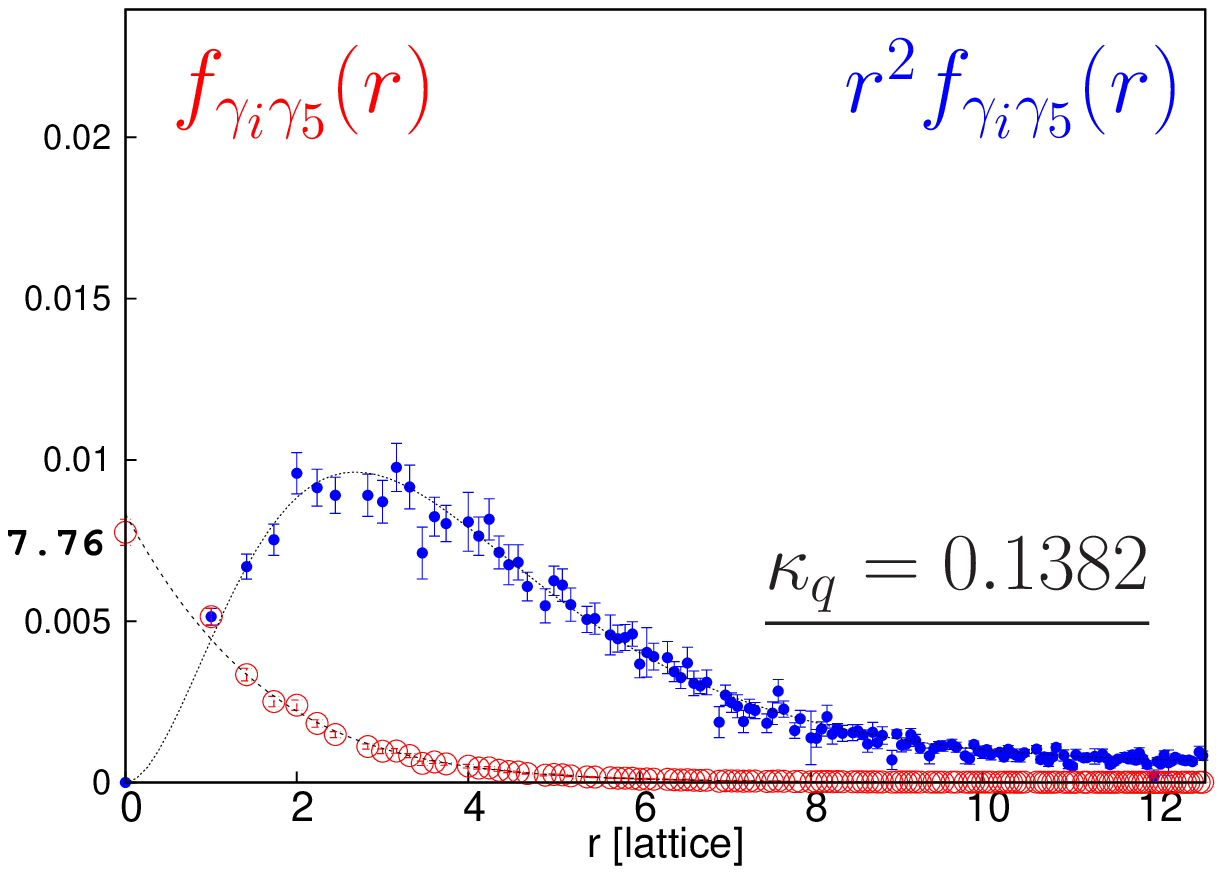}}\\
      && \\
 \hline    && \\
\resizebox{66mm}{!}{\includegraphics{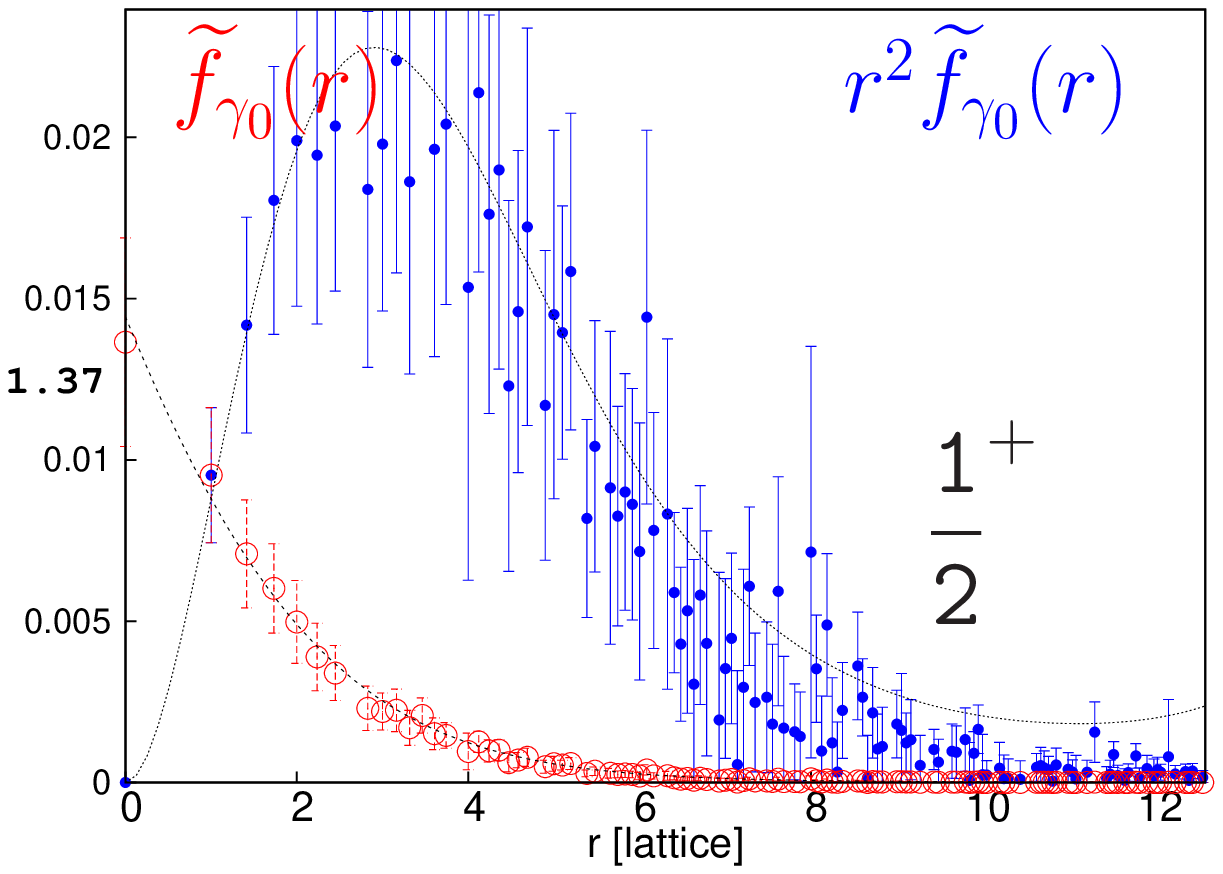}} & \resizebox{66mm}{!}{\includegraphics{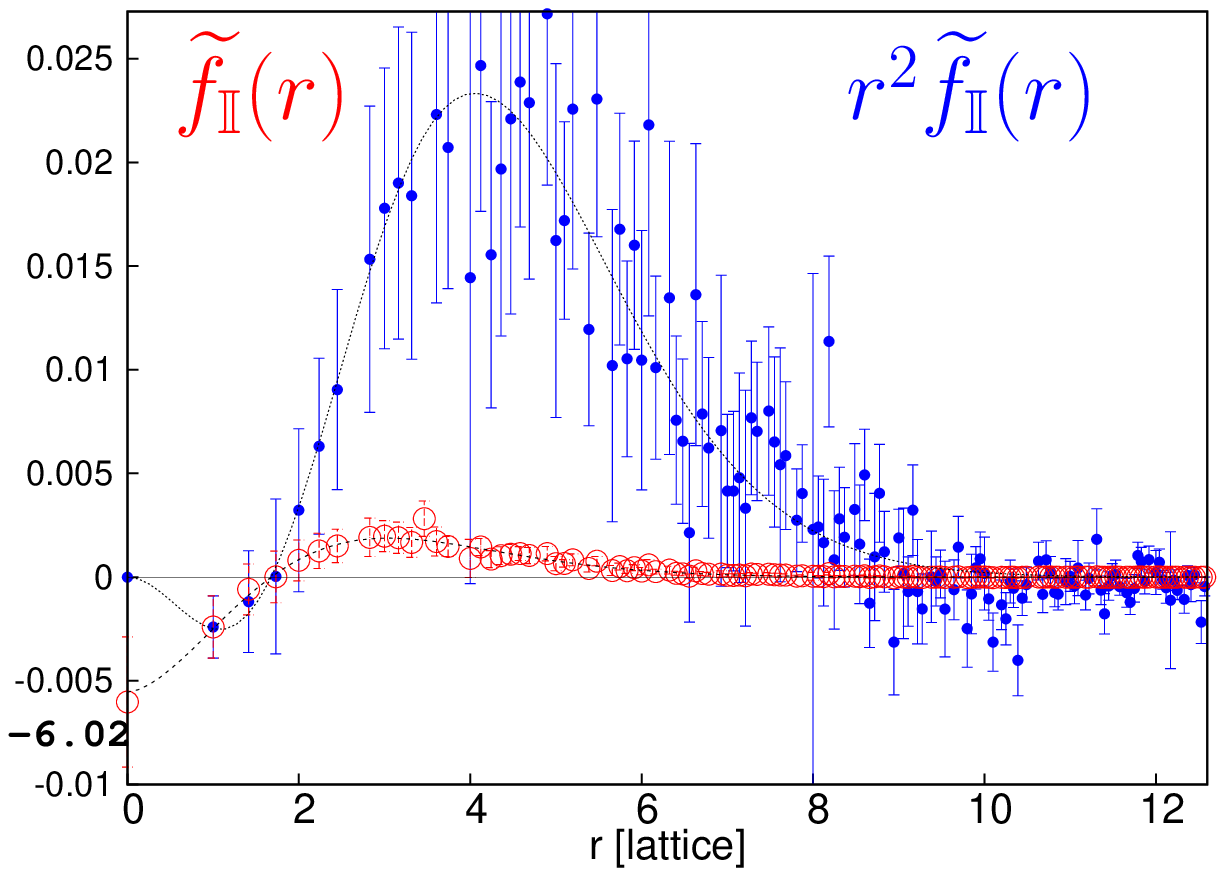}}& \resizebox{66mm}{!}{\includegraphics{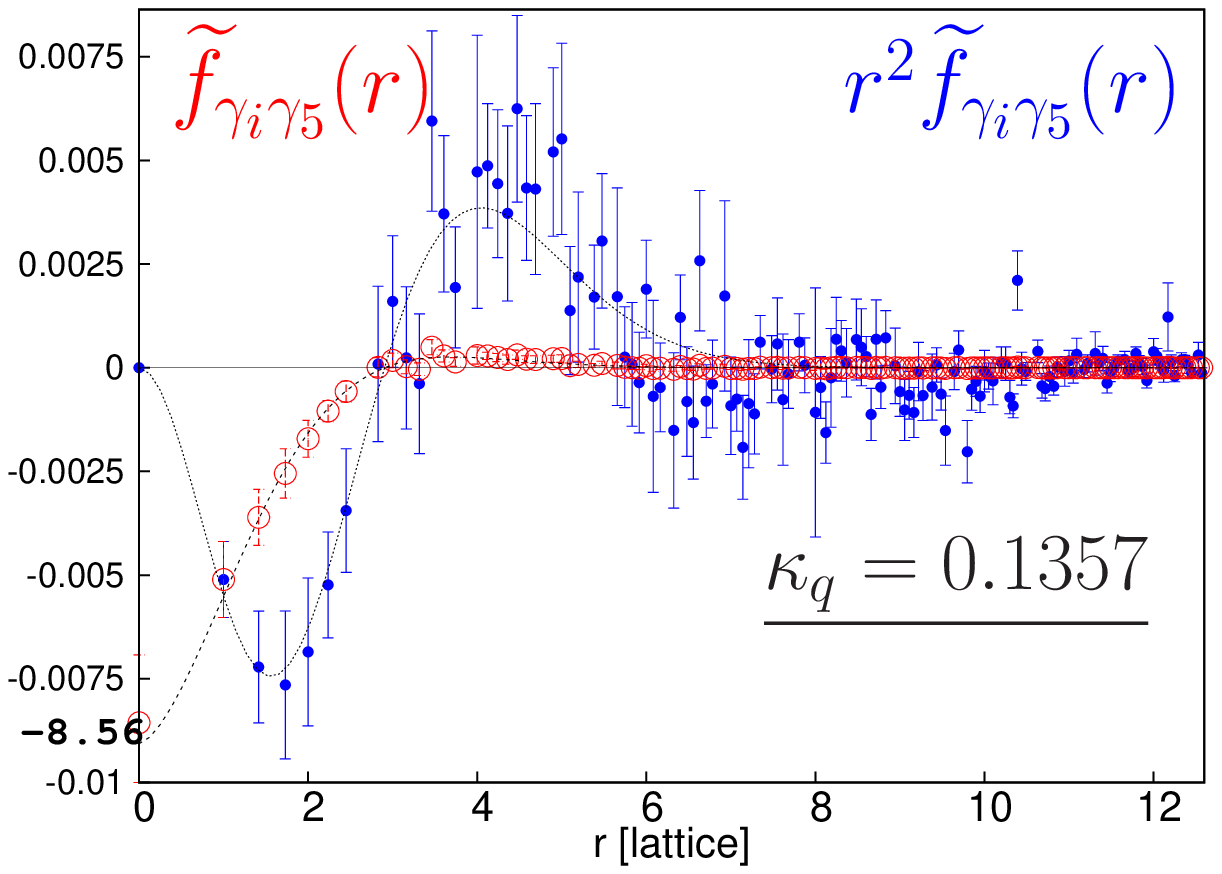}}\\
\resizebox{66mm}{!}{\includegraphics{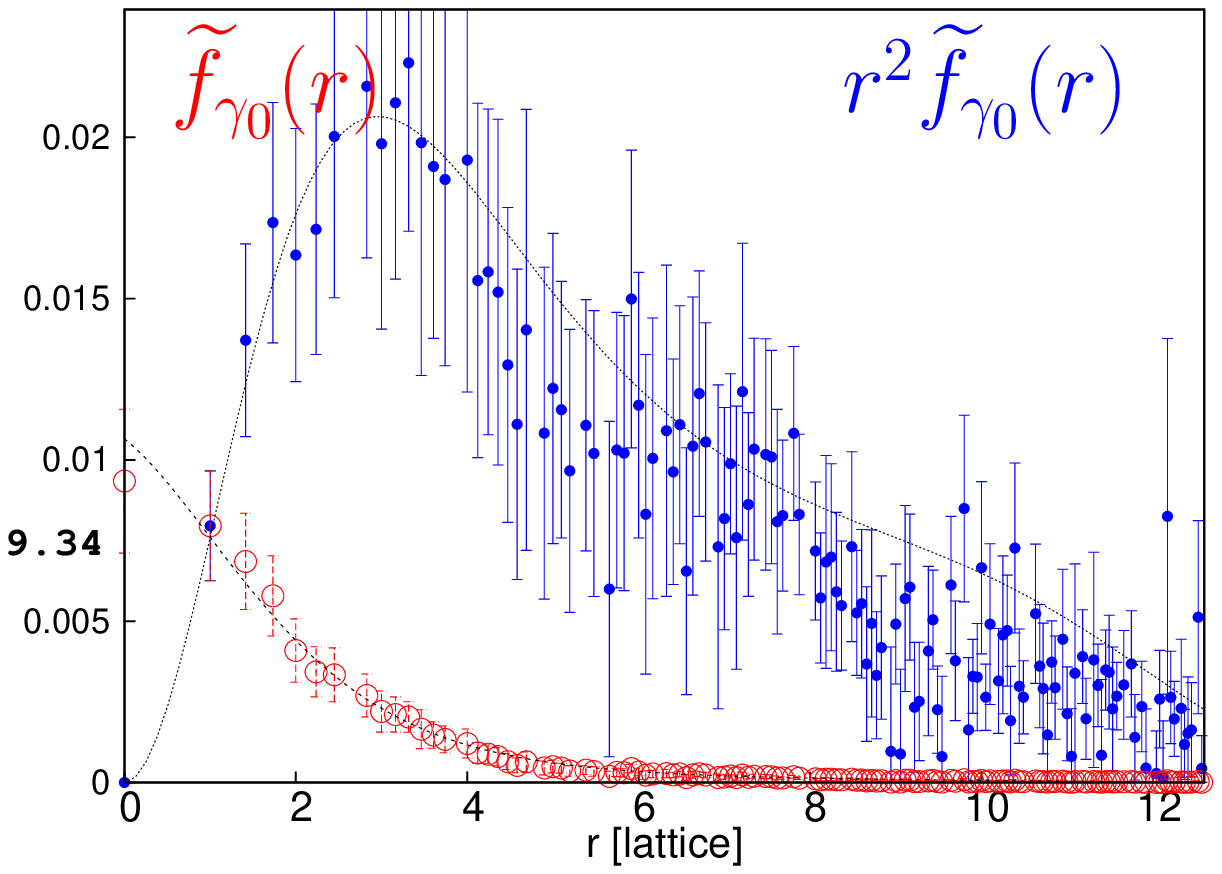}} & \resizebox{66mm}{!}{\includegraphics{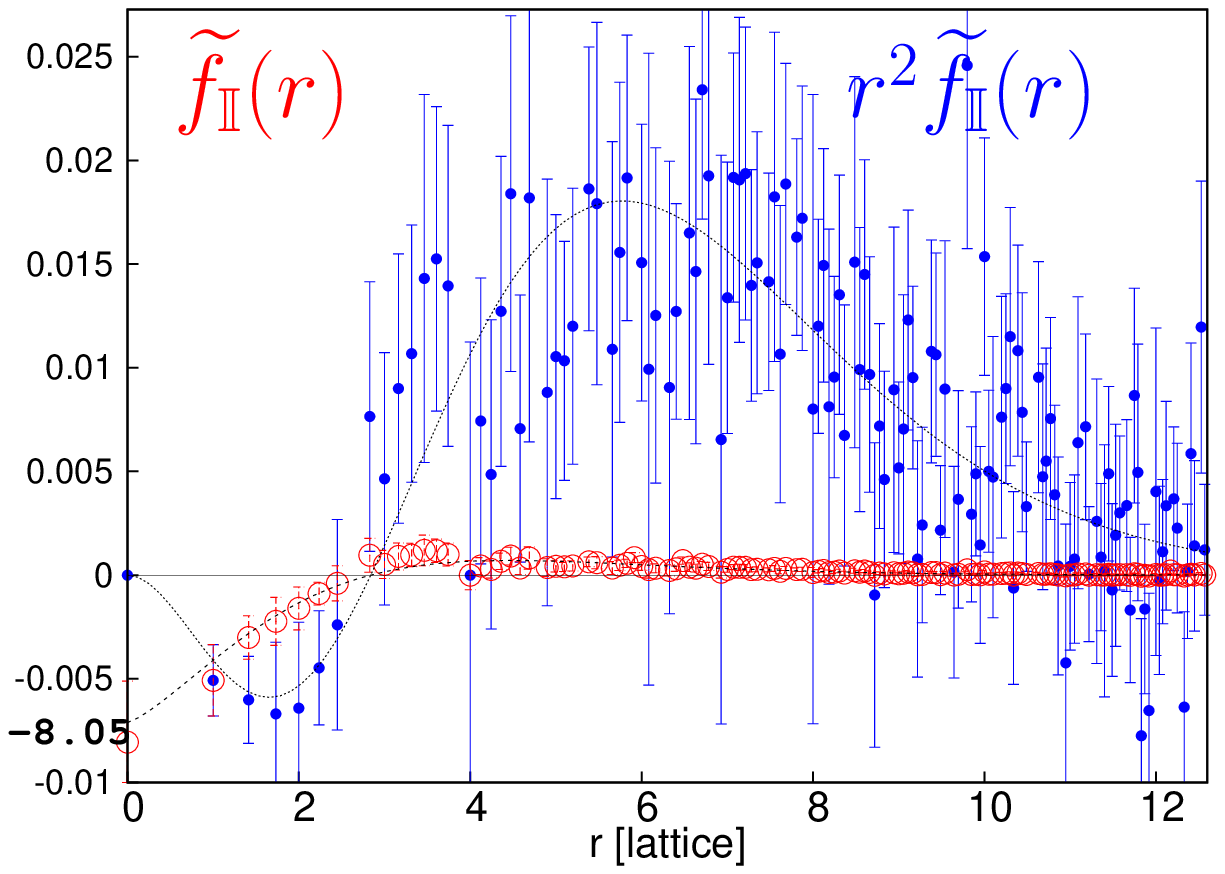}}& \resizebox{66mm}{!}{\includegraphics{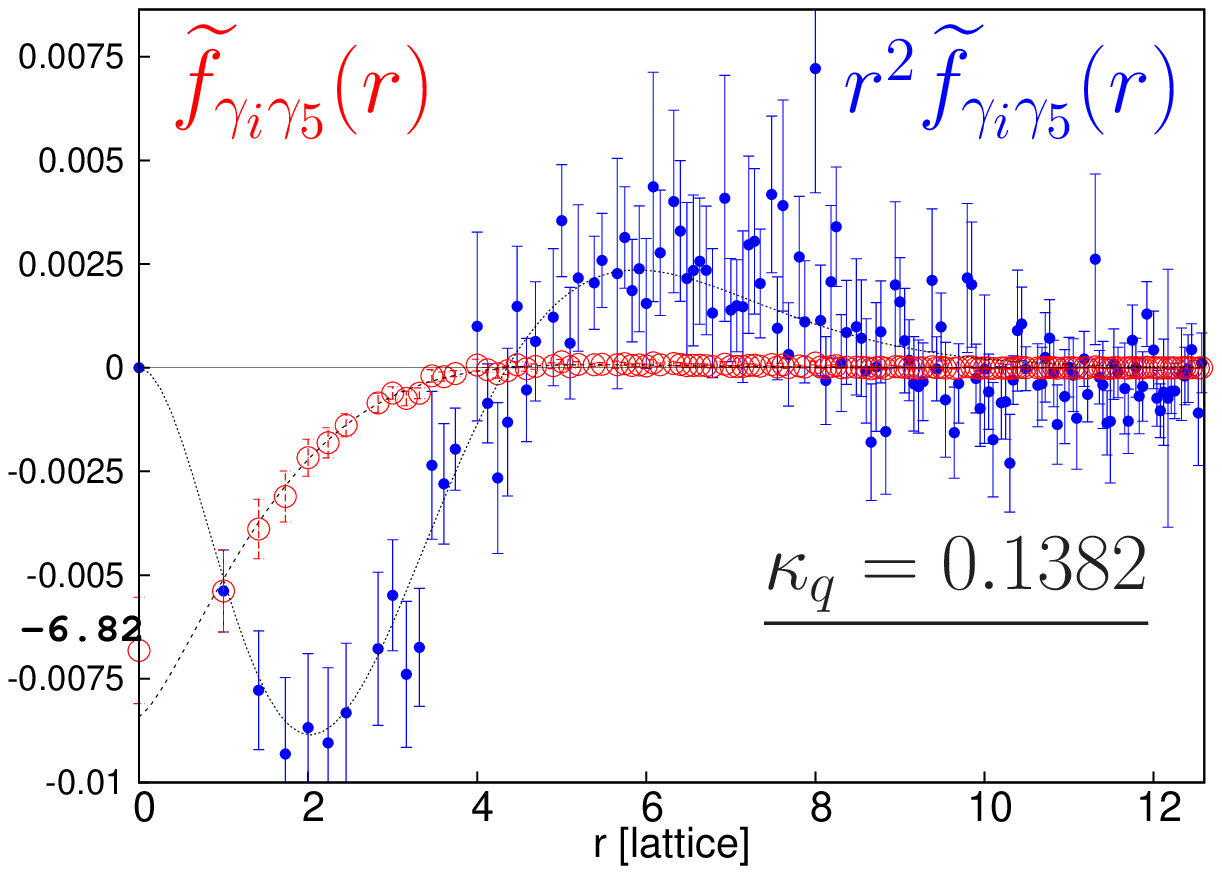}}\\
 \end{tabular}
\caption{\label{fig:A1}\footnotesize{\sl 
Distributions obtained with the heavy-quark actions FAT-6: upper set of two rows  correspond to the external lowest states, and the lower set  to 
othe excited states. In the upper row in each set the light-quark is our least light one ($\kappa_q=0.1357$), while in the lower it is our lightest ($\kappa_q=0.1382$) .
The black lines are obtained from the fits to the distributions $f_\Gamma (r)$.
}}

\end{figure}

\begin{figure}
\vspace*{-1.7cm}
\hspace*{-18mm}
\begin{tabular}{c c c}
\resizebox{66mm}{!}{\includegraphics{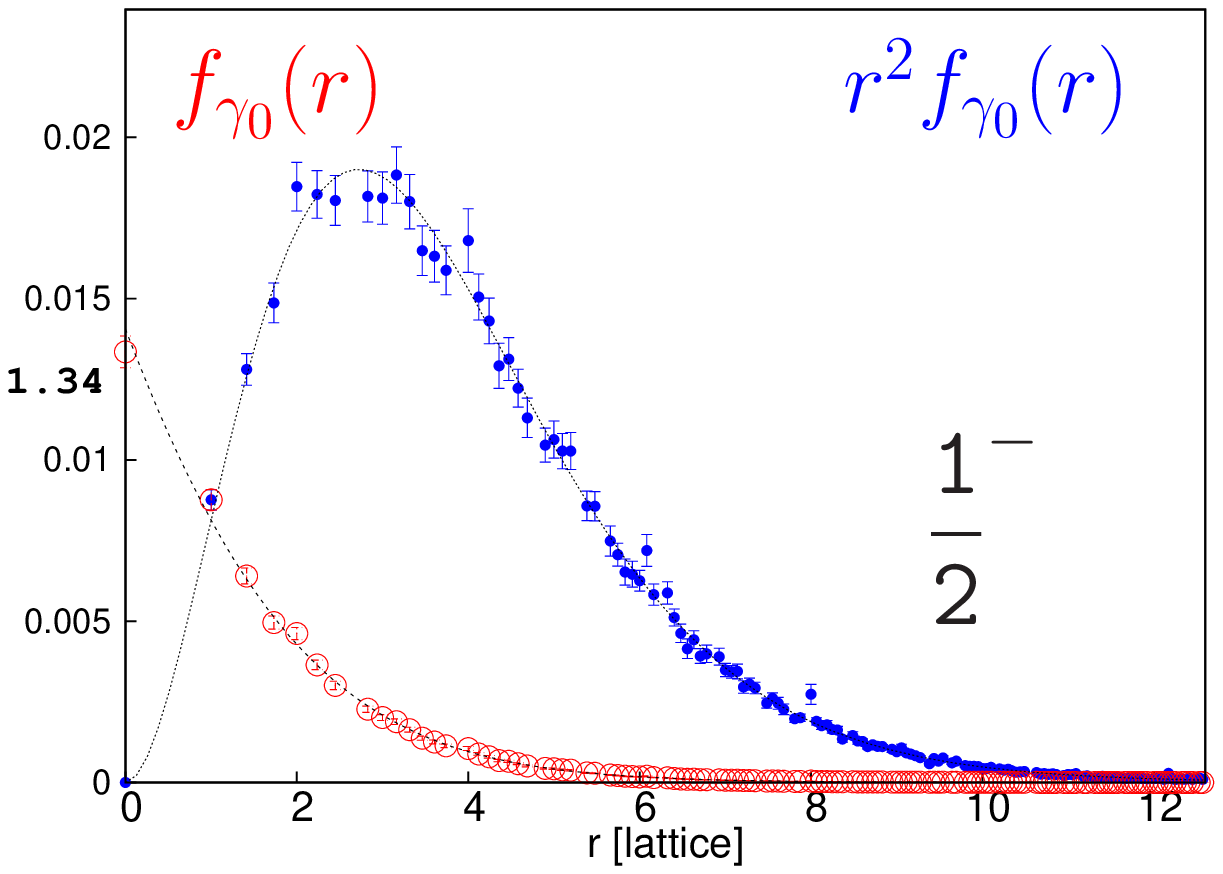}} & \resizebox{66mm}{!}{\includegraphics{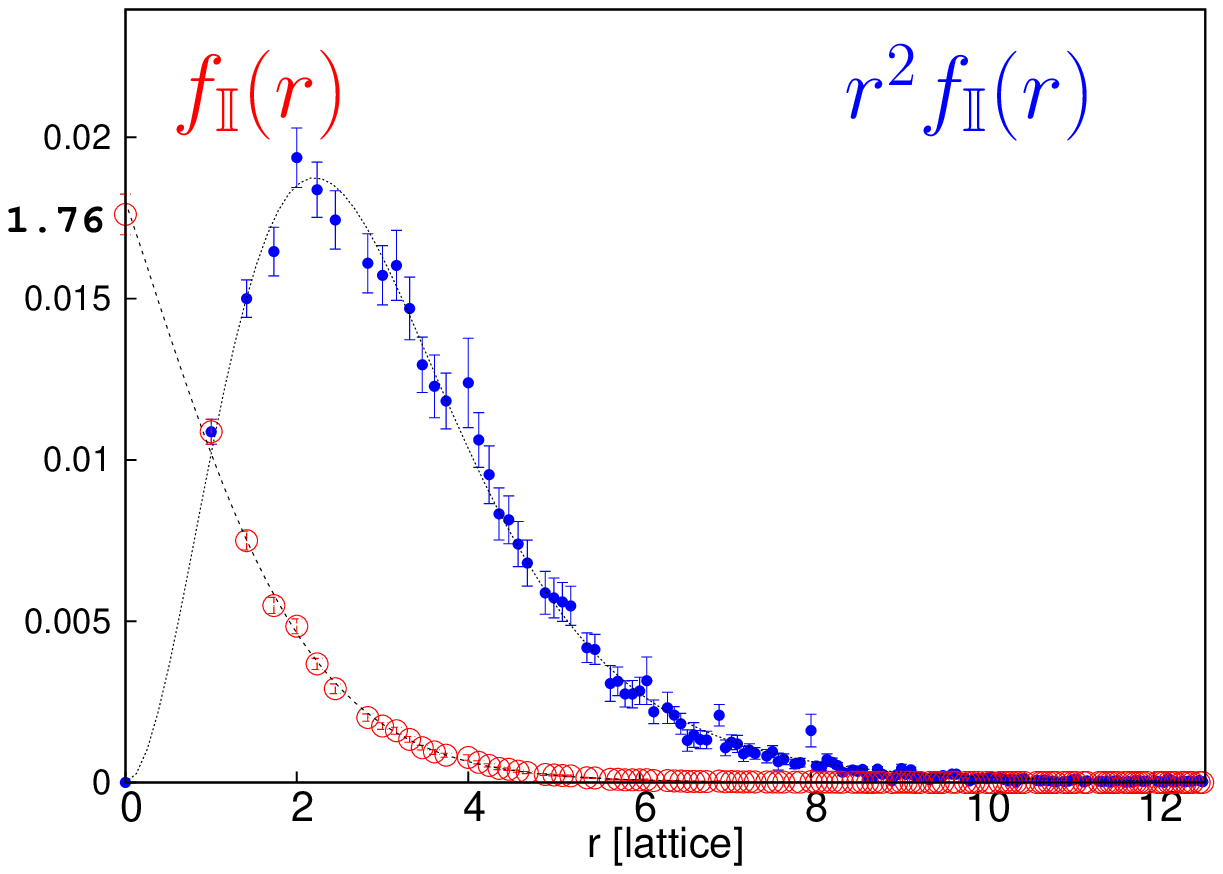}}& \resizebox{66mm}{!}{\includegraphics{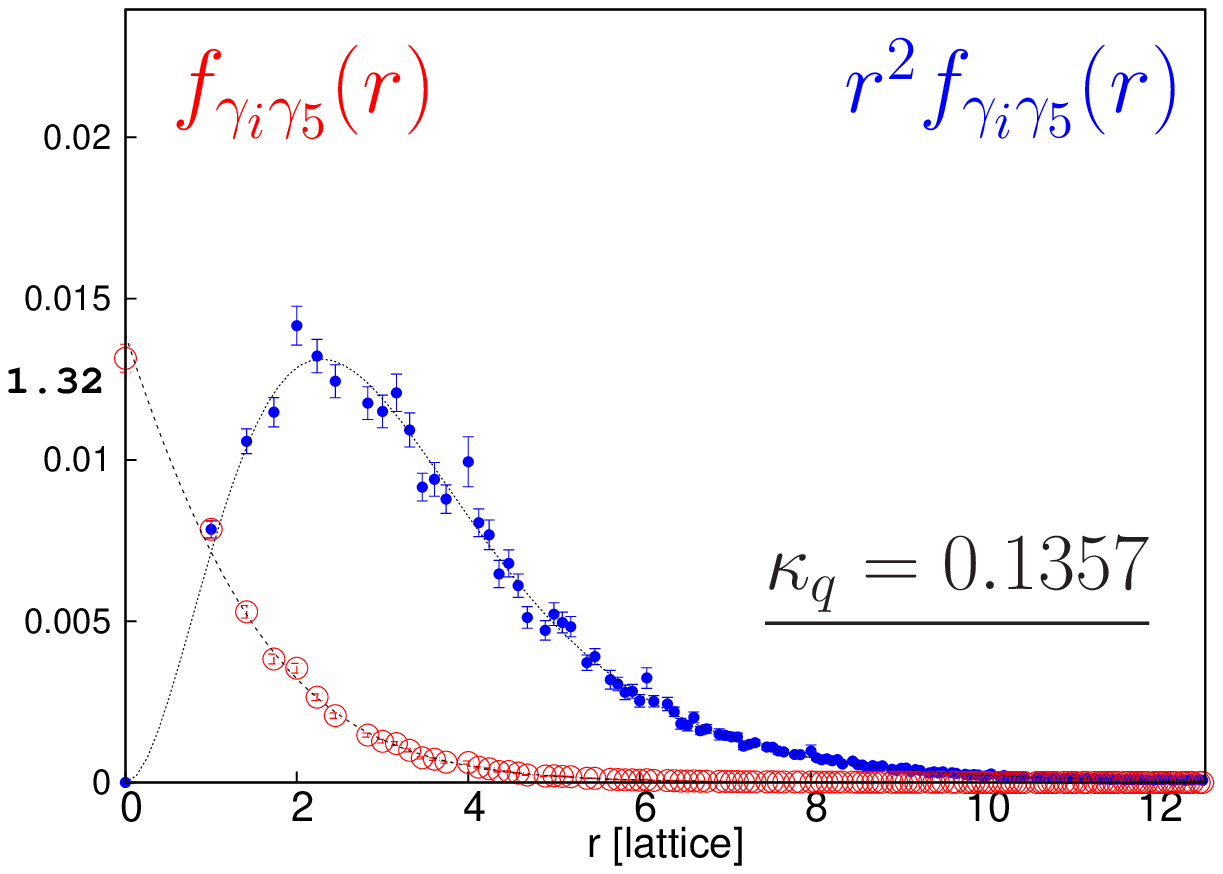}}\\
\resizebox{66mm}{!}{\includegraphics{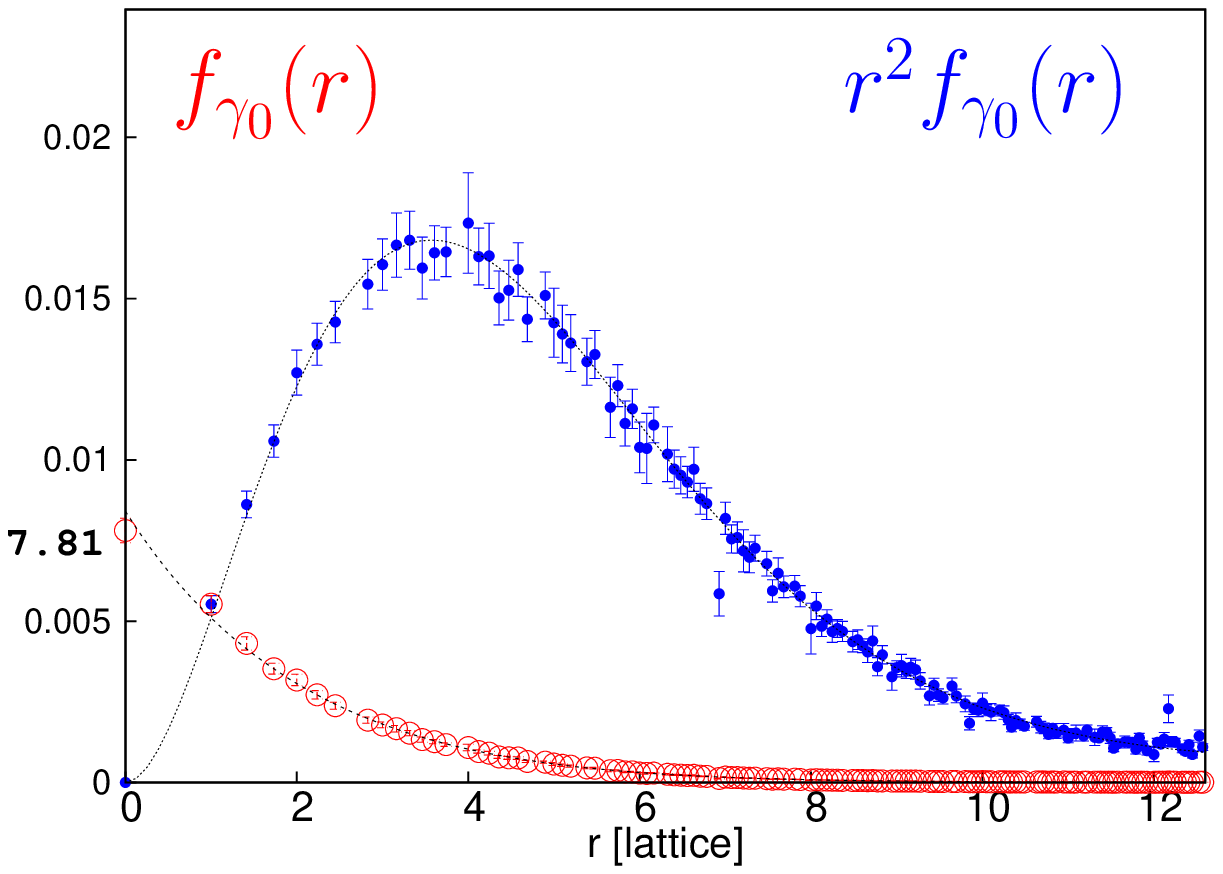}} & \resizebox{66mm}{!}{\includegraphics{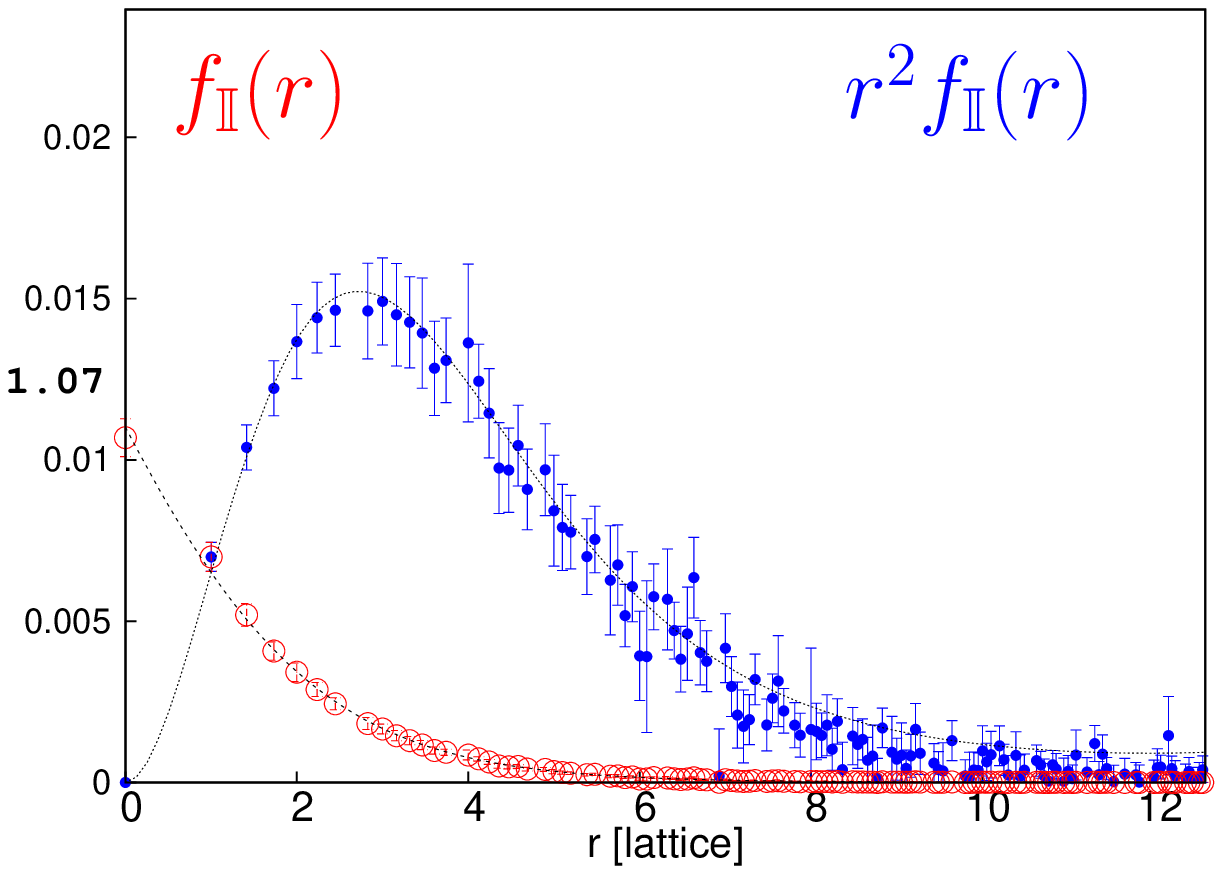}}& \resizebox{66mm}{!}{\includegraphics{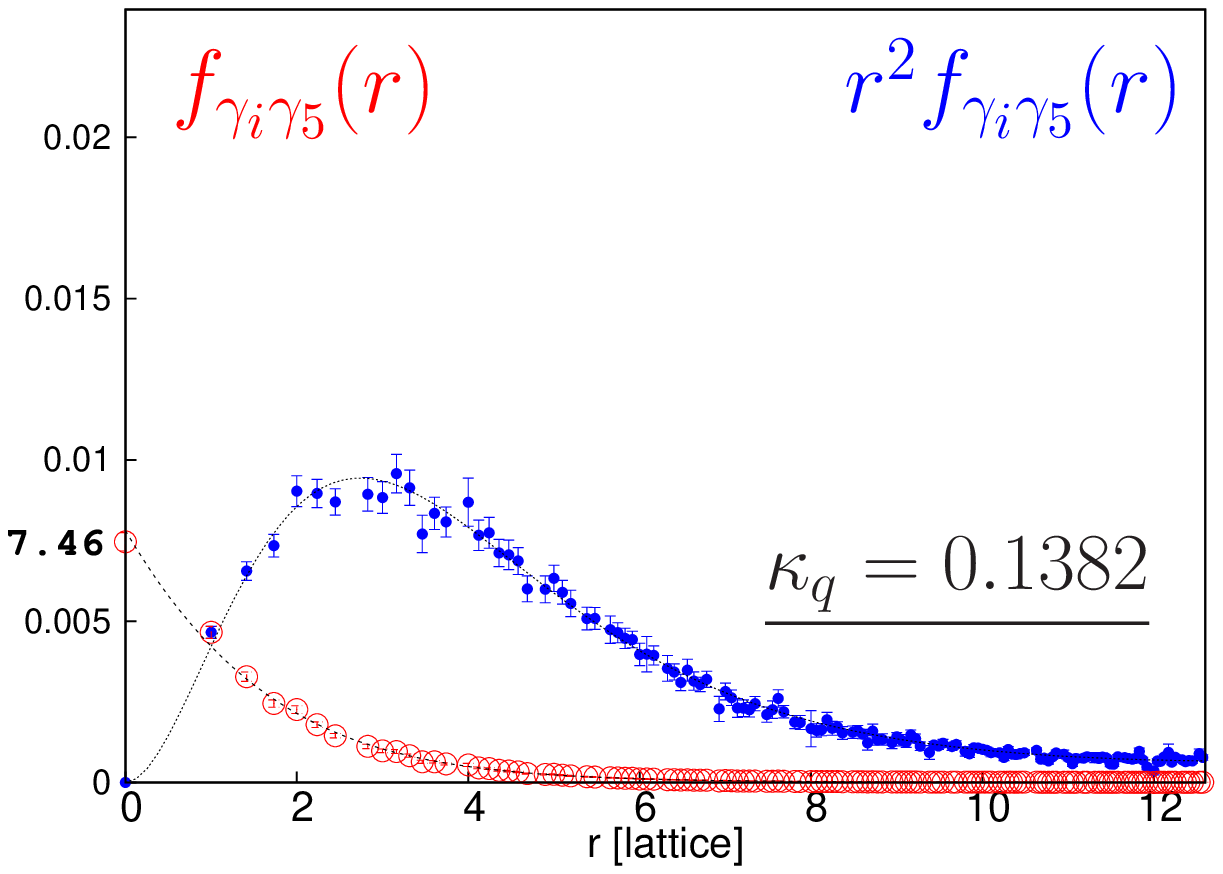}}\\
      && \\
 \hline    && \\
\resizebox{66mm}{!}{\includegraphics{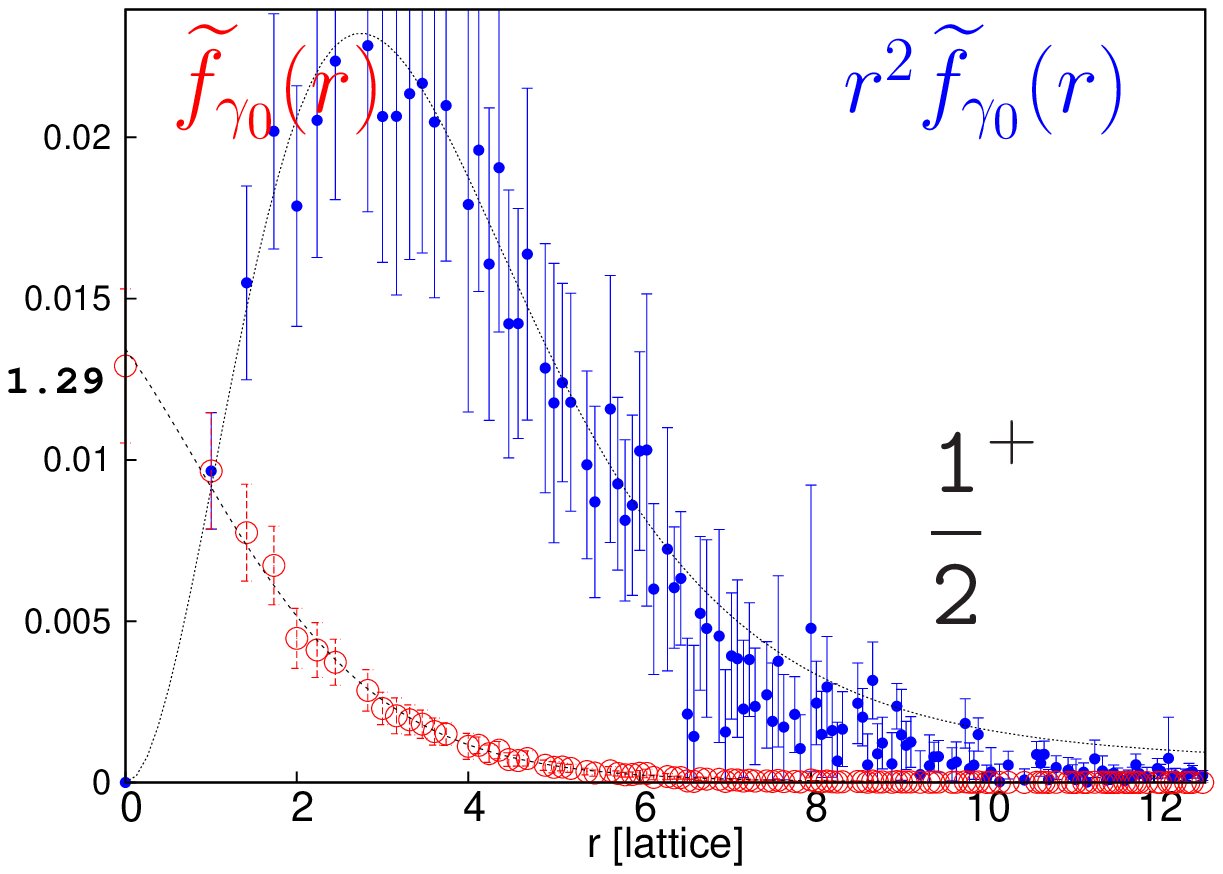}} & \resizebox{66mm}{!}{\includegraphics{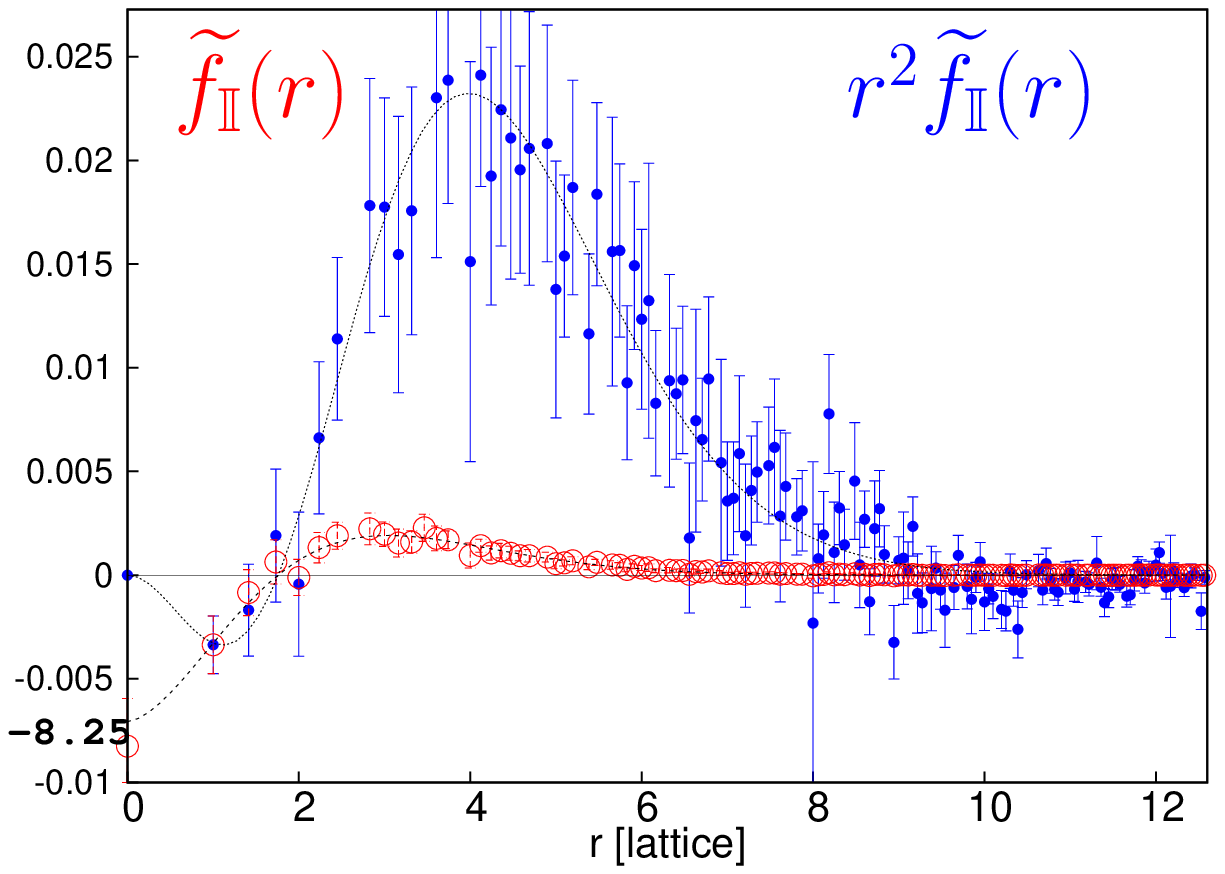}}& \resizebox{66mm}{!}{\includegraphics{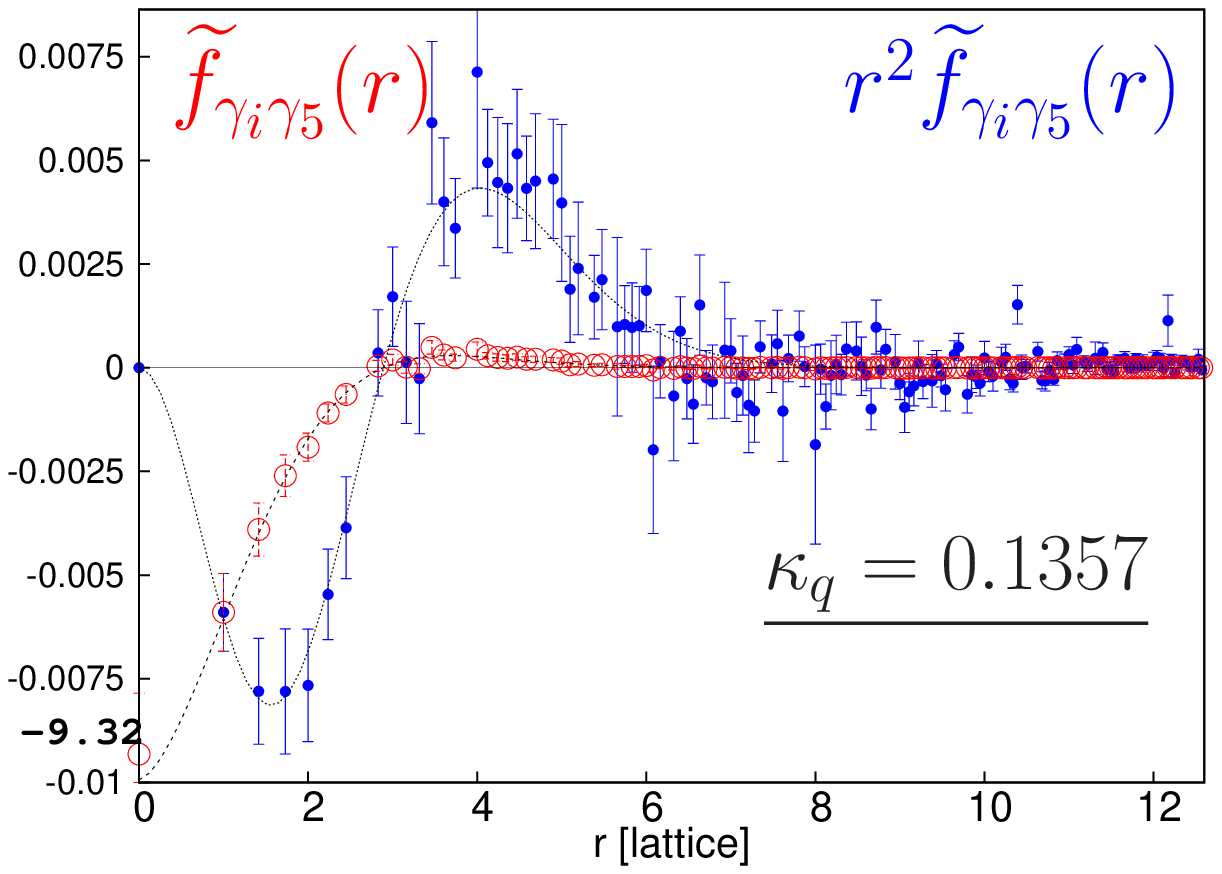}}\\
\resizebox{66mm}{!}{\includegraphics{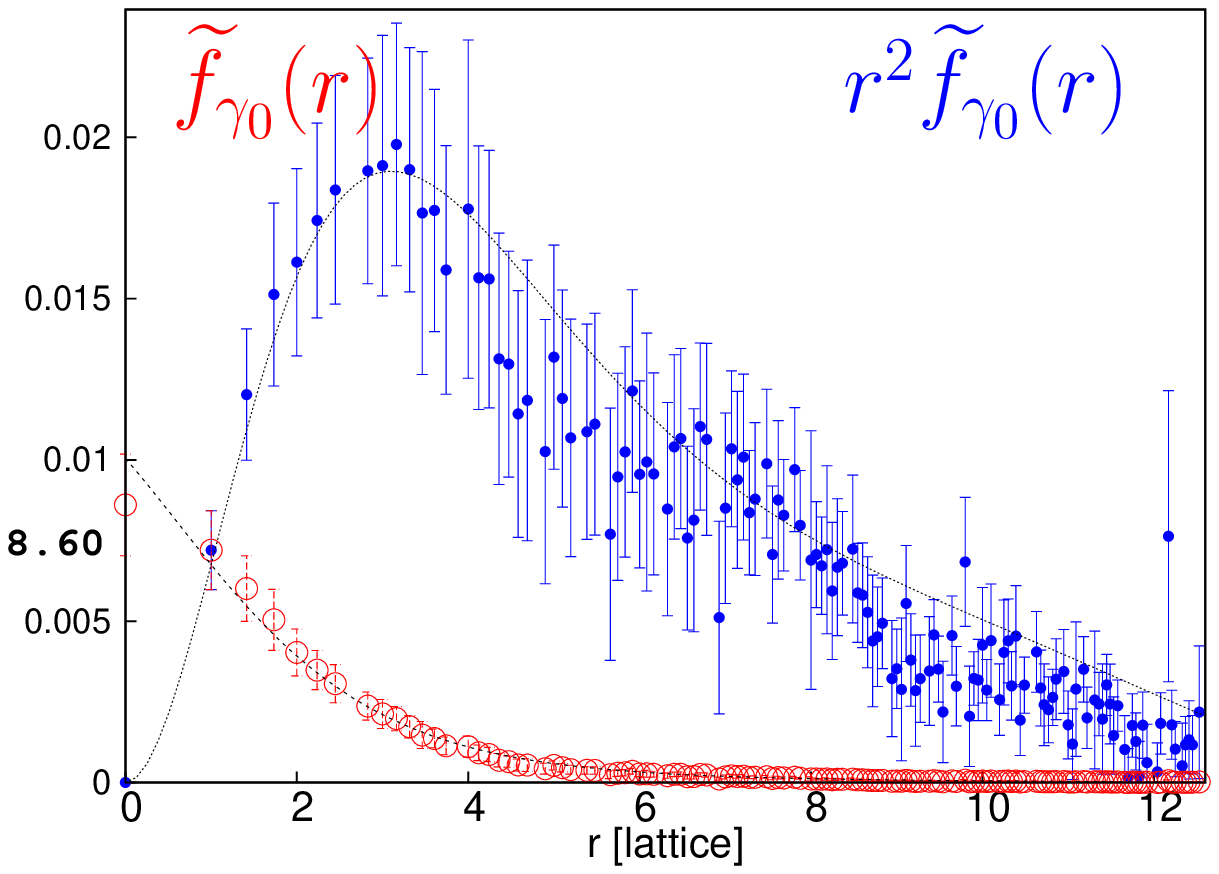}} & \resizebox{66mm}{!}{\includegraphics{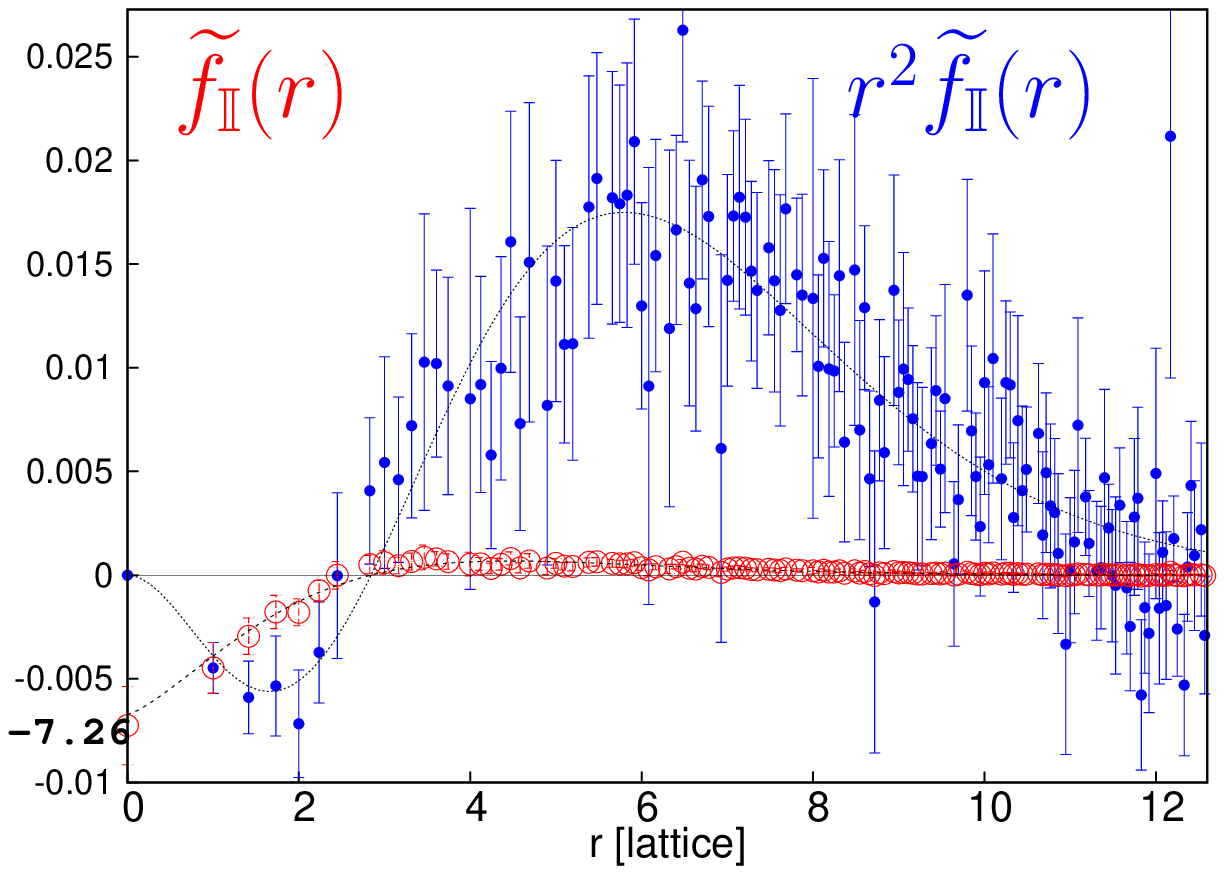}}& \resizebox{66mm}{!}{\includegraphics{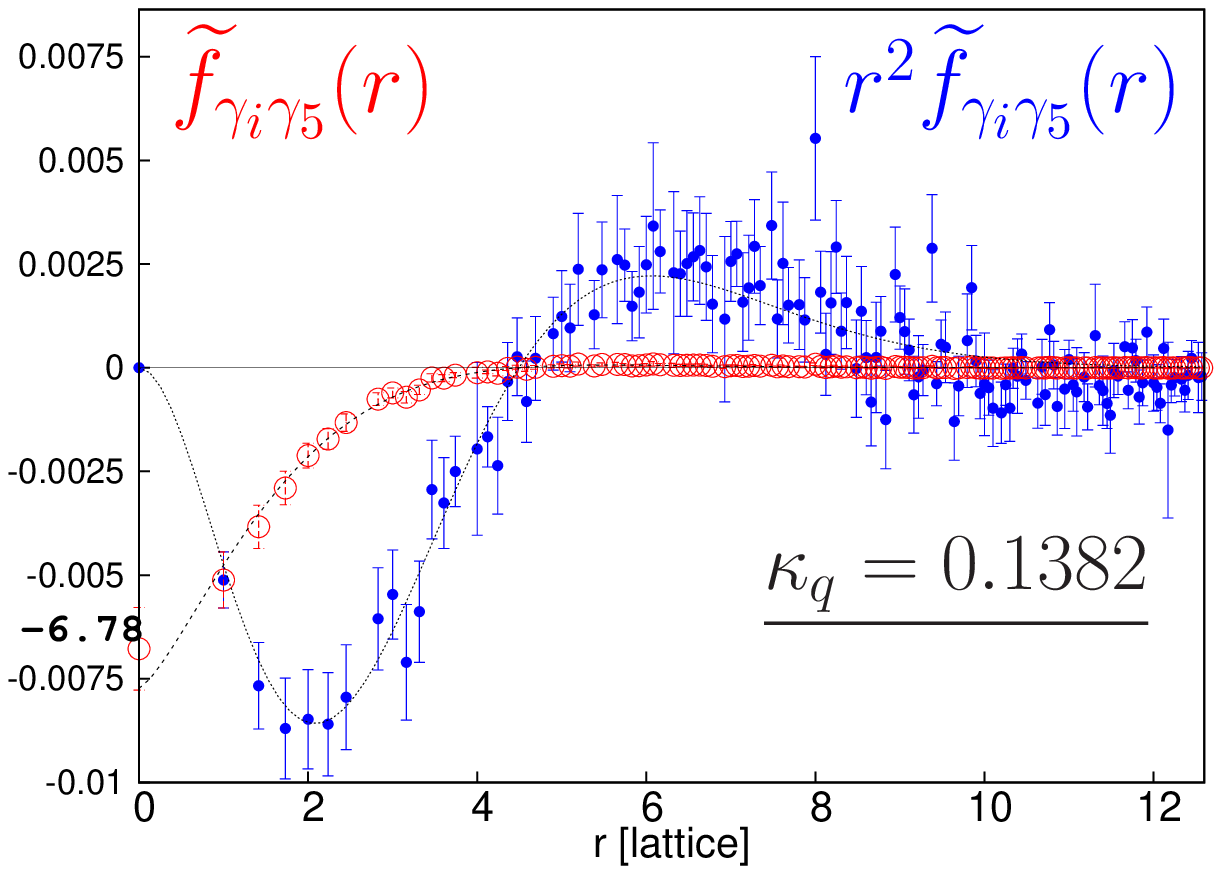}}\\
 \end{tabular}
\caption{\label{fig:A2}\footnotesize{\sl 
The same as in Fig.~\ref{fig:A1} but with the heavy-quark actions HYP-1. 
}}

\end{figure}

\begin{figure}
\vspace*{-1.7cm}
\hspace*{-18mm}
\begin{tabular}{c c c}
\resizebox{66mm}{!}{\includegraphics{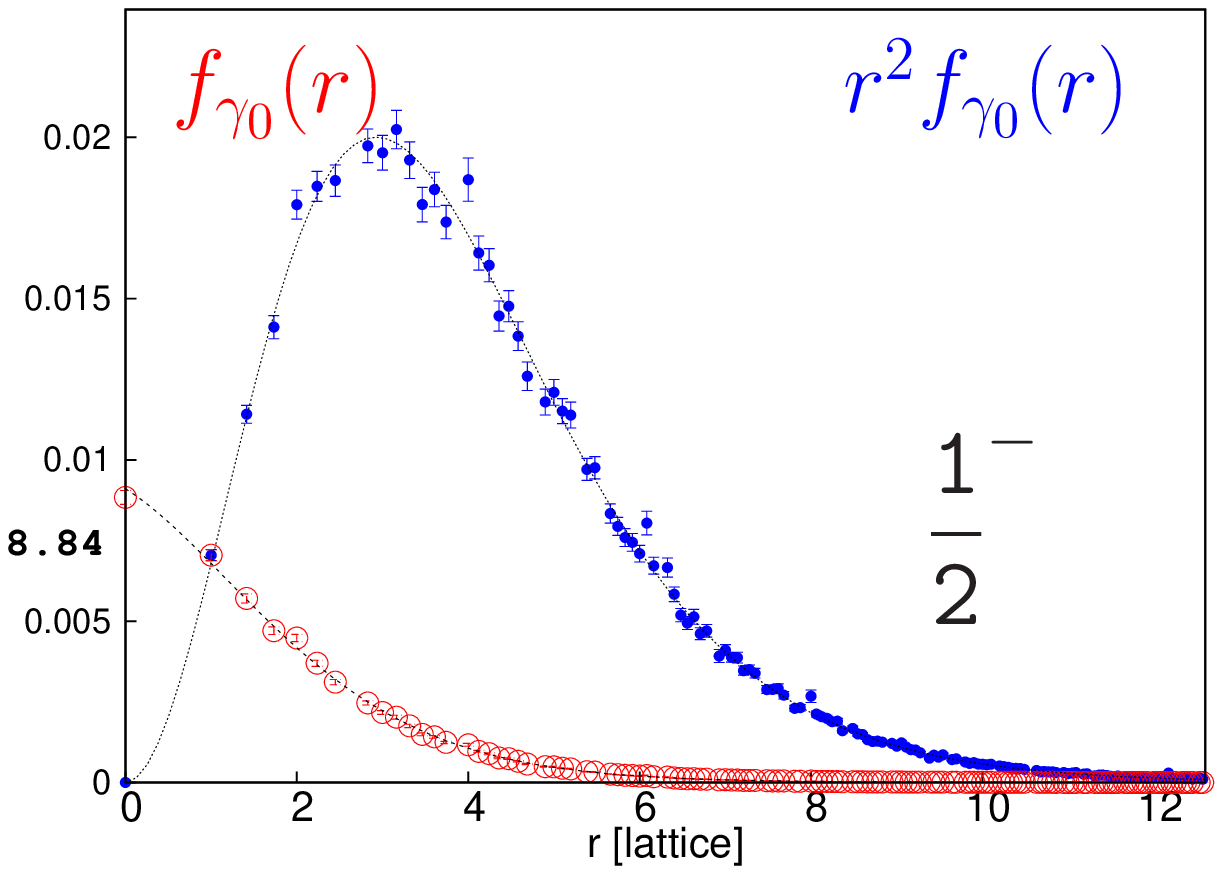}} & \resizebox{66mm}{!}{\includegraphics{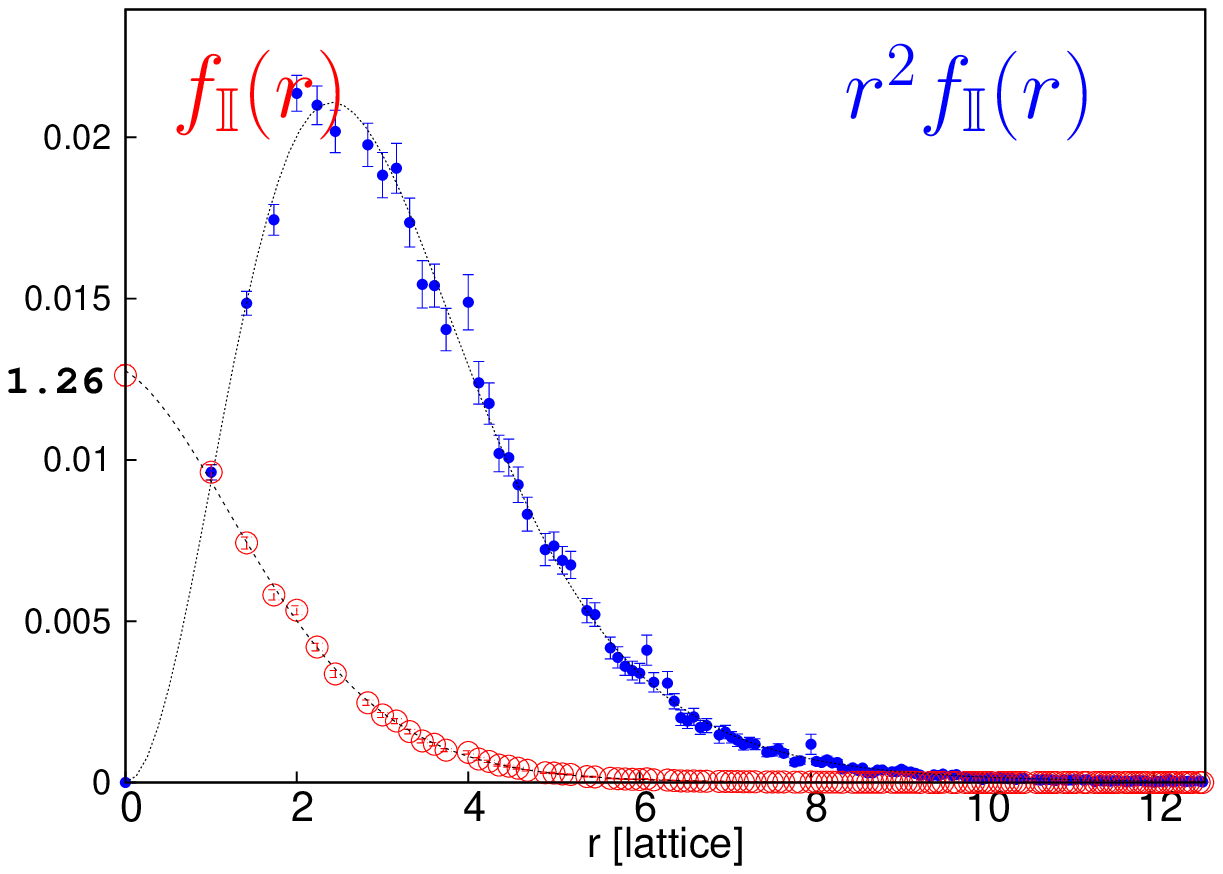}}& \resizebox{66mm}{!}{\includegraphics{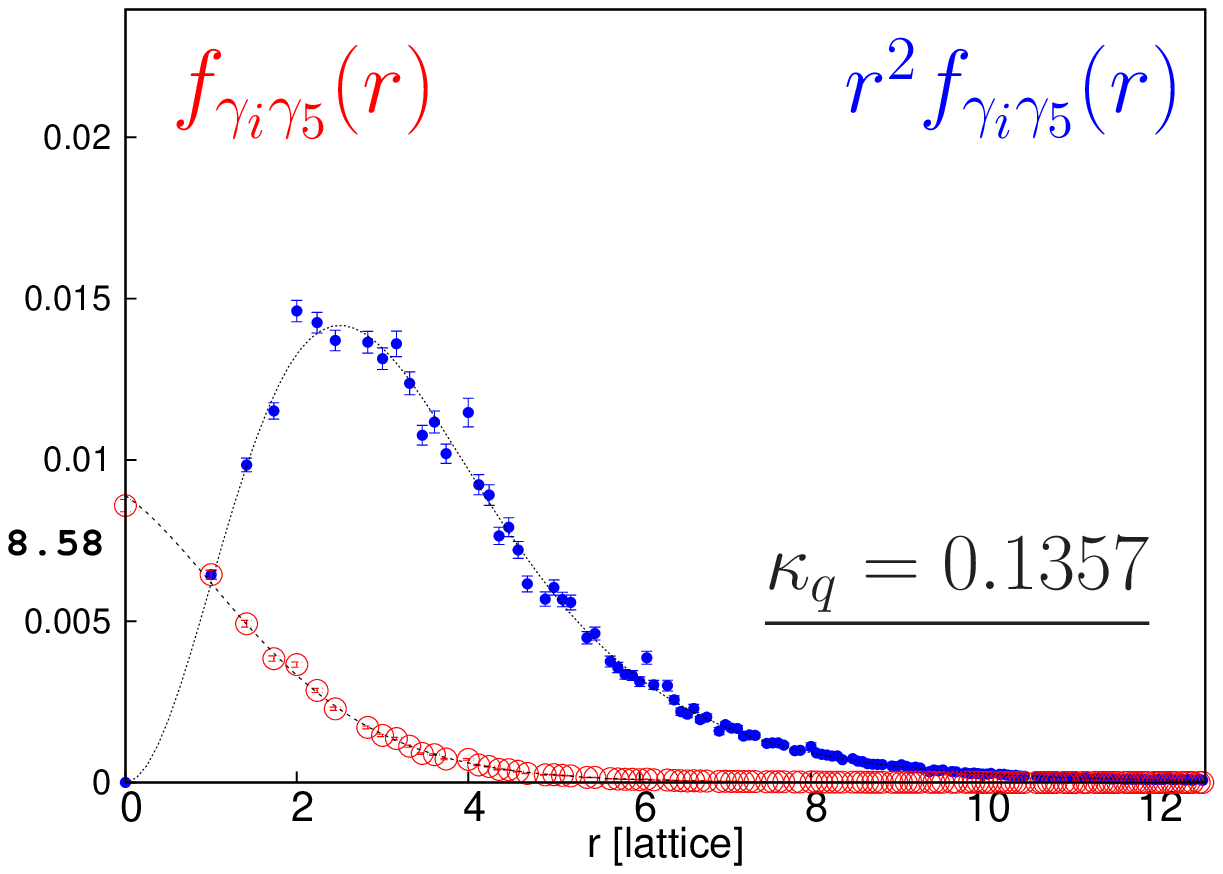}}\\
\resizebox{66mm}{!}{\includegraphics{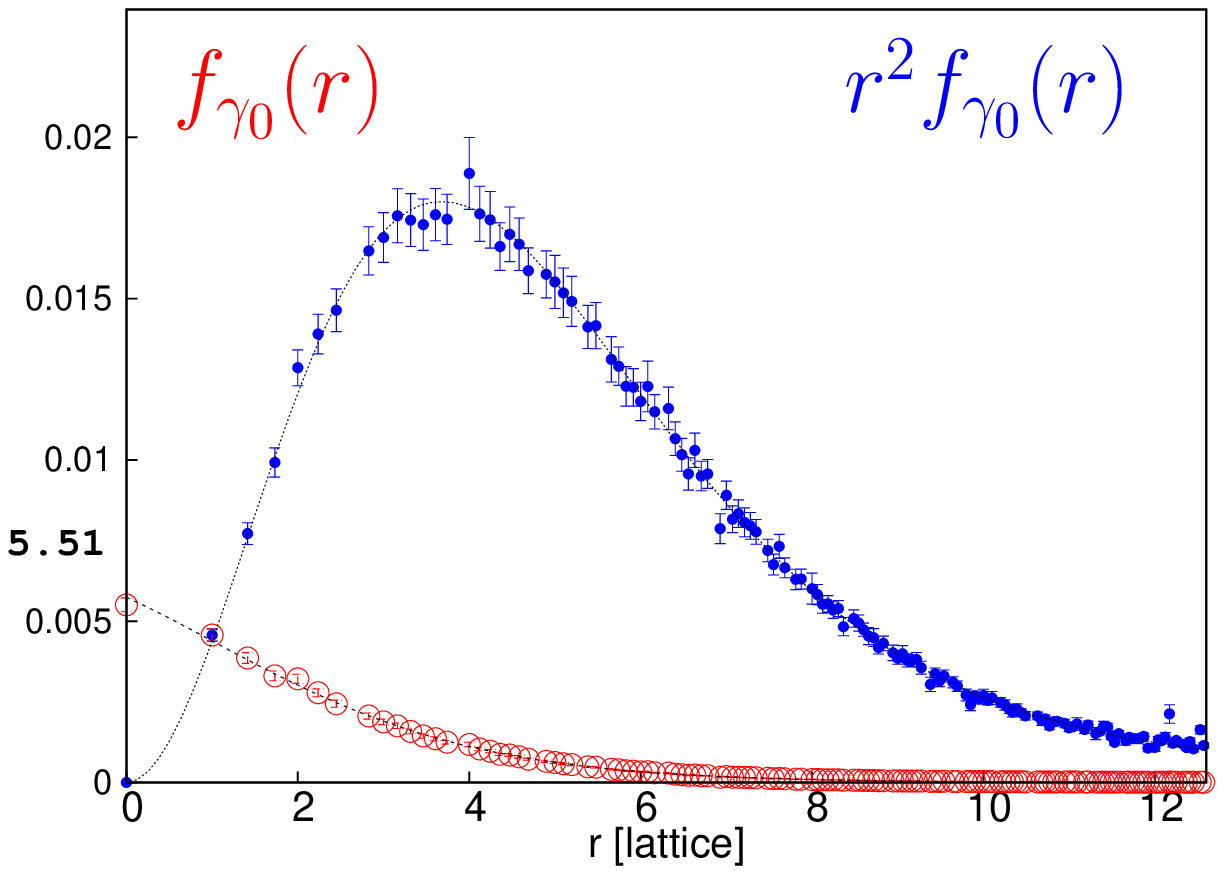}} & \resizebox{66mm}{!}{\includegraphics{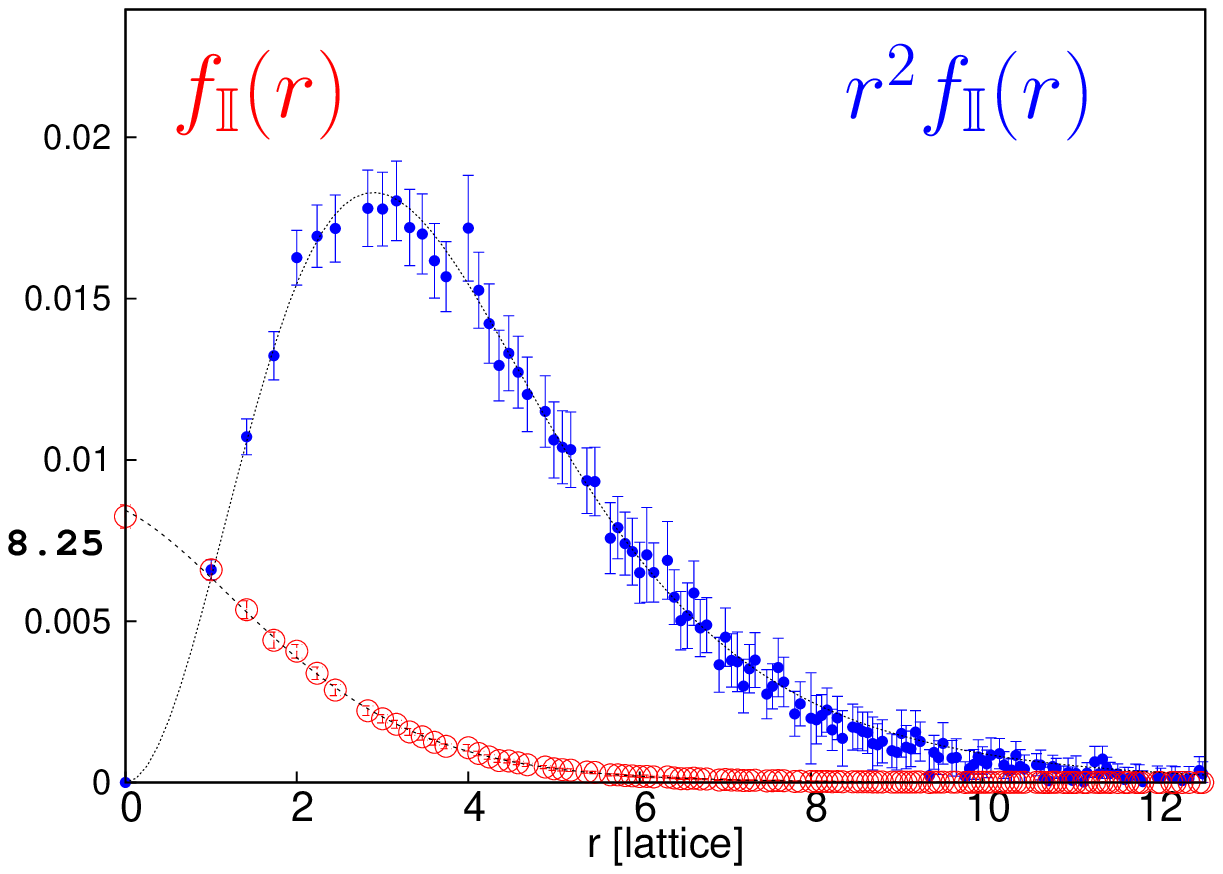}}& \resizebox{66mm}{!}{\includegraphics{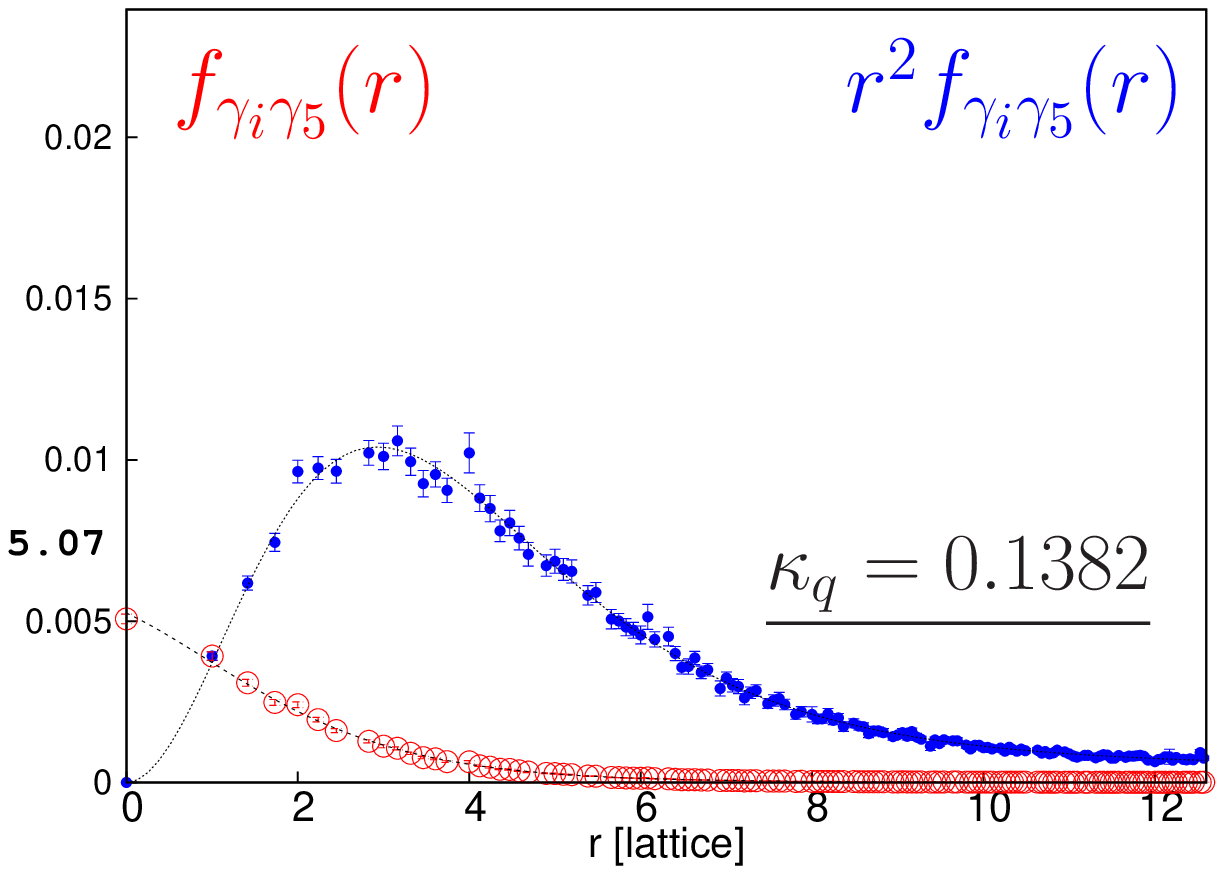}}\\
      && \\
 \hline    && \\
\resizebox{66mm}{!}{\includegraphics{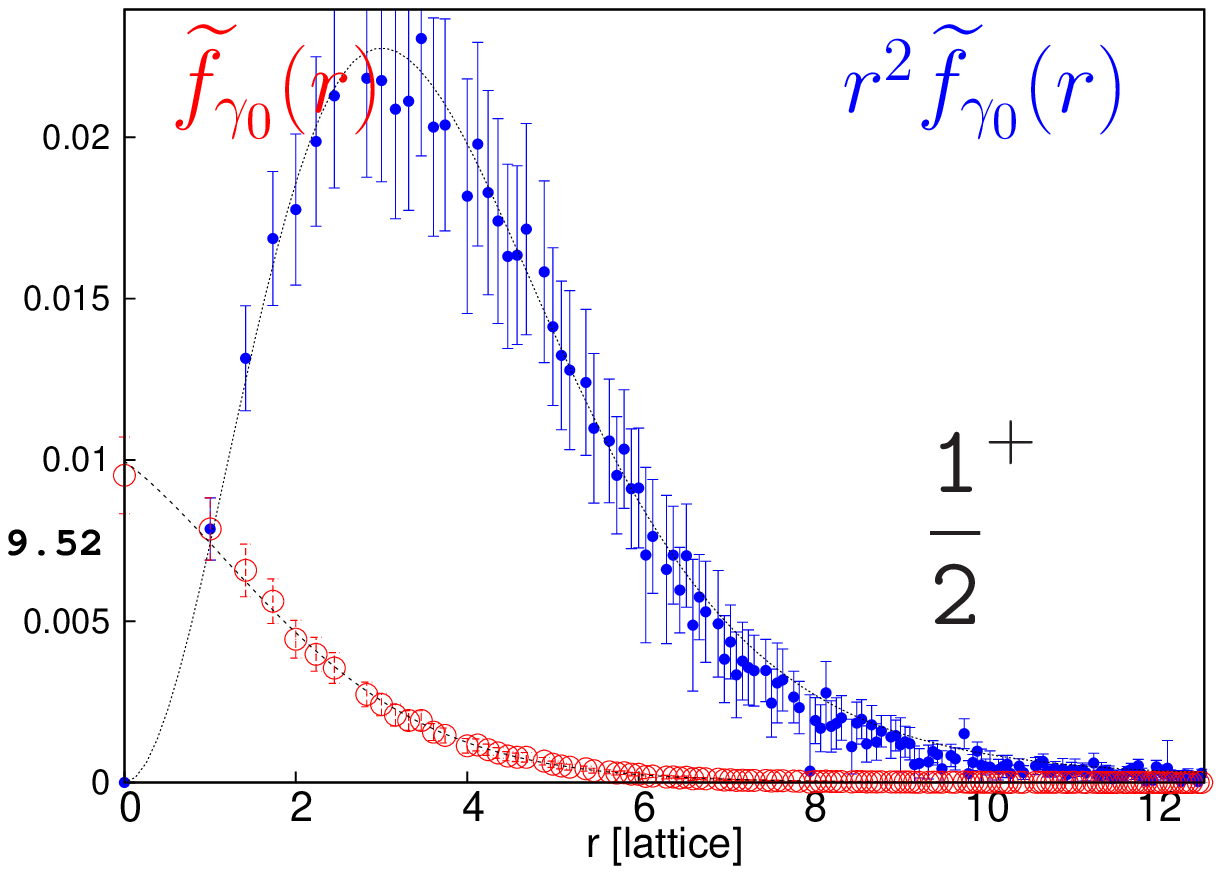}} & \resizebox{66mm}{!}{\includegraphics{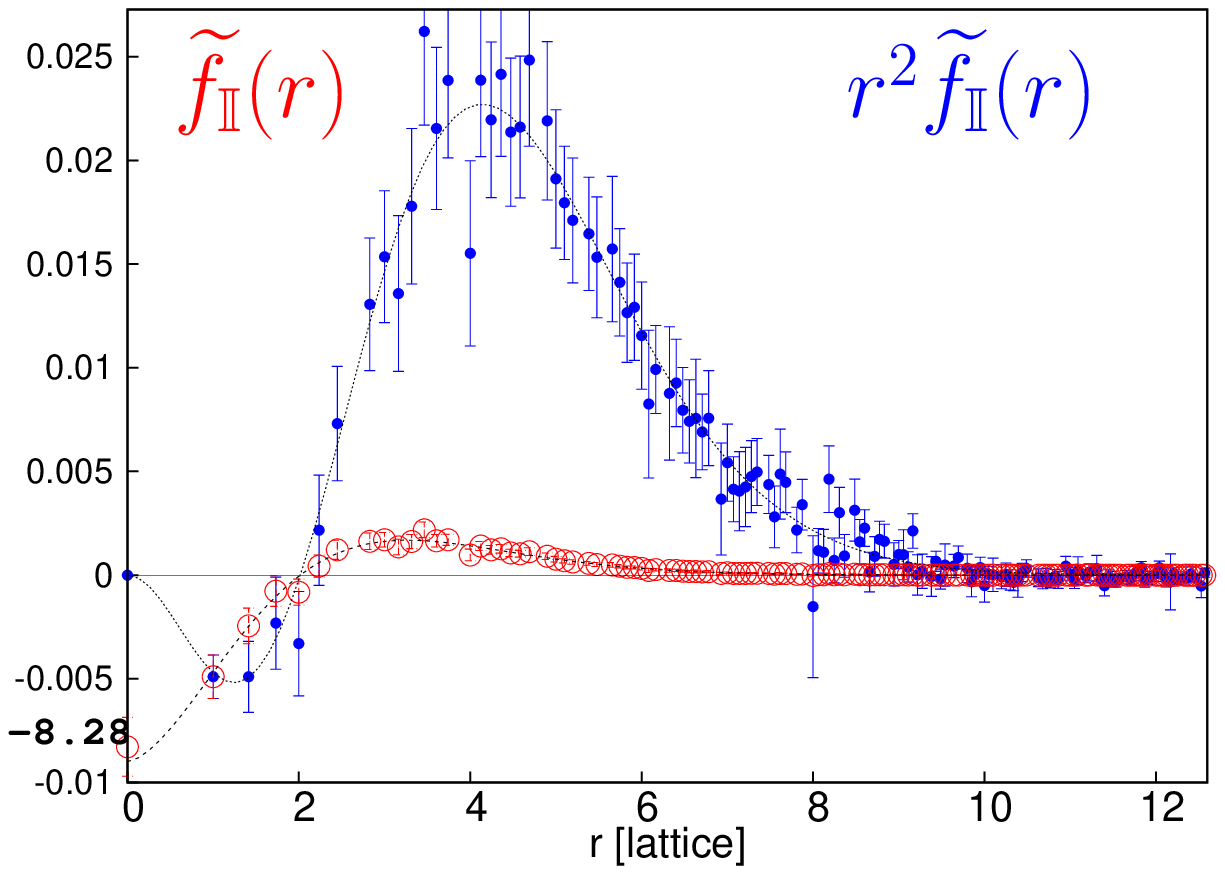}}& \resizebox{66mm}{!}{\includegraphics{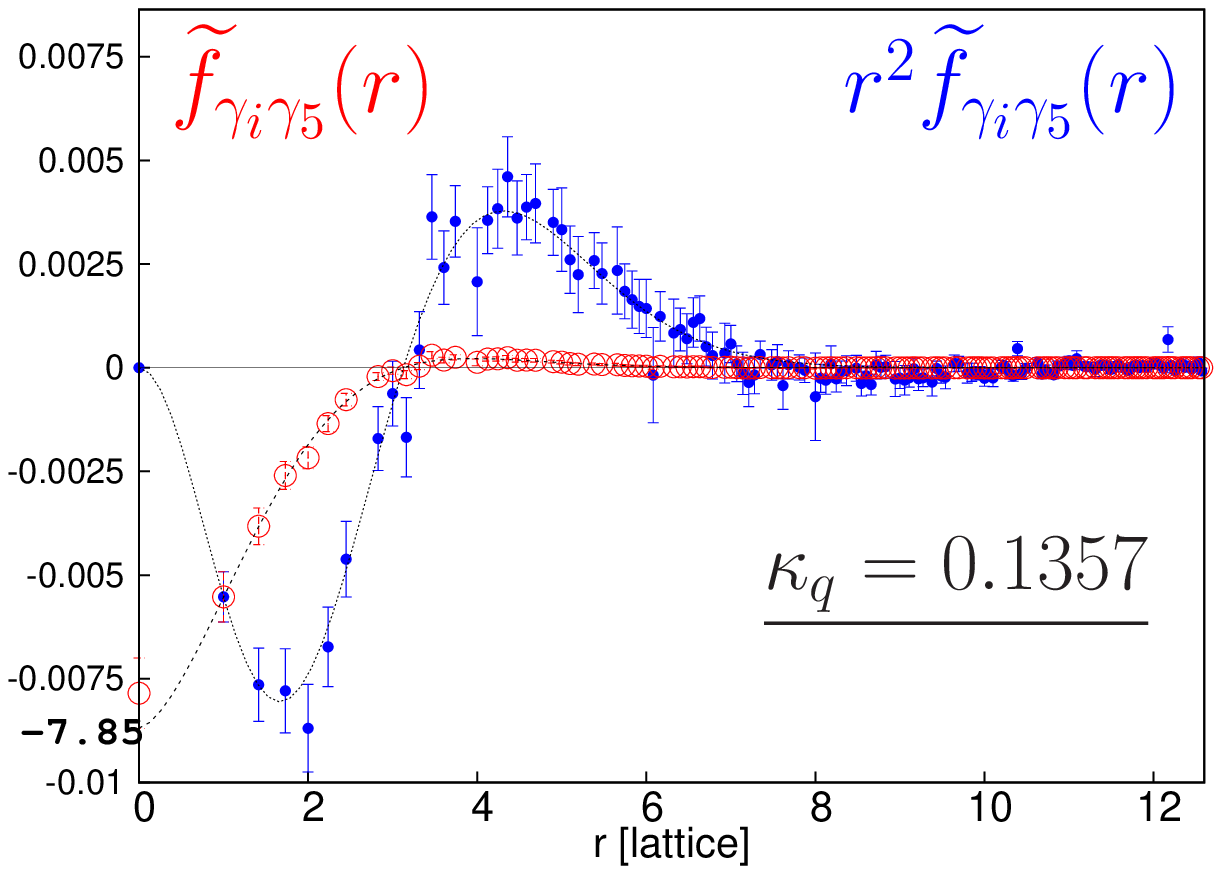}}\\
\resizebox{66mm}{!}{\includegraphics{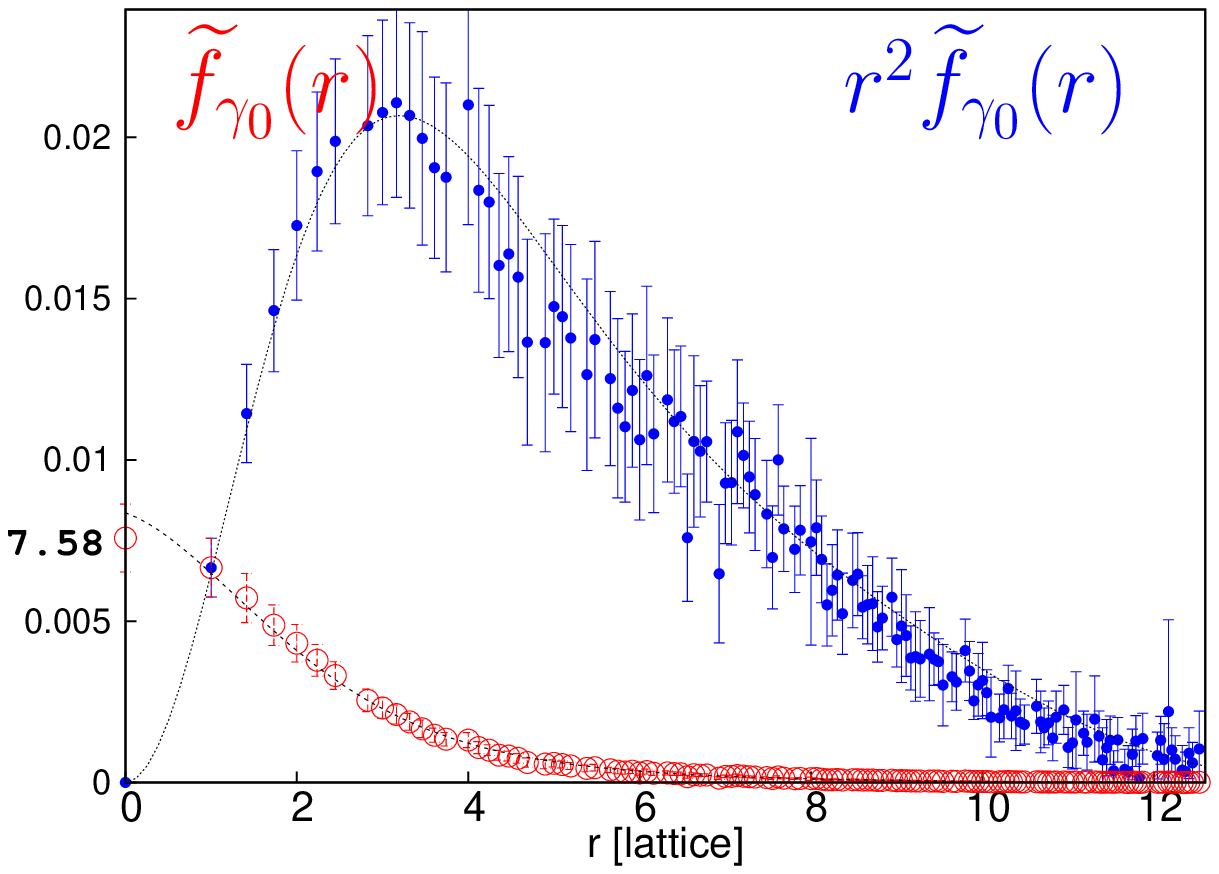}} & \resizebox{66mm}{!}{\includegraphics{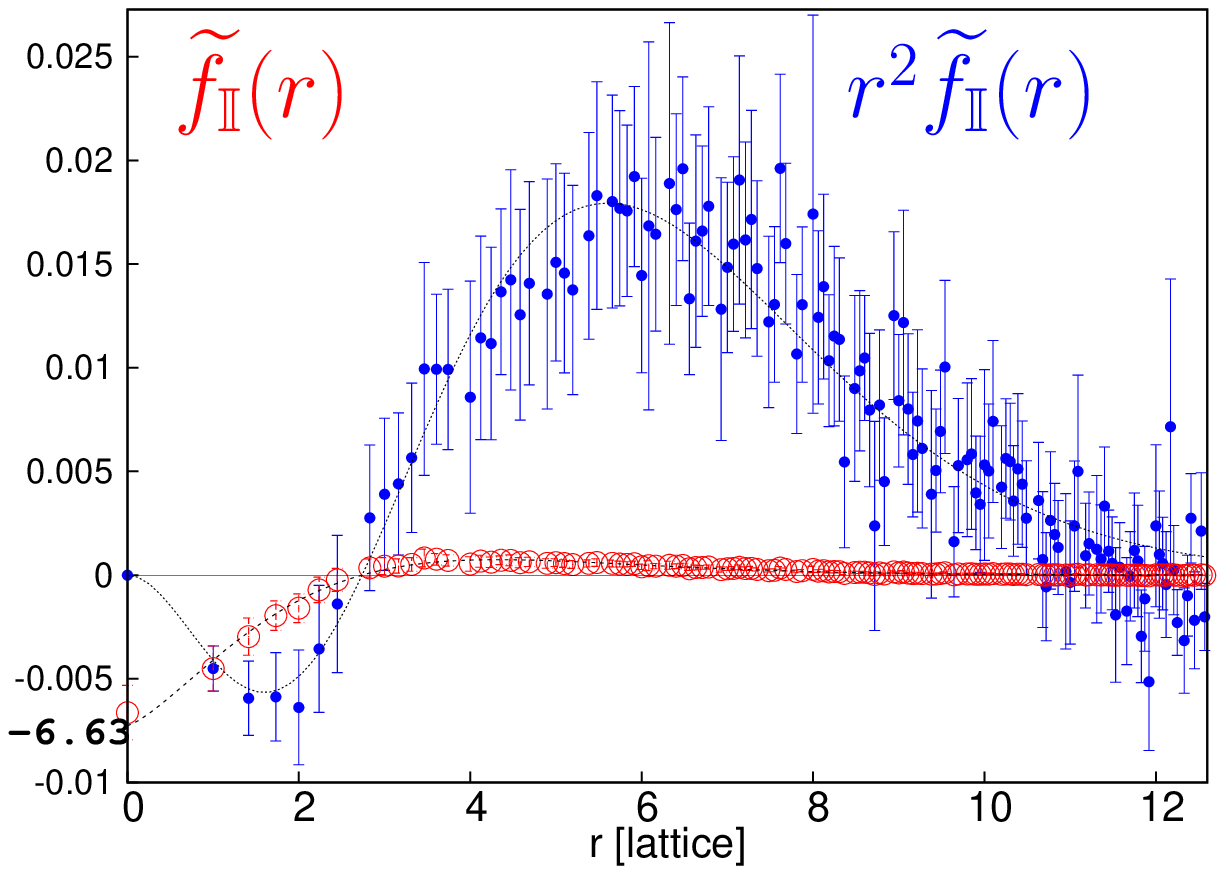}}& \resizebox{66mm}{!}{\includegraphics{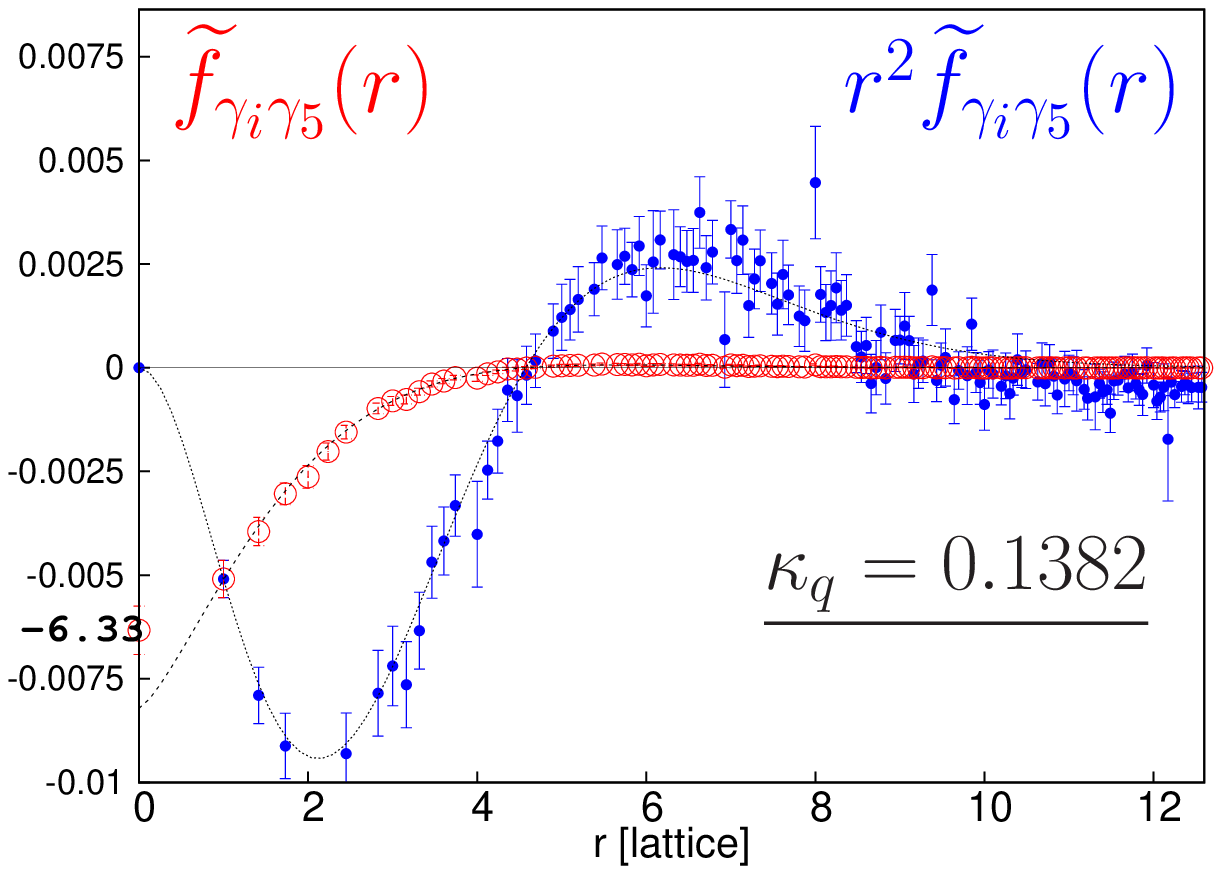}}\\
 \end{tabular}
\caption{\label{fig:A3}\footnotesize{\sl 
The same as in Fig.~\ref{fig:A1} but with the heavy-quark actions HYP-2$^2$.  }}

\end{figure}

\newpage 
\hspace{-1.5cm}
\begin{minipage}[!]{17cm}
\section*{Appendix~A}
In this appendix we present the plots for all distributions $f_\Gamma (r)$. Instead of plotting the results obtained with all 5 heavy-quark actions, we choose 3 of them: FAT-6 (Fig.~\ref{fig:A1}), HYP-1 (Fig.~\ref{fig:A2}) 
and HYP-2$^2$ (Fig.~\ref{fig:A3}). 
The reason is that the numerical results obtained with HYP-1$^2$ are very close to those obtained with HYP-2, and those with HYP-2$^2$ are better in quality than those with 
HYP-2 only  in the situations in which the external state is orbitally excited, i.e. $(1/2)^+$. Furthermore, to get a honest assessment of the quality of our results, we  illustrate the distributions for our largest ($\kappa_q=0.1357$) 
and our smallest  ($\kappa_q=0.1382$) light-quark mass. The plots are made for the distributions in lattice units and without accounting for the renormalization factors. To convert them to physical units, one should 
multiply the ``$x$" axis by $a= 0.0995(4)$~fm, and the ``$y$" axis by $1/a^3$ for $f_\Gamma(r)$, and by $1/a$ for $r^2 f_\Gamma(r)$. \\

When $f_\Gamma (r)$ is strictly positive, we fit its logarithm to a polynomial of degree four, namely $\log[f_\Gamma (r)]=c_0 + c_1 r + c_2 r^2 + c_3 r^3$. Instead, in the cases in which $f_\Gamma (r)$ changes the  sign,  
we fit to the forms listed below (the guiding principle is parsimony of parameters).
\setlength\tabcolsep{5pt}
\begin{longtable}{c|cc|cc}
$\kappa_q$
 &\multicolumn{2}{c|}{$\mathtt{S}\ \left(\frac{1}{2}\right)^+$}&\multicolumn{2}{c}{$\mathtt{A}\ \left(\frac{1}{2}\right)^+$}\\
\hline
0.1357
 &\multirow{2}*{$  ( c_0 + c_1 r^{2} )$}&\multirow{2}*{$\exp( -c_2 r^{\framebox{\scriptsize 1.5}} )$}&\multirow{2}*{$  ( c_0 + c_1 r^{3.5} )$}&\multirow{2}*{$\exp( -c_2 r^{\framebox{\scriptsize 1.6}} )$}\\
0.1374&&&&\\
\hline
0.1367
 &\multirow{2}*{$  ( c_0 + c_1 r^{2} )$}&\multirow{2}*{$\exp( -c_2 r^{\framebox{\scriptsize 1.25}} )$}&\multirow{2}*{$  ( c_0 + c_1 r^{3.5} )$}&\multirow{2}*{$\exp( -c_2 r^{\framebox{\scriptsize 1.35}} )$}\\
0.1382&&&&\\
\end{longtable}

\setlength\tabcolsep{1pt}
The coefficients from the fit within the interval $r\in [0,12]$ for HYP-2$^2$ are listed below, and the fitted curves are also displayed in figs.~\ref{fig:A1},~\ref{fig:A2} and~\ref{fig:A3}.
{\small
\begin{longtable}{c|r@{}lc|r@{}lc|r@{}lc|r@{}lc|r@{}lc|r@{}lc||c}
&\multicolumn{9}{c|}{$\left(\frac{1}{2}\right)^-$}&&\multicolumn{9}{c}{$\left(\frac{1}{2}\right)^+$}\\
\hline
$\mathtt{\kappa_q}$&
\multicolumn{3}{c}{$\mathtt{V}$}& \multicolumn{3}{c}{$\mathtt{S}$}& \multicolumn{3}{c|}{$\mathtt{A}$}&
\multicolumn{3}{c}{$\mathtt{V}$}& \multicolumn{3}{c}{$\mathtt{S}$}& \multicolumn{3}{c}{$\mathtt{A}$}\\[0.125ex]
\hline
\multirow{4}*{0.1357}
 &$   -4.70$&$   (2)$&$        $ &$   -4.36$&$   (3)$&$        $ &$   -4.72$&$  
(2)$&$        $ &$   -4.61$&$  (13)$&$        $ &$   -8.99$&$ (178)$&$ 10^{-3}$ &$ 
 -8.69$&$ (101)$&$ 10^{-3}$ & $c_0$\\
 &$   -1.81$&$  (19)$&$ 10^{-1}$ &$   -1.27$&$  (32)$&$ 10^{-1}$ &$   -2.06$&$ 
(21)$&$ 10^{-1}$ &$   -1.97$&$  (97)$&$ 10^{-1}$ &$   +2.21$&$  (43)$&$ 10^{-3}$ &$
  +1.61$&$  (26)$&$ 10^{-4}$ & $c_1$\\
 &$   -1.22$&$   (8)$&$ 10^{-1}$ &$   -2.03$&$  (19)$&$ 10^{-1}$ &$   -1.75$&$ 
(10)$&$ 10^{-1}$ &$   -1.04$&$  (50)$&$ 10^{-1}$ &$   +3.66$&$  (19)$&$ 10^{-1}$ &$
  +4.33$&$  (21)$&$ 10^{-1}$ & $c_2$\\
 &$   +9.07$&$ (111)$&$ 10^{-3}$ &$   +1.76$&$  (33)$&$ 10^{-2}$ &$   +1.68$&$ 
(15)$&$ 10^{-2}$ &$   +5.90$&$ (819)$&$ 10^{-3}$ &          &        &           & 
        &        &           & $c_3$\\
\hline
\multirow{4}*{0.1367}
 &$   -4.86$&$   (3)$&$        $ &$   -4.50$&$   (3)$&$        $ &$   -4.87$&$  
(3)$&$        $ &$   -4.75$&$  (13)$&$        $ &$   -8.09$&$ (173)$&$ 10^{-3}$ &$ 
 -7.51$&$ (110)$&$ 10^{-3}$ & $c_0$\\
 &$   -1.95$&$  (16)$&$ 10^{-1}$ &$   -1.81$&$  (34)$&$ 10^{-1}$ &$   -2.48$&$ 
(21)$&$ 10^{-1}$ &$   -2.94$&$ (120)$&$ 10^{-1}$ &$   +1.72$&$  (37)$&$ 10^{-3}$ &$
  +9.27$&$ (237)$&$ 10^{-5}$ & $c_1$\\
 &$   -9.92$&$  (62)$&$ 10^{-2}$ &$   -1.65$&$  (19)$&$ 10^{-1}$ &$   -1.39$&$  
(9)$&$ 10^{-1}$ &$   -7.14$&$ (511)$&$ 10^{-2}$ &$   +5.24$&$  (21)$&$ 10^{-1}$ &$ 
 +5.59$&$  (28)$&$ 10^{-1}$ & $c_2$\\
 &$   +7.01$&$  (86)$&$ 10^{-3}$ &$   +1.49$&$  (33)$&$ 10^{-2}$ &$   +1.28$&$ 
(12)$&$ 10^{-2}$ &$   +4.27$&$ (745)$&$ 10^{-3}$ &          &        &           & 
        &        &           & $c_3$\\
\hline
\multirow{4}*{0.1374}
 &$   -4.98$&$   (3)$&$        $ &$   -4.59$&$   (3)$&$        $ &$   -5.05$&$  
(2)$&$        $ &$   -4.82$&$   (9)$&$        $ &$   -6.46$&$ (139)$&$ 10^{-3}$ &$ 
 -6.07$&$  (64)$&$ 10^{-3}$ & $c_0$\\
 &$   -1.71$&$  (18)$&$ 10^{-1}$ &$   -1.33$&$  (34)$&$ 10^{-1}$ &$   -1.91$&$ 
(24)$&$ 10^{-1}$ &$   -3.34$&$ (943)$&$ 10^{-2}$ &$   +1.18$&$  (37)$&$ 10^{-3}$ &$
  +6.05$&$  (85)$&$ 10^{-5}$ & $c_1$\\
 &$   -9.51$&$  (79)$&$ 10^{-2}$ &$   -1.81$&$  (23)$&$ 10^{-1}$ &$   -1.55$&$ 
(12)$&$ 10^{-1}$ &$   -1.66$&$  (50)$&$ 10^{-1}$ &$   +3.21$&$  (41)$&$ 10^{-1}$ &$
  +3.31$&$  (15)$&$ 10^{-1}$ & $c_2$\\
 &$   +6.94$&$ (107)$&$ 10^{-3}$ &$   +1.80$&$  (44)$&$ 10^{-2}$ &$   +1.56$&$ 
(17)$&$ 10^{-2}$ &$   +1.46$&$  (83)$&$ 10^{-2}$ &          &        &           & 
        &        &           & $c_3$\\
\hline
\multirow{4}*{0.1382}
 &$   -5.16$&$   (4)$&$        $ &$   -4.77$&$   (4)$&$        $ &$   -5.25$&$  
(3)$&$        $ &$   -4.78$&$  (16)$&$        $ &$   -7.27$&$ (182)$&$ 10^{-3}$ &$ 
 -8.21$&$  (79)$&$ 10^{-3}$ & $c_0$\\
 &$   -2.07$&$  (28)$&$ 10^{-1}$ &$   -1.57$&$  (46)$&$ 10^{-1}$ &$   -2.37$&$ 
(25)$&$ 10^{-1}$ &$   -1.15$&$  (94)$&$ 10^{-1}$ &$   +9.69$&$ (244)$&$ 10^{-4}$ &$
  +4.01$&$  (63)$&$ 10^{-5}$ & $c_1$\\
 &$   -6.41$&$ (106)$&$ 10^{-2}$ &$   -1.40$&$  (27)$&$ 10^{-1}$ &$   -1.20$&$ 
(12)$&$ 10^{-1}$ &$   -1.57$&$  (47)$&$ 10^{-1}$ &$   +4.29$&$  (32)$&$ 10^{-1}$ &$
  +4.70$&$  (21)$&$ 10^{-1}$ & $c_2$\\
 &$   +3.53$&$ (139)$&$ 10^{-3}$ &$   +1.24$&$  (53)$&$ 10^{-2}$ &$   +1.12$&$ 
(19)$&$ 10^{-2}$ &$   +1.96$&$  (74)$&$ 10^{-2}$ &          &        &           & 
        &        &           & $c_3$\\
\hline
\end{longtable}
}

\end{minipage}

\setlength\tabcolsep{0pt}

\newpage

\begin{minipage}[!]{17cm}
\section*{Appendix~B}
In this appendix we give the explicit values for the distributions $f_\Gamma(r)$ and  $\widetilde f_\Gamma(r)$ which we call $S$, $V$ and $A$ for $\Gamma = \mathbb{I}$, $\gamma_0$ and $\gamma_i \gamma_5$  
respectively. We first present those obtained for the lowest heavy-light states, i.e. $j^P=(1/2)^-$, and then for the orbitally excited ones $j^P=(1/2)^+$. For each $r$ we give simultaneously the values for all the light-quarks 
($\kappa_q$) used in this work. We stress again that the light valence and the sea quarks are degenerate in mass. Finally, we need to explain how the tables are to be read. The first point (obtained at $r=0$) is to be multiplied by $10^{-3}$. When the shading changes an extra factor $10^{-1}$ should be inserted (e.g. for $\kappa_q=0.1357$, $f_{\mathbb{I}} (0)=12.6(3)\times 10^{-3}$,   $f_{\mathbb{I}} (16)=9.30(53)\times 10^{-4}$, 
$f_{\mathbb{I}} (35)=9.91(83)\times 10^{-5}$, $f_{\mathbb{I}} (68)=9.0(1.4)\times 10^{-6}$ and so on). The results listed here correspond to the heavy-quark action, which we called in the text HYP-2$^2$. Similar tables for other actions can be obtained from the authors. All results are given in lattice units.
\end{minipage}
\addtolength\hoffset{-45pt}
{
\setlength\tabcolsep{2.00pt}
\scriptsize
\begin{longtable}{r|r@{}lr@{}lr@{}l|r@{}lr@{}lr@{}l|r@{}lr@{}lr@{}l|r@{}lr@{}lr@{}l}
\multicolumn{2}{l}{$\left(\frac{1}{2}\right)^-$\ $\mathtt{\kappa_q}$}&\multicolumn{5}{c}{$\mathtt{0.1357}$}&\multicolumn{6}{c}{$\mathtt{0.1367}$}&\multicolumn{6}{c}{$\mathtt{0.1374}$}&\multicolumn{6}{c}{$\mathtt{0.1382}$}\\[0.125ex]
\hline
\multicolumn{1}{l|}{$\mathtt{r^2}$}&
\multicolumn{2}{c}{$\mathtt{S}$}& \multicolumn{2}{c}{$\mathtt{V}$}& \multicolumn{2}{c}{$\mathtt{A}$}&
\multicolumn{2}{c}{$\mathtt{S}$}& \multicolumn{2}{c}{$\mathtt{V}$}& \multicolumn{2}{c}{$\mathtt{A}$}&
\multicolumn{2}{c}{$\mathtt{S}$}& \multicolumn{2}{c}{$\mathtt{V}$}& \multicolumn{2}{c}{$\mathtt{A}$}&
\multicolumn{2}{c}{$\mathtt{S}$}& \multicolumn{2}{c}{$\mathtt{V}$}& \multicolumn{2}{c}{$\mathtt{A}$}\\[0.125ex]
\hline
   0  & \m12.6  &  (3)  & \m8.84  & (22)  & \m8.58  & (19)  & \m10.8  &  (3)  & \m7.51  & (18)  & \m7.24  & (17)  & \m9.98  & (30)  & \m6.69  & (18)  & \m6.24  & (15)  & \m8.25  & (36)  & \m5.51  & (22)  & \m5.07  & (15) \\
   1  & \m9.62  & (24)  & \m7.05  & (17)  & \m6.44  & (14)  & \m8.30  & (20)  & \m6.08  & (14)  & \m5.53  & (13)  & \m7.76  & (22)  & \m5.46  & (14)  & \m4.78  & (11)  & \m6.60  & (31)  & \m4.57  & (19)  & \m3.92  & (13) \\
   2  & \m7.43  & (19)  & \m5.71  & (14)  & \m4.92  & (10)  & \m6.50  & (17)  & \m5.00  & (12)  & \m4.29  & (10)  & \m6.12  & (18)  & \m4.54  & (12)  & \m3.74  &  (9)  & \m5.36  & (28)  & \m3.86  & (17)  & \m3.09  & (11) \\
   3  & \m5.81  & (16)  & \m4.70  & (12)  & \m3.84  &  (8)  & \m5.18  & (14)  & \m4.18  & (11)  & \m3.39  &  (8)  & \m4.92  & (16)  & \m3.85  & (11)  & \m2.99  &  (7)  & \m4.41  & (25)  & \m3.31  & (15)  & \m2.48  &  (9) \\
\end{longtable}
}
\newpage 
\addtolength\hoffset{45pt}

\end{document}